\documentclass[11pt,a4paper]{article}
\pdfoutput=1 
             
\usepackage{graphicx}
             
\usepackage[utf8x]{inputenc}	
\usepackage[english]{babel}
\usepackage{paralist}
\usepackage{multirow}
\usepackage{bbm}
\usepackage{amsmath}
\numberwithin{equation}{section}
\usepackage{amssymb}
\usepackage{amsthm}
\usepackage{MnSymbol}
\usepackage{mathtools}		
\usepackage{dsfont}		
\usepackage{stmaryrd}		
\usepackage{cite}		
\usepackage[svgnames]{xcolor}
\usepackage{enumerate}
\usepackage{setspace}
\usepackage{booktabs}
\usepackage{tikz}
\usetikzlibrary{shapes}
\usetikzlibrary{positioning}
\usetikzlibrary{chains}
\usetikzlibrary{arrows,fit,decorations.pathreplacing, shapes.misc}
\usetikzlibrary{decorations.markings}
\tikzstyle{every picture}+=[remember picture]
\tikzstyle{na} = [baseline=-.5ex]
\usepackage {todonotes}
\usepackage{tikz-3dplot}
\usepackage{pgfplots}
\usepackage[font=small,labelfont=bf]{caption}
\usepackage[font=small,labelfont=bf]{subcaption}
\usepackage[pdftex,breaklinks,colorlinks=true,urlcolor=RoyalBlue,linkcolor=blue,
citecolor=blue]{hyperref}

\def\beq{\begin{equation}}
\def\eeq{\end{equation}}
\def\beqn{\begin{eqnarray}}
\def\eeqn{\end{eqnarray}}

\def\={\!&=&}
\def\+{\!&+&}
\def\-{\!&-&}

\newcommand{\PE}{\mathrm{PE}}
\newcommand{\diff}{\mathrm{d}}
\newcommand{\im}{\mathrm{i}}

\newcommand{\Ncal}{\mathcal{N}}

\newcommand{\R}{\mathbb{R}}
\newcommand{\T}{\mathbb{T}}

\newcommand{\surm}{\mathrm{SU}}
\newcommand{\urm}{\mathrm{U}}

\newcommand{\orm}{\mathrm{O}}
\newcommand{\sorm}{\mathrm{SO}}
\newcommand{\sprm}{\mathrm{Sp}}

\newcommand{\surmL}{\mathfrak{su}}

\newcommand{\sormL}{\mathfrak{so}}

\newcommand{\str}{\mathrm{str}}
\newcommand{\pert}{\mathrm{pert}}

\newcommand{\sh}{\mathrm{sh}}
\newcommand{\ch}{\mathrm{ch}}

\newcommand{\am}{a_-}
\newcommand{\ap}{a_+}
\newcommand{\px}{\partial_x}
\newcommand{\be}{\begin{equation}}
\newcommand{\ee}{\end{equation}}
\newcommand{\ep}{\epsilon}
\def\QQ{q_\phi}

\newcommand{\thone}{\theta_1}
\def\ns#1{
	\node[circle, draw, fill=white,minimum size=0.6cm] at (#1){};
	\node[cross out, draw,minimum size=0.425cm] at (#1){};
}

\catcode`@=11
\allowdisplaybreaks

\begin{document}

\begin{titlepage}
\setcounter{page}{0}
\begin{flushright}KIAS-Q21002\vspace{10mm}\end{flushright}
\begin{center}

{\Large\bf 
E-string Quantum Curve
}

\vspace{15mm}

{\large Jin Chen${}^{1}$},\ 
{\large Babak Haghighat${}^{1}$},\
{\large Hee-Cheol Kim${}^{2,3}$},\ 
{\large Marcus Sperling${}^{1}$},\  and \ 
{\large Xin Wang${}^{4}$}
\\[5mm]
\noindent ${}^{1}${\em Yau Mathematical Sciences Center, Tsinghua University}\\
{\em Haidian District, Beijing, 100084, China}\\
{Email: {\tt jinchen@mail.tsinghua.edu.cn},} \\  
{{\tt babakhaghighat@tsinghua.edu.cn},} \\  
{{\tt msperling@mail.tsinghua.edu.cn }}
\\[5mm]
\noindent ${}^{2}${\em Department of Physics, POSTECH}\\
{\em  Pohang 790-784, Korea}\\
{Email: {\tt heecheol@postech.ac.kr}}
\\[5mm]
\noindent ${}^{3}${\em Asia Pacific Center for 
Theoretical Physics, POSTECH}\\
{\em Pohang 37673, Korea}
\\[5mm]
\noindent ${}^{4}${\em Quantum Universe Center, Korea Institute for Advanced Study}\\
{\em Seoul 02455, Korea}\\
{Email: {\tt wxin@kias.re.kr}} 
\\[5mm]
\vspace{15mm}

\begin{abstract}
In this work we study the quantisation of the Seiberg-Witten curve for the E-string theory compactified on a two-torus. We find that the resulting operator expression belongs to the class of elliptic quantum curves. It can be rephrased as an eigenvalue equation with eigenvectors corresponding to co-dimension 2 defect operators and eigenvalues to co-dimension 4 Wilson surfaces wrapping the elliptic curve, respectively. Moreover, the operator we find is a generalised version of the van Diejen operator arising in the study of elliptic integrable systems. Although the microscopic representation of the co-dimension 4 defect only furnishes an $\sorm(16)$ flavour symmetry in the UV, we find an enhancement in the IR to representations in terms of affine $E_8$ characters. Finally, using the Nekrasov-Shatashvili limit of the E-string BPS partition function, we give a path integral derivation of the quantum curve.
\end{abstract}

\end{center}

\end{titlepage}
{\baselineskip=12pt
{\footnotesize
\tableofcontents
}
}

%
\section{Introduction}
Quantum field theories in four dimensions with $\mathcal{N}=2$ supersymmetry belong to a special class of theories for which the vacua in the IR are fully characterised by a holomorphic function known as the prepotential. The prepotential serves to compute the effective gauge coupling of the theory and receives classical, one-loop, and instanton contributions. In particular, instanton contributions give rise to an infinite series and, in general, can only be computed by means of sophisticated techniques involving integrals over their moduli spaces \cite{Nekrasov:2002qd}. Before explicit instanton computations became tractable, another way to solve for the quantum vacua of the theory was presented in the seminal work of Seiberg and Witten \cite{Seiberg:1994rs} which provides a very elegant and geometric interpretation of the prepotential in terms of periods of an algebraic curve. In this framework, masses of BPS magnetic monopoles and dyons are expressed in terms linear combinations of IR gauge fugacities and suitable dual variables which in turn are computed in terms of integrals of a meromorphic one-form, denoted by $\lambda_{\textrm{SW}}$, over one-cycles of the algebraic Seiberg-Witten curve
\begin{equation}
    H(p,x) = 0, \quad \lambda_{\textrm{SW}} = p\, \diff x.
\end{equation}

It turns out that this picture can be lifted to string theory where the corresponding $\mathcal{N}=2$ supersymmetric theories are obtained through compactifications of string theory on Calabi-Yau 3-folds and where the algebraic Seiberg-Witten curve has a physical manifestation in the geometry of the mirror-dual of the Calabi-Yau \cite{Katz:1996fh}, where it is known as the mirror curve. It was later found that the Calabi-Yau framework provides a natural setting for the ``quantisation" of the Seiberg-Witten curve \cite{Aganagic:2003qj} where it was observed that branes wrapping special Lagrangian 3-cycles in the Calabi-Yau intersect the mirror curve at points and, hence, their classical phase space is given by the total space of the curve. One can quantise this classical phase space by defining the holomorphic symplectic two-form
\begin{equation}
    \omega = \diff \lambda_{\textrm{SW}} = \diff p \wedge \diff x \,,
\end{equation}
which gives rise to the canonical commutation relations
\begin{equation}
    [\hat{p},\hat{x}] = i \hbar\,.
\end{equation}
In this picture, the partition function of the brane becomes a quantum wave-function annihilated by the operator version of the mirror curve, denoted by $\widehat{H}$. From the point of view of the supersymmetric Yang-Mills theory, the Calabi-Yau compactification together with the Lagrangian brane gives rise to co-dimension $2$ defect operators in the gauge theory and the partition function of the brane encodes expectation values of such operators. Another breakthrough was provided by \cite{Nekrasov:2009rc} where the authors realised that the Hamiltonian interpretation of the Seiberg-Witten curve in the presence of defects arises in the Nekrasov-Shatashvili (NS) limit where one of the deformation parameters in the Nekrasov partition function is sent to zero. Connecting to this work, it was shown in \cite{Aganagic:2011mi} that the quantisation of mirror curves can be performed in the framework of topological string theory by using the WKB approximation which leads to ``quantum periods" admitting a series expansion in powers of $\hbar$. Several such quantum curves were subsequently studied in detail in the case of compactifications on toric Calabi-Yau manifolds \cite{Kashaev:2015kha}. In cases where the toric Calabi-Yau gives rise to a gauge theory, such quantum curves were further connected to surface defect partition functions in \cite{Gaiotto:2014ina}.

A particularly interesting class of quantum curves is the class of ``elliptic quantum curves" which arise in compactifications of 6d superconformal field theories (SCFTs) with $8$ supercharges on elliptic curves \cite{Chen:2020jla}. Compactifications of 6d SCFTs give rise to a wide plethora of four and five-dimensional quantum field theories where the 6d origin sheds light on many dualities between the lower dimensional theories \cite{Gaiotto:2009we,Benini:2009mz,Bah:2012dg,Gaiotto:2015usa,Razamat:2016dpl,Ohmori:2015pua,Ohmori:2015pia,Kim:2017toz,Bah:2017gph,Kim:2018bpg,Razamat:2018gro,Kim:2018lfo,Ohmori:2018ona,Razamat:2019mdt,Chen:2019njf,Pasquetti:2019hxf,Razamat:2019ukg,Razamat:2020bix,Hwang:2021xyw} and thus it is expected that elliptic quantum curves mirror these dualities. In particular, compactifications on an elliptic curve can be viewed as two successive circle compactifications where the intermediate 5d theory is of KK type and often admits a gauge theory interpretation \cite{Ganor:1996pc,Tachikawa:2011ch,Gaiotto:2015una,Bergman:2015dpa,Hayashi:2015fsa,Zafrir:2015rga,Hayashi:2015zka,Ohmori:2015tka,Yonekura:2015ksa,Kim:2015jba,Hayashi:2015vhy,Hayashi:2019yxj}. Hence one can once again ask the question whether expectation values/partition functions of defect operators play the role of eigenvalues of a quantum mirror curve. Moreover, 6d SCFTs admit a  classification in terms of an F-theory compactification on elliptic Calabi-Yau manifolds \cite{Heckman:2013pva,Heckman:2015bfa} and hence it is expected that the study of quantum curves in this context sheds further light on topological string theory as well as  (refined) Ooguri-Vafa invariants \cite{Ooguri:1999bv,Aganagic:2012hs}.

In this work we focus on the E-string theory \cite{Witten:1995gx,Ganor:1996mu,Seiberg:1996vs}, which belongs to the class of 6d SCFTs whose corresponding 5d KK theories admit a description in terms of (quiver) $\surm(2)$ gauge theories. For such theories, after a further circle compactification to four dimensions, one obtains a Seiberg-Witten curve whose moduli space is the one of a suitable $\surm(2)_{\mathbb{C}}$-bundle over the elliptic curve \cite{Donagi_1996,Nekrasov:2012xe}. In fact, in such a case the Seiberg-Witten curve itself can be constructed from the determinant section of such a bundle \cite{Haghighat:2017vch,Haghighat:2018dwe}. Restricting to the class of 5d KK theories with $\surm(2)$ gauge group, it is expected that the corresponding quantum curve acting on the defect wave-function $\widetilde\Psi$ takes the form
\begin{equation} \label{eq:introq-curve}
    \left(Y^{-1} + \mathcal{P}(x) Y\right) \widetilde{\Psi}(x) \sim \mathcal{W}^S_{\hbar}(x) \widetilde \Psi(x),
\end{equation}
where we have defined the shift operator $Y \equiv e^{-\hat p}$ and $\mathcal{P}(x)$ is a (meromorphic) Jacobi-form with elliptic argument $x$. Moreover, $\mathcal{W}^S_{\hbar}$ admits an interpretation as a Wilson surface expectation value \cite{Nekrasov:2013xda,Nekrasov:2015wsu,Kim:2016qqs,Kimura:2017auj,Agarwal:2018tso}. The above form of the quantum curve was confirmed for the case of two M5 branes probing a $\mathbb{C}^2/\mathbb{Z}_k$ singularity in \cite{Chen:2020jla} and can be motivated by dividing both sides of the equation by $\widetilde\Psi(x)$, leading to an equation of the form
\begin{equation} \label{eq:introiWeyl}
    \mathcal{Y}(x+\hbar)^{-1} + \mathcal{P}(x) \mathcal{Y}(x) \sim \mathcal{W}^S_{\hbar} \,,
\end{equation}
where the function $\mathcal{Y}(x)$ is specified by a suitable ratio of wave-functions. The left-hand side of \eqref{eq:introiWeyl} is invariant under ``iWeyl" reflections of $\surm(2)$ \cite{Nekrasov:2012xe} (see also \cite{Haouzi:2020yxy} for a nice exposition) and can be interpreted as the saddle-point in the NS-limit $\epsilon_2 \to 0$ of the tensor branch BPS partition function of the 6d SCFT on $\mathbb{T}^2 \times \mathbb{R}^4_{\epsilon_1,\epsilon_2}$ upon identification of $\epsilon_1$ with $\hbar$.

The Seiberg-Witten curve of the E-string theory was derived in \cite{Eguchi:2002fc}. However, the form of the curve presented in \cite{Eguchi:2002fc} is not directly suitable for quantisation as the canonical variables $x$ and $p$ (with $x$ being elliptic) do not appear explicitly. When restricting the $E_8$ flavour symmetry to $\sorm(16)$ and then further to the diagonal $\sorm(8)$ subgroup, the E-string SCFT can be viewed as conformal matter arising from an M5 brane probing a $D_4$ singularity \cite{DelZotto:2014hpa}. Using recent results about elliptic genera of strings in 6d minimal SCFTs \cite{Kim:2014dza,Haghighat:2014vxa,Kim:2015fxa}, the Seiberg-Witten curve for the $D_4$ conformal matter theory was computed by means of a saddle point analysis in the thermodynamic limit \cite{Haghighat:2018dwe}. As discussed in Section \ref{sec:quantum_curves}, the structure of the classical curve suggests a quantisation in terms of the van Diejen operator which has appeared in the study of elliptic integrable systems \cite{MR1275485,MR2153340,MR3313680,MR4120359}. In fact, the van Diejen operator has already appeared in the study of the E-string theory in \cite{Nazzal:2018brc}, let us therefore explain the connection to our work in the following. The co-dimension $2$ defect operators which are relevant in the study of the quantum curve arise in the E-string SCFT as extended over $\mathbb{T}^2 \times \mathbb{R}^2$ and localised at a point on the orthogonal $\mathbb{R}^2$. Viewing $\mathbb{R}^2$ as a decompactified two-sphere, in the NS-limit the space $\mathbb{T}^2 \times \mathbb{R}^2_{\epsilon_1}$ can be identified with $S^1 \times S^3$ by using the Hopf-fibration description of $S^3$ as a circle-bundle over the two-sphere. In the IR, the theory on the defect then flows to a 4d $\mathcal{N}=1$ SCFT whose superconformal index on $S^1 \times S^3$ can be identified with the defect partition function. Indeed it was shown in \cite{Nazzal:2018brc} that the superconformal index of a suitable E-string compactification preserving the diagonal $\sorm(8)$ flavour subgroup satisfies the van Diejen difference equation. In our work, we find that the defect partition function with unrestricted $E_8$ fugacities satisfies a generalised version of the van Diejen equation found in \cite{Nazzal:2018brc} and can indeed be brought into the form \eqref{eq:introq-curve}. We then prove this using the path integral formalism in the NS-limit.

The remainder of the paper is organised as follows: Section \ref{sec:E-string+partition_fct} reviews the partition function for the E-string theory and the Higgs mechanism within partition functions. Thereafter, a co-dimension 2 defect is introduced via Higgsing with a position dependent vacuum expectation value. Subsequently, the normalised defect partition function is evaluated up to order $q_\phi^2$. Co-dimension 4 defects, in form of Wilson surfaces in 6d and Wilson lines in 5d, are discussed in Section \ref{sec:Wilson_surface_line}, as these are relevant for the eigenvalues of the quantum curves. After these preliminary considerations, the detailed study of the E-string quantum curve is presented in Section \ref{sec:quantum_curves}. To begin with, the instructive M-string example is reviewed. Thereafter, the classical E-string curve is quantised. The resulting quantum curve can be rewritten in a form that allows an identification with the van Diejen difference equation. Lastly, the van Diejen difference equation is derived from a path integral formulation of the elliptic genera of the E-string theory in Section \ref{sec:path_integral}.
Appendix \ref{app:background} provides relevant notations and computational details. Appendix \ref{app:OV_invariant} presents Ooguri-Vafa invariants.


%
\section{E-string: Partition function and defects via Higgsing}
\label{sec:E-string+partition_fct}
The description of $6$d $\Ncal=(1,0)$ SCFTs on their tensor branch relies on three types of supermultiplets: tensor, hyper, and vector multiplets. Each tensor multiplet admits self-dual strings, with tension proportional to the vacuum expectation value of the real scalar field in the tensor multiplet. One of the simplest $6$d $\Ncal=(1,0)$ theories is engineered in M-theory from an M5 brane approaching an M9 plane, leading to the well-known small $E_8$ instanton \cite{Ganor:1996mu,Seiberg:1996vs,Intriligator:1997kq,Blum:1997mm}. The self-dual strings are realised by M2 branes connecting the M5 and the M9. Since the M9 wall supports an $E_8$ global symmetry, the strings are conventionally called E-strings and have been intensively studied \cite{Klemm:1996hh,Minahan:1998vr,Eguchi:2002fc,Eguchi:2002nx,Kim:2014dza,Kim:2015fxa,DelZotto:2018tcj,Gu:2019pqj}. 

For the considerations of this paper, it is convenient to use the dual brane realisation in Type IIA string theory of M5 branes near an M9 plane on a A-type singularity \cite{Hanany:1997gh,Brunner:1997gk,Hanany:1997sa}.
This becomes useful as it embeds the E-string theory into the family of $\sprm(n)$ gauge theories with $N_f=2n+8$ flavours, where the theories are related via the Higgs mechanism, see for instance \cite{DelZotto:2018tcj,Bourget:2019aer}. In particular, the Higgsing 
\begin{align}
 \text{ 6d $\sprm(1)$ theory with $10$ flavours} \quad \longrightarrow \quad
\text{E-string theory}
\label{eq:E-string_Higgsing}
\end{align}
is an essential step for analysing the E-string theory with co-dimension 2 defects.

\begin{figure}
\centering
\begin{subfigure}{0.4\textwidth}
\begin{tikzpicture}
    \draw[thick,dashed] (2,1.5)--(2,-1.5);
    \foreach \i in {1,...,8}
    {
    \draw (2-0.15*\i,1.5)--(2-0.15*\i,-1.5);
    }
    \draw (-2,0.1)--(2,0.1);
    \draw (-2,0.05)--(2,0.05);
    \node at (0,-0.05) {$\cdots$};
    \node at (-1.75,-0.05) {$\cdots$};
    \draw (-2,-0.1)--(2,-0.1);
    \draw[red,thick] (-1,0.2)--(2,0.2);
    \ns{-1,0};
    \draw[->] (3,0)--(4,0);
    \draw[->] (3,0)--(3,1);
    \node at (4.2,0.2) {$\scriptstyle{x^6}$};
    \node at (3.2,1.2) {$\scriptstyle{x^{7,8,9}}$};
    \node at (-1,-0.5) {\small{NS5}};
    \node at (-1.75,0.5) {\small{$2n$ D6}};
    \node at (0.25,0.5) {\small{\textcolor{red}{$k$ D2}}};
    \draw[decoration={brace,mirror,raise=30pt},decorate,thick](0.55,-0.5) -- node[below=30pt] {\small{$8$ D8 + O$8^-$}  } (2.15,-0.5);
    \end{tikzpicture}
    \caption{}
\end{subfigure}
\hfill
\begin{subfigure}{0.5\textwidth}
\begin{tabular}{c|cc|cccc|c|cccc}
 \toprule 
\multirow{2}{*}{IIA}  & \multicolumn{2}{c|}{$\T^2$} & 
\multicolumn{4}{c|}{$\R^4_{\epsilon_1,\epsilon_2}$} & & \\
  & 0 & 1 & 2 & 3& 4 & 5 & 6  & 7 & 8 & 9 \\ \midrule
 NS5 & $\bullet$ & $\bullet$  & $\bullet$ & $\bullet$& $\bullet$ & 
$\bullet$&  \\
 D6 & $\bullet$ & $\bullet$ & $\bullet$ & $\bullet$ & $\bullet$ & 
$\bullet$ & $\bullet$
 \\
 D8/O$8^-$ & $\bullet$ & $\bullet$ & $\bullet$ & $\bullet$ & $\bullet$ & 
$\bullet$ & & $\bullet$ & $\bullet$ & $\bullet$
 \\ \midrule
 D2 & $\bullet$ & $\bullet$ & & & & & $\bullet$ & \\
  \bottomrule 
\end{tabular}
\caption{}
\end{subfigure}
\caption{Type IIA brane setup comprised of D6-D8-NS5 branes in the presence of an orientifold 8-plane leads to a 6d $\Ncal=(1,0)$ $\sprm(n)$ gauge theory with $2n+8$ fundamental flavours and one tensor multiplet. In the limit $n\to0$, the E-string theory is recovered. The addition of D2 branes captures the dynamics of the self-dual strings.}
\label{fig:branes}
\end{figure}
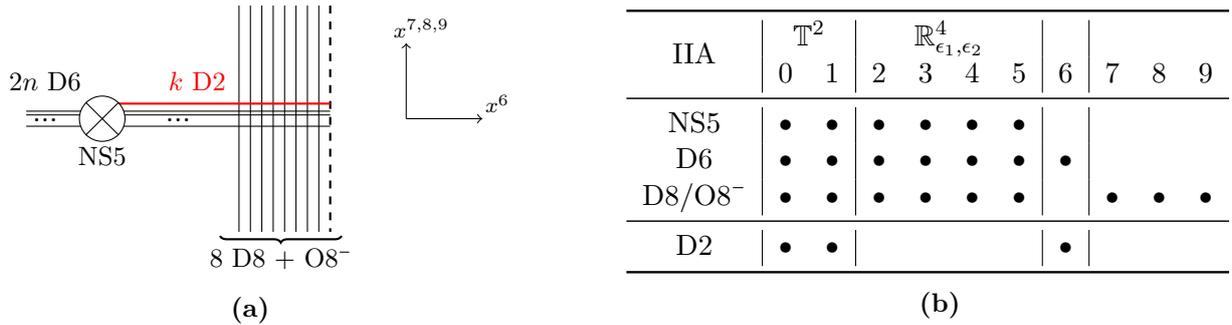

Consider a single M5 at some finite distance away from the M9 on an $A_{2n-1}$ singularity, which is dual to the Type IIA configuration displayed in Figure \ref{fig:branes}. The brane configuration gives rise to the following tensor branch description
\begin{align}
    \raisebox{-.5\height}{
    \begin{tikzpicture}
	\tikzstyle{gauge} = [circle, draw,inner sep=3pt];
	\tikzstyle{flavour} = [regular polygon,regular polygon sides=4,inner 
sep=3pt, draw];
	\node (g1) [gauge,label=below:{$\sprm(n)$}] {};
	\node (f1) [flavour,above of=g1, label=above:{$\sorm(4n+16)$}] {};
	\draw (g1)--(f1);
	\end{tikzpicture}
    } \quad + \qquad \text{1 tensor multiplet}
    \label{eq:6d_quiver}
\end{align}
such that the $n\to 0$ limit formally leads to $\sprm(0)$ with $8$ fundamentals, i.e.\ the E-string theory.
%
\subsection{Partition function}
The partition function is evaluated by placing the E-string theory on a $\T^2\times \R^4_{\epsilon_1,\epsilon_2}$ background. The full partition function is then composed of two contributions
\begin{align}
    Z_{6d} = Z_{\pert} \cdot Z_{\str}
\end{align}
wherein $Z_{\pert}$ denotes the perturbative contributions and $Z_{\str}$ accounts for all non-perturbative contributions.
The 6d supermultiplets, which specify the description of the 6d theory on the tensor branch, completely determine $Z_{\pert}$. On the contrary, the non-perturbative part stems from 2d supersymmetric world-volume theories of additional D$2$ branes. The resulting instanton partition function can be expressed as sum over elliptic genera
\begin{align}
    Z_{\str} = 1 + \sum_{k=1}^\infty  q_\phi^k \ Z_k 
    \qquad \text{with} \qquad
    q_\phi = e^{\phi}
\end{align}
where $\phi$ denotes the vacuum expectation value of the scalar field in the tensor multiplet.
\subsubsection{Perturbative contributions}
Consider the perturbative contribution of the $6$d SUSY multiplets of a  $\surm(2)\cong \sprm(1)$ gauge theory with $N_f=10$ fundamental flavours and one tensor multiplet, i.e.\ the $n=1$ case of \eqref{eq:6d_quiver}.
Following \cite{Hayashi:2016abm}, the perturbative single-letter contribution of 
the 6d supermultiplets are given as follows:
\begin{subequations}
\begin{align}
 I_{\mathrm{tensor}} &= 
 -\frac{p+q}{(1-p)(1-q)}
\\
 I_{\mathrm{vector}} &=
 -\frac{(1+p\cdot q)}{(1-p)(1-q)}\,  \chi_{[2]_C}
 \\
  I_{\mathrm{hyper}} &= \frac{\sqrt{p\cdot q}}{(1-p)(1-q)} \,
  \chi_{[1]_C} \cdot \chi_{[1,0,\ldots,0]_D}
\end{align}
\end{subequations}
with $p=e^{\epsilon_1}$, $q=e^{\epsilon_2}$, which are the Cartan generators of the rotations on $\R^4_{\epsilon_1,\epsilon_2}$. Moreover, the appearing characters of $C_1 \cong A_1$ and $D_{10}\cong \sormL(20)$ are given by
\begin{subequations}
\begin{align}
\chi_{[2]_C} &=e^{2\alpha} + 1 + e^{-2\alpha}  = A^2 + 1 + A^{-2}\,, \\
\chi_{[1]_C} &= e^{\alpha} +  e^{-\alpha} = A + A^{-1} \,,\\
\chi_{[1,0,\ldots,0]_D}  &=\sum_{l=1}^{10} \left( e^{\mu_l} +  e^{ -\mu_l} \right)
= \sum_{l=1}^{10} \left( M_l +  M_l^{-1} \right)\,.
\end{align}
\end{subequations}
Here the roots of $\surmL(2)$ are $\pm 2 \alpha$ such that the fundamental weight is simply $\alpha$. For $\sormL(20)$, the weight fugacities are $\mu_l$, $l=1,\ldots,10$.  Lastly, one defines the exponentiated variables $A=e^\alpha$ and $M_l=e^{\mu_l}$.
Then
\begin{align}
 Z_{\mathrm{pert}} = 
\PE \left[ \left( I_{\mathrm{tensor}}  + I_{\mathrm{vector}}  + 
I_{\mathrm{hyper}}\right) \cdot \left( \frac{Q}{1-Q} +\frac{1}{2} \right)\right]
\end{align}
with $Q=e^{2\pi i \tau}$, where $\tau$ denotes the complex structure modulus of $\T^2$.
Then, to summarise
\begin{align}
\begin{aligned}
  Z_{\mathrm{pert}} = 
  \PE \Bigg[
  \frac{1}{(1-p)(1-q)} \left( \frac{Q}{1-Q} +\frac{1}{2} \right)
  \bigg\{
  &-(p+q) - (1+pq) \left( A^2 + 1 + A^{-2}  \right)  \\
  &+ \sqrt{pq}  (A + A^{-1}) \sum_{l=1}^{10} \left( M_l +  
M^{ -1}_l \right)
  \bigg\}
  \Bigg]
  \end{aligned}
\label{eq:6d_perturbative_partition_function}
\end{align}

\paragraph{Comparision to 5d.}
For later convenience in any comparison to 5d instanton partition functions, one may apply a flop transition \cite{Hayashi:2016abm} 
\begin{align}
    A^{-1} \longrightarrow A \,,
\end{align}
for all $A^{-1}$ associated to the ``$\frac{1}{2}$" terms in \eqref{eq:6d_perturbative_partition_function}. It corresponds to assigning $\frac{1}{2}$-BPS boundary conditions to the 5d theory at infinity. After the flop transition, the 6d perturbative partition function turns out to be
\begin{align}
    Z_{\mathrm{pert}}^{\sprm(1)} &= 
  \PE \left[-\frac{(1+pq)\left(A^2+A^{-2}Q\right) }{(1-p)(1-q)(1-Q)}+\frac{\sqrt{pq} \left(A+A^{-1}Q\right)}{(1-p)(1-q)(1-Q)}\sum_{l=1}^{10}\left(M_l+M_l^{-1}\right)\right]\notag\\
  &\cdot\PE\left[-\frac{p+q}{(1-p)(1-q)}\left(\frac{Q}{1-Q}+\frac{1}{2}\right) \right]
  \cdot\PE\left[-\frac{1+pq}{(1-p)(1-q)}\left(\frac{Q}{1-Q}+\frac{1}{2}\right)\right]\,,
  \label{eq:6d_perturbative_partition_function_flopped}
\end{align}
where the contributions from the tensor and the Cartan piece of the vector multiplet have been separated into the second line.

%
%
\subsubsection{Elliptic genus}
\label{sec:partition_function_elliptic_genus}
The non-perturbative contributions stem from the self-dual strings, which are due to D2 branes in the brane configuration, see Figure \ref{fig:branes}. The 2d $\Ncal=(0,4)$ world-volume theory on a stack of $k$ D2 branes has been studied in \cite{Kim:2014dza,Kim:2015fxa} and can be summarised as 
\begin{align}
    \raisebox{-.5\height}{
    \begin{tikzpicture}
	\tikzstyle{gauge} = [circle, draw,inner sep=3pt];
	\tikzstyle{flavour} = [regular polygon,regular polygon sides=4,inner sep=3pt, draw];
	\node (g1) [gauge,label=below:{$\orm(k)$}] {};
	\node (f1) [flavour,above of=g1, label=above:{$\sprm(n)$}] {};
	\node (f2) [flavour,right of=g1, label=right:{$\sorm(2N_f)$}] {};
	\draw (g1)--(f1);
	\draw[dashed] (g1)--(f2);
	\draw (g1) to [out=140,in=220,looseness=10] (g1);
	\node at (-1,0) {\small{sym}};
	\end{tikzpicture}
    } 
    \quad \text{with } N_f=2n+8
    \label{eq:2d_quiver}
\end{align}
and solid/dashed lines denote hypermultiplets/Fermi multiplets, respectively. 

Once the D2 world-volume theory is known, one evaluates the partition function of the 2d $\Ncal=(0,4) $ theory on a torus $\T^2$ with complex structure $\tau$. In other words, the elliptic genus. The prescription for the elliptic genera via supersymmetric localisation has been presented in \cite{Benini:2013nda,Benini:2013xpa}. To begin with, the compact zero modes need to be identified. These originate from $\orm(k)$ flat connections on $\T^2$ and have been detailed in \cite{Kim:2014dza}.
In brief, the flat connections are parametrised by two commuting group elements, which are the Wilson lines along the two circles of $\T^2$.
As $\orm(k)$ is a disconnected group, disconnected sectors for the Wilson lines arise. 
\begin{compactitem}
\item  For $k=2p\geq 4$, there are at most $p$ complex moduli $u_i$ and seven disconnected sectors.
\item $\orm(2)$ has seven sectors: One with continuous complex moduli and six with discrete holonomies.
\item For $k=2p+1\geq 3$ there are at most $p$ complex moduli $u_i$ and in total eight sectors.
\item For $\orm(1)$, there are no continuous moduli, but four sectors with discrete Wilson lines.
\end{compactitem}
The reader is referred to \cite{Kim:2014dza} for the derivation. The $\orm(1)$ and $\orm(2)$ cases are summarised to some extend in Appendix \ref{app:elliptic_genus_E-string}.

The field contents of the 2d GLSM are determined by the degrees of freedom of open strings attached to various branes and their boundary conditions. The D2-D2 modes, in presence of the O8-plane, give rise to the $(0, 4)$ vector and hyper multiplets in the adjoint and symmetric reps. of the 2d $\orm(k)$ gauge group; the D2-D8-O8 modes provides left-moving Majorana-Weyl fermions furnishing the bi-fundamental rep. of $\orm(k) \times \sorm(4n+16)$; at last, the D2-D6-O8 modes provide the (0, 4) half hypers in the bi-fundamental rep. of $\orm(k) \times \sprm(n)$. For more details we refer the readers to \cite{Kim:2014dza}. Keeping the background of zero-modes fixed, the Gaussian path-integral over the massive non-zero modes yields the following 1-loop determinants for the various 2d supermultiplets \cite{Benini:2013xpa}
\begin{subequations}
\begin{align}
    Z_{\mathrm{vec}} &= 
    \prod_{i=1}^r \left( \frac{2\pi \eta^2 \diff u_i}{i} \frac{\theta_1(\epsilon_1 + \epsilon_2)}{i \eta} \right) 
    \prod_{e \in \mathrm{roots}} 
    \frac{\theta_1(e(u)) \theta_1(\epsilon_1 +\epsilon_2 + e(u))}{i^2 \eta^2} \,, \\
    Z_{\sorm(2N_f)\, \mathrm{Fermi}} &=
    \prod_{\rho \in \mathrm{fund}} \prod_{l=1}^{N_f} 
    \frac{\theta_1(\mu_l + \rho(u))}{i \eta} \,,\\
    Z_{\sprm(n)\, \mathrm{hyper}} &=
    \prod_{\rho \in \mathrm{fund}} \prod_{j=1}^{n} 
    \frac{i^2 \eta^2}{\theta_1(\epsilon_1+  \rho(u) \pm \alpha_j)} \,,\\
    Z_{\text{sym hyper}} &=
    \prod_{\rho \in \mathrm{sym}}  
    \frac{i \eta}{\theta_1(\epsilon_1+  \rho(u) )}
    \frac{i \eta}{\theta_1(\epsilon_2+  \rho(u) )} \,,
\end{align}
\end{subequations}
where $\epsilon_1$ and $\epsilon_2$ denote the rotations in the plane $x_{23}$ and $x_{45}$ planes respectively, and $\epsilon_{\pm}=(\epsilon_1\pm \epsilon_2)/2$, corresponding to the fugacities of $\sorm(4)\simeq\surm(2)_{L}\times \surm(2)_R$ symmetry of the plane $x_{2345}$. Besides, the brane set-ups also admit a $\sorm(3)\simeq \surm(2)_{R'}$ symmetry of the plane $x_{789}$. In terms of the 2d $(0, 4)$ GLSM living on the D2 branes, $\surm(2)_R \times \surm(2)_R'$ form the SO(4) R-symmetries. The ``roots" denotes the $\sorm(k)$ roots and ``fund" and ``sym" denote representations of $\sorm(k)$. The appearing functions $\eta\equiv \eta(\tau)$, $\theta_i(z)\equiv \theta_i(\tau|z)$ are defined in Appendix \ref{app:theta_fct}.
The ``rank" $r$ equals the number of continuous complex parameters $u_i$.
Since $\orm(k)$ is a disconnected group, the integration of one-loop determinants over the compact zero-modes needs to proceed with great care.  As a result\cite{Kim:2014dza}, the elliptic genus is computed by multiplying all 1-loop determinants for a given discrete sector, then integrating over the continuous parameter $u$, and, finally, summing over all disconnected sectors of the flat connections. In detail
\begin{align}
    \sum_{I} \frac{1}{|W_I|} \frac{1}{(2\pi i)^r} \oint Z_{\mathrm{1-loop}}^{(I)}
    \;,\qquad 
    Z_{\mathrm{1-loop}}^{(I)} 
    \coloneqq
     Z_{\mathrm{vec}}^{(I)} \cdot
      Z_{\sorm(2N_f)\, \mathrm{Fermi}}^{(I)} \cdot
       Z_{\sprm(n)\, \mathrm{hyper}}^{(I)} \cdot
        Z_{\text{sym hyper}}^{(I)}
\end{align}
and the appearing contour integrals are evaluated via the Jeffrey-Kirwan residue prescription \cite{Jeffrey1995}.

In view of \eqref{eq:E-string_Higgsing}, it is sufficient to consider the $2$d $\Ncal=(0,4)$ gauge theory for the self-dual strings of the 6d $\sprm(1)$ theory with $N_f=10$ and the E-string theory itself. To illustrate the above generic expressions, it is instructive to detail the one-loop determinant for the elliptic genera of the 6d $\sprm(1)$ theory and the E-string theory. One finds
\begin{align}
 Z_{\mathrm{1-loop}}^{\sprm(1),N_f} &=
 \prod_{i=1}^k (2\pi \diff u_i \thone(2\epsilon_+) )
 \prod_{e\in \mathrm{root}} \frac{\thone(e(u)) \thone(2\epsilon_+ +e(u))  }{\eta^2}
 \cdot 
 \prod_{\rho\in \mathrm{sym}} \frac{\eta^2}{\thone(\epsilon_{1,2}+\rho(u) )}
 \label{eq:elliptic_genus_decompose}\\
 &\phantom{\prod_{i=1}^k (2\pi \diff u_i \thone(2\epsilon_+) )}
 \cdot
 \prod_{\rho\in \mathrm{fund}} \left( \frac{\eta^2}{\thone(\epsilon_+ +\rho(u) \pm
\alpha)}  \prod_{l=1}^{10} \frac{\thone(\mu_l +\rho(u))}{\eta} \right)  \,,
\notag
\\
 Z_{\mathrm{1-loop}}^{\mathrm{E-str}} &=\prod_{i=1}^k (2\pi \diff u_i \thone(2\epsilon_+) )
 \prod_{e\in \mathrm{root}} \frac{\thone(e(u)) \thone(2\epsilon_+ +e(u))  }{\eta^2}
 \cdot \prod_{\rho\in \mathrm{sym}} \frac{\eta^2}{\thone(\epsilon_{1,2}+\rho(u) )} 
 \label{eq:E-string_1-loop_det}\\
 &\phantom{\prod_{i=1}^k (2\pi \diff u_i \thone(2\epsilon_+) ) }
 \cdot
 \prod_{\rho\in \mathrm{fund}}  \prod_{l=1}^{8} \frac{\thone(\mu_l +\rho(u))}{\eta}  \,.
 \notag
\end{align}
\paragraph{Example.}
The $k=1$ elliptic genus for the E-string theory has been computed in \cite{Kim:2014dza} 
and reads 
\begin{align}
 Z_{k=1}^{\mathrm{E-str}} = 
 -\frac{\Theta(\tau,\mu_l)}{\eta^6 \theta_1 (\epsilon_1) \theta_1 (\epsilon_2)}
 \quad \text{with} \quad 
  \Theta(\tau,\mu_l) = \frac{1}{2}\sum_{n=1}^4 \prod_{l=1}^8 
\theta_n(\tau,\mu_l) \,.
\label{eq:ell_genus_E-string_k=1}
\end{align}
For convenience, the details of the derivation is summarised in Appendix \ref{app:elliptic_genus_E-string}.
The $k=2$ elliptic genus faces the subtlety that there are seven sectors of $\orm(2)$ Wilson lines to take into account, as alluded above. The reader is referred to Appendix \ref{app:elliptic_genus_E-string} for a brief revision of the calculation.
%
%
\subsection{Higgs mechanism in partition functions}
To begin with, consider 6d $\surm(2)\cong\sprm(1)$ with $10$ fundamental flavours, such that 
the hypermultiplet may be written as
\begin{align}
 \text{hyper}=\left(Y_\alpha^m,(\tilde{Y}_n^\beta )^\dagger \right) 
\quad \text{with}\quad
 Y_\alpha^m \in \mathbf{2} \otimes \overline{\mathbf{10}} \;,\quad 
 \tilde{Y}_n^\beta \in \overline{\mathbf{2}} \otimes \mathbf{10}
\;.
\end{align}
However, the 10 hypermultiplets of $\surm(2)$, which at first glance have a $\urm(10)$ flavour symmetry, can be arranged into 20 half-hypermultiplets which transform in the fundamental representation of an enhanced $\sorm(20)$ flavour symmetry. The half-hypermultiplets are defined as 
\begin{align}
 X_{\alpha}^I = (Y_\alpha^m, J_{\alpha \beta} \tilde{Y}_{m}^{\beta})
 \quad \text{such that}\quad
 \{I\}\equiv(\{m\},\{m+10\})\,,
\end{align}
and $J_{\alpha \beta}= \epsilon_{\alpha \beta}$ is the invariant 
$\surm(2)\cong \sprm(1)$ anti-symmetric tensor.
The mesonic gauge invariant operator composed of the $X_\alpha^I$ can be written as 
\begin{align}
 \mathcal{M}^{[IJ]} = X_{\alpha}^I X_{\beta}^J J^{\alpha \beta} 
 = \begin{pmatrix}
    Y_{\alpha}^{m} J^{\alpha \beta} Y_{\beta}^n & - Y_\alpha^m 
\tilde{Y}_n^{\alpha} \\
     \tilde{Y}_m^\alpha Y_{\alpha}^n &  \tilde{Y}_m^\alpha J_{\alpha \beta} 
\tilde{Y}_n^\beta
   \end{pmatrix}
   \,.
\end{align}
If, for instance, $\langle \mathcal{M}^{[9,10]} \rangle\neq 0$, which breaks the 
6d gauge group completely, then the global symmetry
charges of this operator are given by
\begin{align}
  \mathcal{M}^{[9,10]} : \qquad 
  \sqrt{pq} e^{-m_9} \cdot \sqrt{pq} e^{-m_{10}} = pq \cdot e^{-\mu_9 
-\mu_{10}} \;,
  \label{eq:VEV_meson}
\end{align}
because each hypermultiplet has charges $(\tfrac{1}{2},\tfrac{1}{2})$ under 
$\sorm(4)\cong \surm(2) \times \surm(2)$. Moreover, note that the relevant hypermultiplets need 
to have opposite $\surm(2)$ gauge fugacity; therefore, there are at least two consistent VEVs for the hypermultiplets
\begin{align}
 \langle X_{1}^{9}\rangle\neq 0\, , \; \langle X_{2}^{10}\rangle \neq 0
 \quad \text{or} \quad 
 \langle X_{2}^{9}\rangle \neq 0\, , \;  \langle X_{1}^{10}\rangle \neq 0 
 \,.
 \label{eq:VEV_hypers}
\end{align}
Following \cite{Gaiotto:2012xa,Gaiotto:2014ina}, the 
Higgsing \eqref{eq:VEV_meson} implies a condition on the fugacities
\begin{align}
 pq \cdot e^{-\mu_9 -\mu_{10}} =1 \,,
 \label{eq:pole_Higgs}
\end{align}
which realises the Higgsing on the level of the partition function as discussed in the next subsections.

\subsubsection{Perturbative contributions}
At the level of the partition function, imposing \eqref{eq:pole_Higgs} is expected to reduce the $\sprm(1)$ partition function to the E-string theory partition function up to Goldstone boson contributions, which are due to the breaking of flavour symmetry from $\sorm(20)$ to $\sorm(16)$. 
As \eqref{eq:pole_Higgs} implies only one condition on three fugacities, one may parametrise this condition in a convenient fashion to ease computations; for instance, 
\begin{align}
    \begin{cases}
    \alpha&=\mu \,,\\
    \mu_9&=\mu+\epsilon_+\,, \\
    \mu_{10}&=-\mu+\epsilon_+ \,,
    \end{cases}
\label{eq:normal_higgsing_equation}
\end{align}
where $\mu$ is a free variable.
Next, imposing \eqref{eq:normal_higgsing_equation} to perturbative contributions \eqref{eq:6d_perturbative_partition_function_flopped} leads to
\begin{align}
    Z_{\mathrm{pert}}^{\sprm(1)}
  %
  &=\PE\left[\frac{\sqrt{pq}\left(M+M^{-1}Q\right)}{(1-p)(1-q)(1-Q)}\sum_{i=1}^8\left(M_i+M_i^{-1}\right)\right]\cdot\PE\left[\frac{1+pq}{(1-p)(1-q)}\left(\frac{Q}{1-Q}+\frac{1}{2}\right)\right]\notag\\
  &\qquad \cdot\PE\left[-\frac{p+q}{(1-p)(1-q)}\left(\frac{Q}{1-Q}+\frac{1}{2}\right)\right]\,,
\end{align}
where $M\equiv e^\mu$ corresponds to the fugacity of Goldstone boson modes. Therefore, one identifies
\begin{align}
    Z_{\mathrm{GB}}=\PE\left[\frac{\sqrt{pq}\left(M+M^{-1}Q\right)}{(1-p)(1-q)(1-Q)}\sum_{i=1}^8\left(M_i+M_i^{-1}\right)\right]\cdot\PE\left[\frac{1+pq}{(1-p)(1-q)}\left(\frac{Q}{1-Q}+\frac{1}{2}\right)\right]
\label{eq:Goldstone_part}
\end{align}
as the Goldstone boson contribution. 
On the other hand, consider the Higgsing 
\eqref{eq:E-string_Higgsing} where the tensor branch description of the E-string theory is an empty gauge theory with no matter content. The Higgs branch of the $\sprm(1)$ gauge theory with 10 flavours is known to be the closure of the minimal nilpotent orbit of $D_{10}\cong\sormL(20)$, which has quaternionic dimension $2 \cdot 10 -3 =17$. Since there are no anomaly-free $\urm(1)$ gauge groups in 6d, any consistent assignment of VEVs breaks the $\sprm(1)$ gauge group fully. Hence, there is a space of VEVs of complex dimension $34$ along which the $\sprm(1)$ theory is higgsed to the tensor branch description of the E-string theory, see for instance \cite{Bourget:2019aer}.
Consequently, the appearing $2+2\cdot(8+8)$ massless chiral multiplets in \eqref{eq:Goldstone_part} parametrise the closure of the $\sormL(20)$ minimal nilpotent orbit.

Consequently, removing the free fields from the Goldstone modes $Z_{\mathrm{GB}}$ leads to the result
\begin{align}
Z_{\mathrm{pert}}^{\sprm(1)} \xrightarrow{\;\eqref{eq:normal_higgsing_equation}\;}  Z_{\mathrm{pert}}^{\mathrm{E-str}} \cdot  Z_{\mathrm{GB}}
\quad 
\text{with}
\quad 
  Z_{\mathrm{pert}}^{\mathrm{E-str}}
  &=\PE \left[
  \frac{ -(p +q) }{(1-p)(1-q)} \left( \frac{Q}{1-Q} +\frac{1}{2} \right)
   \right] \,.
  \label{eq:E-string_part-fct}
\end{align}
One recognises \eqref{eq:E-string_part-fct} as the contribution of a single 
tensor multiplet, which is consistent since the only 6d supermultiplet in the 
E-string theory is the tensor multiplet.
\subsubsection{Elliptic genus}
\label{sec:NH_elliptic_genus_Sp1}
Consider a Higgsing process towards the E-string theory. As a naive check, one may inspect the relation of the one-loop determinants  
\begin{align}
  Z_{\mathrm{1-loop}}^{\sprm(1)} =
   Z_{\mathrm{1-loop}}^{\mathrm{E-str}}
  \cdot \prod_{\rho\in \mathrm{fund}}  \frac{ \thone(\mu_9 +\rho(u)) \thone(\mu_{10} +\rho(u)) }{ \thone(\epsilon_+ +\rho(u) +\alpha) \thone(\epsilon_+ +\rho(u) -\alpha) }
\end{align}
and for a consistent Higgsing, one would impose 
\begin{align}
 \forall \rho \in \mathrm{fund}: \qquad   \frac{ \thone(\mu_9 +\rho(u)) \thone(\mu_{10} +\rho(u)) }{ \thone(\epsilon_+ +\rho(u) +\alpha) \thone(\epsilon_+ +\rho(u) -\alpha) } =1
\label{eq:mass deformation}
\end{align}
which can be realised for two choices
\begin{alignat}{6}
 &(A): & \qquad 
 \mu_9 &= \epsilon_+ + \alpha \;, & \quad \mu_{10} &= \epsilon_+ - \alpha &
 \quad 
 &\Leftrightarrow &
 \quad 
 \alpha&=\mu_9- \epsilon_+ \;, & \quad \mu_{10} &= 2\epsilon_+ -\mu_9
  \label{eq:Higgs_no_defect}
 \\
 &(B): & \qquad 
 \mu_9 &= \epsilon_+ -\alpha \;, & \quad \mu_{10} &= \epsilon_+ + \alpha &
 \quad 
 &\Leftrightarrow &
 \quad 
 \alpha&=-\mu_9+ \epsilon_+ \;,  & \quad \mu_{10} &= 2\epsilon_+ -\mu_9
\end{alignat}
which coincides with the two VEV assignments in 
\eqref{eq:VEV_hypers}, and, of course, is compatible with \eqref{eq:pole_Higgs}.
The first choice \eqref{eq:Higgs_no_defect} is consistent with the parametrisation \eqref{eq:normal_higgsing_equation}, while the second choice would require a sign change  to $\mu = - \alpha$.

The simplification of \eqref{eq:mass deformation} can be interpreted in terms of a mass deformation corresponding to a quadratic $J$-term interaction that couples two of the half-hypers with two of the Fermis. One may also understand this mass deformation in terms of the brane set-ups for the 6d theory. First, the higgsing procedure is visualized in the following way: Starting from brane system Figure \ref{fig:branes}, one can move a pair of D8 branes across the pair of half-NS5 branes along the $x_6$ direction from left and right respectively; There would be new D6 branes stretched between the pair of half-NS5 branes and D8 branes, once the D8s go across the half-NS5s; The new created D6 branes are fixed on the half-NS5 branes and immobilized. However they can rejoin with the D6 brane in the segment of the pair of half-NS5 branes, and lift along the direction of $x_{789}$; A normal higgsing procedure is to move the one whole D6 and two D8s to infinity. On the other hand, focus on the worldsheet theory of the D2 branes along $x_{016}$, which describes the instanton strings. The above higgsing procedure will introduce a tree level $J$-term superpotential between the half hypers $q$ from D2-D6 (within the pair of half-NS5s), and Fermis $\chi$ from the D2-NS5-D6 branes, as $J=m \chi q$ in the language of (0, 2) SUSY, where $m$ is the mass proportional to $(x_7+{\rm i}x_8)/l_s^{2}$ and $l_s$ is the string length. With the D6 and D8s going to infinity, the hypers and Fermis go to very heavy and decouple from the system as they should be.
%
\subsection{Defects via Higgsing}
The partition function of a theory with surface defects of type $(r,s)$ can be obtained from the partition function of the theory without defect via a Higgsing procedure with a position dependent VEV \cite{Gaiotto:2012xa,Gaiotto:2014ina}. The Higgsing procedure was recently studied in the brane web diagram in \cite{Kim:2020npz}. 
Recalling the mesonic VEV \eqref{eq:pole_Higgs} that realises the desired Higgsing \eqref{eq:E-string_Higgsing} of the $\sprm(1)$ theory to the E-string, the E-string theory with a co-dimension 2 defect can be constructed by the following position dependent VEV
\begin{align}
 (p^r q^s) \cdot pq \cdot e^{-\mu_9 -\mu_{10}} =1 \,.
 \label{eq:pole_Higgs+defect}
\end{align}
In the following, a co-dimension 2 defect of type $(r,s)=(0,1)$ is considered.

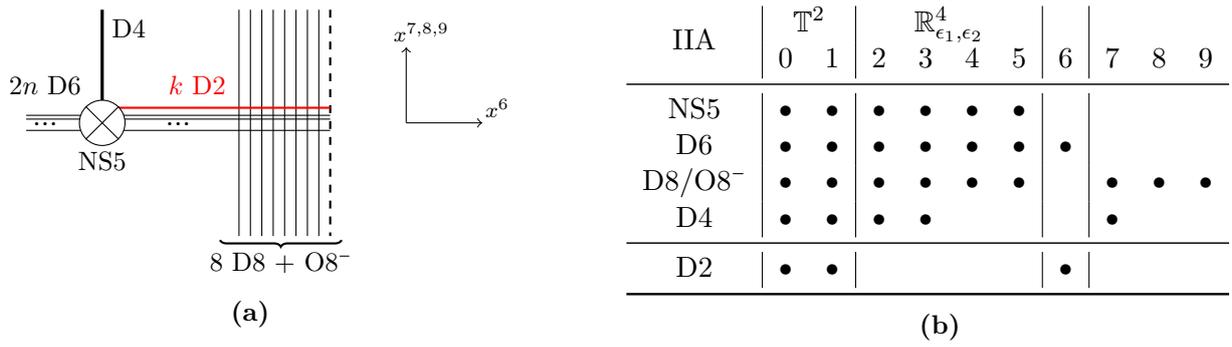
\begin{figure}
\centering
\begin{subfigure}{0.4\textwidth}
\begin{tikzpicture}
    \draw[thick,dashed] (2,1.5)--(2,-1.5);
    \foreach \i in {1,...,8}
    {
    \draw (2-0.15*\i,1.5)--(2-0.15*\i,-1.5);
    }
    \draw (-2,0.1)--(2,0.1);
    \draw (-2,0.05)--(2,0.05);
    \node at (0,-0.05) {$\cdots$};
    \node at (-1.75,-0.05) {$\cdots$};
    \draw (-2,-0.1)--(2,-0.1);
    \draw[red,thick] (-1,0.2)--(2,0.2);
    \draw[very thick] (-1,0)--(-1,1.5);
    \ns{-1,0};
    \draw[->] (3,0)--(4,0);
    \draw[->] (3,0)--(3,1);
    \node at (4.2,0.2) {$\scriptstyle{x^6}$};
    \node at (3.2,1.2) {$\scriptstyle{x^{7,8,9}}$};
    \node at (-1,-0.5) {\small{NS5}};
    \node at (-1.75,0.5) {\small{$2n$ D6}};
    \node at (-0.65,1.25) {\small{D4}};
    \node at (0.25,0.5) {\small{\textcolor{red}{$k$ D2}}};
    \draw[decoration={brace,mirror,raise=30pt},decorate,thick](0.55,-0.5) -- node[below=30pt] {\small{$8$ D8 + O$8^-$}  } (2.15,-0.5);
    \end{tikzpicture}
    \caption{}
\end{subfigure}
\hfill
\begin{subfigure}{0.5\textwidth}
\begin{tabular}{c|cc|cccc|c|cccc}
 \toprule 
\multirow{2}{*}{IIA}  & \multicolumn{2}{c|}{$\T^2$} & 
\multicolumn{4}{c|}{$\R^4_{\epsilon_1,\epsilon_2}$} & & \\
  & 0 & 1 & 2 & 3& 4 & 5 & 6  & 7 & 8 & 9 \\ \midrule
 NS5 & $\bullet$ & $\bullet$  & $\bullet$ & $\bullet$& $\bullet$ & 
$\bullet$&  \\
 D6 & $\bullet$ & $\bullet$ & $\bullet$ & $\bullet$ & $\bullet$ & 
$\bullet$ & $\bullet$
 \\
 D8/O$8^-$ & $\bullet$ & $\bullet$ & $\bullet$ & $\bullet$ & $\bullet$ & 
$\bullet$ & & $\bullet$ & $\bullet$ & $\bullet$
 \\
  D4 & $\bullet$ & $\bullet$ & $\bullet$ & $\bullet$ & & 
 & & $\bullet$ &  & 
 \\\midrule
 D2 & $\bullet$ & $\bullet$ & & & & & $\bullet$ & \\
  \bottomrule 
\end{tabular}
\caption{}
\end{subfigure}
\caption{Addition of a co-dimension 2 defect can be realised via a D4 brane. For a $(r,s)=(0,1)$ type defect, the D4 occupies $\R^2_{\epsilon_1} \subset \R^4_{\epsilon_1,\epsilon_2}$ and is point-like in $\R^2_{\epsilon_2}$.  The resulting world-volume theory is 6d $\sprm(n)$ gauge theory with $2n+8$ flavours in the presence of a defect. In the limit $n\to0$, the E-string theory with defects is recovered. Again, the addition of D2 branes captures the dynamics of the self-dual strings.}
\label{fig:branes+defect}
\end{figure}

Equivalently, the inclusion of such a co-dimension 2 defect can be realised in the Type IIA brane configuration via an additional D4 brane as indicated in Figure \ref{fig:branes+defect}. For the E-string case, the open strings between D2-D4 give rise to a pair of bosonic and fermionic $\Ncal=(0,2)$ multiplets charged under the 2d $\orm(k)$ gauge group. These are the defect contributions to the elliptic genus.

For the theory in the presence of a defect, one may wish to compute the partition function
\begin{align}
  Z_{6d}^{\mathrm{def}} = Z_{\pert}^{\mathrm{def}} \cdot Z_{\str}^{\mathrm{def}}  \,,
\end{align}
which again is composed of perturbative and non-perturbative contributions.  A particularly useful quantity is the normalised defect partition function
\begin{align}
    \widetilde{Z}_{6d}^{\mathrm{def}} = \frac{Z_{6d}^{\mathrm{def}}}{Z_{6d}}
    = \widetilde{Z}_{\pert}^{\mathrm{def}}  \cdot \left(1 + \sum_{k=1}^{\infty} q_\phi^k \, \widetilde{Z}_k^{\mathrm{def}} \right)
\end{align}
and the $q_\phi$ series expansion is detailed in Appendix \ref{app:normalised_partition_fct}.
\subsubsection{Perturbative contributions}
\label{sec:perturbative_defect_higgs}
To begin with, consider the perturbative part \eqref{eq:6d_perturbative_partition_function_flopped}. One may parametrise the defect Higgsing equation \eqref{eq:pole_Higgs+defect} as follows
\begin{align}
    \begin{cases}
    \alpha &=\mu+\frac{\epsilon_2}{2} \,,\\
    \mu_9 &=\mu+\epsilon_++\frac{\epsilon_2}{2} \,,\\
    \mu_{10} &=-\mu+\epsilon_++\frac{\epsilon_2}{2} \,.
    \end{cases}
    \label{eq:defect_higgsing_equation}
\end{align}
Next, inserting \eqref{eq:defect_higgsing_equation} into \eqref{eq:6d_perturbative_partition_function_flopped}, after some algebra, one finds
\begin{align}
\begin{aligned}
    Z_{\mathrm{pert}}^{\sprm(1)}&\longrightarrow Z_{\mathrm{pert}}^{\mathrm{E-str}}\cdot
    \PE\left[\frac{M^2-pM^{-2}Q}{(1-p)(1-Q)}\right]\cdot
    \PE\left[\frac{\sqrt{pq}\left(\sqrt{q} M+(\sqrt{q}M)^{-1}Q\right)}{(1-p)(1-q)(1-Q)}\sum_{i=1}^8\left(M_i+M_i^{-1}\right)\right]\, \\
    &\equiv Z_{\mathrm{pert}}^{\mathrm{E-str}}\cdot Z_{\mathrm{pert}}^{\mathrm{extra}} \,,
    \end{aligned}
\end{align}
where all terms independent on gauge and flavour fugacities have been dropped, as these are irrelevant to the rest of the discussion. The resulting extra term $Z_{\mathrm{pert}}^{\mathrm{extra}}$ corresponds to the contributions from both the co-dimension 2 defect and Goldstone bosons. Therefore, one has to further strip off the Goldstone boson part \eqref{eq:Goldstone_part} in order to derive the genuine defect contribution. In more detail, one computes
\begin{align}
\begin{aligned}
Z_{\mathrm{pert}}^{\mathrm{E-str+def}}&=Z_{\mathrm{pert}}^{\mathrm{extra}}\big|_{M_i\rightarrow \sqrt{q} M_i}\cdot \frac{1}{Z_{\mathrm{GB}}} \\
&=\PE\left[\frac{M^2-pM^{-2}Q}{(1-p)(1-Q)}\right]\cdot
\PE\left[\sum_{i=1}^8\frac{\sqrt{pq}\left(-M M_i+(M M_i q)^{-1}Q\right)}{(1-p)(1-Q)}\right]\,,
\end{aligned}
\end{align}
and the shift $M_i\rightarrow \sqrt{q} M_i$ is a normalisation which allows for a comparison with the partition function without defect. Moreover, one may introduce a defect parameter
\begin{align}
    x\equiv-\mu_{9}+2\epsilon_+=-\mu+\frac{\epsilon_1}{2}\,,\quad {\rm or}\, \quad X=e^x=\frac{\sqrt{p}}{M}\,.
\end{align}
Finally, the perturbative defect contribution in the NS-limit, $q\rightarrow 1$, reads
\begin{align}
\begin{aligned}
    \widetilde{Z}_{\mathrm{pert}}^{\mathrm{E-str+def}}(X)&=Z^{\text{E-str+def}}_{\text{class}}~\PE\left[\frac{p X^{-2}-X^2Q}{(1-p)(1-Q)}\right]
    \cdot\PE\left[\sum_{i=1}^8\frac{-p X^{-1} M_i+X M_i^{-1}Q}{(1-p)(1-Q)}\right] \\
    &=Z^{\text{E-str+def}}_{\text{class}}~\left(\prod_{j=0}^\infty\prod_{k=0}^\infty\frac{1-X^2 p^j Q^{k+1}}{1-X^{-2}p^{j+1} Q^k}\right)\cdot \prod_{i=1}^8\left(\prod_{j=0}^\infty\prod_{k=0}^\infty\frac{1-M_i X^{-1} p^{j+1} Q^{k}}{1-M_i^{-1}X p^{j} Q^{k+1}}\right)\\
    &=Z^{\text{E-str+def}}_{\text{class}}~\frac{\Gamma_e(p X^{-2};\, p, Q)}{\prod_{i=1}^8 \Gamma_{e}(p M_i X^{-1};\,  p, Q)}\,,
    \end{aligned}
\label{eq:6d_perturbative_partition_function_with_defect}
\end{align}
where $\Gamma_e(z;\, p, Q)$ denotes the elliptic gamma function. In addition, the classical contribution to the prepotential, denoted by $ Z^{\text{E-str-def}}_{\text{class}} = \exp\left(\frac{\mathcal{F}_{\textrm{class}}}{\epsilon_1 \epsilon_2}\right)$, has been included as explained below.

Note that \eqref{eq:6d_perturbative_partition_function_with_defect} without the factor $Z^{\text{E-str-def}}_{\text{class}}$ is not invariant under the Weyl symmetry $\mu_l\rightarrow -\mu_l$, which suggests that one is missing some classical terms in the above considerations. The one-loop contribution in 5d, which is part of the classical prepotential, can be reconstructed from the expression
\begin{align}
\begin{aligned}
\frac{\widetilde{Z}_{\mathrm{pert}}^{\mathrm{E-str+def}}(X)}{Z^{\text{E-str+def}}_{\text{class}}}&=
\PE
\left[
\frac{\sqrt{pq}}{(1-p)(1-q)(1-Q)}\left(\sqrt{\frac{p}{q}}X^{-2}
+\sum_{i=1}^{8}qXM_i
+Q \left(\sqrt{\frac{q}{p}}X^{2}+\sum_{i=1}^{8}\frac{1}{qXM_i} \right)\right)
\right]\\
&\times\PE\left[\frac{\sqrt{pq}}{(1-p)(1-q)(1-Q)}\left(-\sqrt{pq}X^{-2}-\sum_{i=1}^{8}XM_i+Q\left(-\frac{X^2}{\sqrt{pq}}-\sum_{i=1}^{8}\frac{1}{XM_i}\right)\right)\right],
\end{aligned}
\end{align}
such that following the notation in \cite{Gu:2020fem}, one expects that the one-loop contribution is given by
\begin{align}
\begin{aligned}
\exp\left(\frac{\mathcal{F}_{\textrm{one-loop}}}{\epsilon_1 \epsilon_2}\right)=\exp\left[-\frac{1}{12\ep_1\ep_2}\left((-2x+\ep_-)^3-(-2x+\ep_+)^3+\sum_{i=1}^8(\mu_i+x+\ep_2)^3-(\mu_i+x)^3\right)\right] \,.
\end{aligned}
\label{6ddefectclassical}
\end{align}
In the NS limit $\ep_2\rightarrow 0$, the $x$ dependent term in \eqref{6ddefectclassical} becomes
\be
\label{6ddefectclassicalNS}
\exp\left[\frac{1}{\ep_1}\left( -\frac{x}{2} \left(\ep_1+\sum_{i=1}^8\mu_i \right) \right)\right]\,.
\ee
As discussed in \cite{Gu:2020fem}, the 6d $\sprm(1)$ theory with 10 fundamental flavours has the tree-level contribution
\be
Z^{\sprm(1)}_{\text{tree}}=\exp \left[
\frac{1}{\ep_1\ep_2}
\left(\phi+\frac{\tau}{2}\right)
\left(\frac{1}{4}(2\alpha)^2-\frac{1}{2}\sum_{l=1}^{10}\mu_l^2\right) \right].
\ee
Applying the same Higgsing method as above, including the  subtraction of the Goldstone boson part, and recalling  \eqref{6ddefectclassicalNS}, one finds the classical contribution to the defect partition function in the NS-limit
\be
Z^{\text{E-str-def}}_{\text{class}} = \exp\left(\frac{\mathcal{F}_{\textrm{tree}}+\mathcal{F}_{\textrm{one-loop}}}{\epsilon_1}\right)
=\exp 
\left[
\frac{1}{\epsilon_1}
\left(x\phi-\frac{1}{2}x\left(-\tau+\ep_1+\sum_{i=1}^8\mu_i\right)\right) \right],
\ee
where all $x$ independent terms have been neglected.

%
%
\subsubsection{Elliptic genus}
\label{sec:DH_elliptic_genus_Sp1}
As a next step, apply the defect Higgsing \eqref{eq:defect_higgsing_equation} to the elliptic genus. Starting from \eqref{eq:elliptic_genus_decompose}, the relevant piece to consider is
\begin{align}
 \prod_{\rho\in \mathrm{fund}} \frac{ \thone(\mu_9 +\rho(u)) \thone(\mu_{10} +\rho(u)) }{ \thone(\epsilon_+ +\rho(u) +\alpha) \thone(\epsilon_+ +\rho(u) -\alpha) }
   = \prod_{\rho\in \mathrm{fund}} \frac{  \thone(x +\rho(u) + \epsilon_2) }{ 
\thone(x +\rho(u) ) }
\end{align}
and the 1-loop determinant for the E-string with co-dimension 2 defect of type $(r,s)=(0,1)$ becomes
\begin{align}
  Z_{\mathrm{1-loop}}^{\mathrm{E-str}+\mathrm{def}}=
Z_{\mathrm{1-loop}}^{\mathrm{E-str}}
  \cdot  \prod_{\rho\in \mathrm{fund}} \frac{  \thone(x +\rho(u) + \epsilon_2) 
}{ \thone(x +\rho(u) ) } \,.
\label{eq:1-loop_det_E-string+defect}
\end{align}
One can also understand this result from the brane realization in Figure \ref{fig:branes+defect}, and the higgsing procedure involving a co-dimension two defect. In this case, we attach a D4 brane along $x_{01239}$ between one side of the half-NS5 brane and the whole D6 brane moving to infinity. In the presence of the D4 brane, the open strings from D2-half-NS5-D4 introduce a $(0, 2)$ chiral and a $(0, 2)$ Fermi, and thus break the $(0, 4)$ SUSY down to $(0, 2)$. On the other side of the half-NS5 node, since there is no D4 brane introduced, a pair of half hyper and Fermi decouple from the 2d theory via the mass deformation as the usual higgings procedure. Therefore, the new defect contributions in \eqref{eq:1-loop_det_E-string+defect} are due to a $\Ncal=(0,2)$ chiral and $\Ncal=(0,2)$ Fermi multiplets, both of which transform under the fundamental representation of the $\orm(k)$ gauge group. 

The full E-string partition function with defect then takes the form,
\begin{equation}
    Z^{\mathrm{E-str+def}} = Z^{\mathrm{E-str+def}}_{\mathrm{pert}} Z^{\mathrm{E-str+def}}_{\mathrm{str}},
\end{equation}
where 
\begin{equation}
    Z^{\mathrm{E-str+def}}_{\mathrm{str}} = 1 + \sum_{k=1}^{\infty} q_{\phi}^k~Z^{\mathrm{E-str+def}}_k, \quad Z^{\mathrm{E-str+def}}_k = \sum_{I} \frac{1}{|W_I|} \frac{1}{(2\pi i)^{r(k)}} \oint Z^{\mathrm{E-str}+\mathrm{def}, (I)}_{\mathrm{1-loop},k}.
\end{equation}
Finally, in order to compute the NS limit $\epsilon_2 \to 0$, one needs the normalised partition function $\widetilde Z^{\mathrm{E-str}+\mathrm{def}}$ as defined in Appendix \ref{app:normalised_partition_fct}. Once the defect contribution from the additional 2d multiplets is understood, one can proceed to evaluate the 1 and 2-string genera. 

\paragraph{1-string.}
The $k=1$ string genus can be computed similarly to the 1-string genus of the E-string theory because there are only discrete sectors without continuous parameters. One finds  
\begin{align}
 Z_{k=1}^{\mathrm{E-str}+\mathrm{def}} = 
 -\frac{\widetilde{\Theta}(\tau,\mu_l,x)}{\eta^6 \theta_1 (\epsilon_1) \theta_1 
(\epsilon_2)}
 \quad \text{with} \quad 
  \widetilde{\Theta}(\tau,\mu_l,x) = \frac{1}{2}\sum_{I=1}^4 \left( \frac{  
\theta_I(x + \epsilon_2) }{ \theta_I(x  ) } \cdot \prod_{l=1}^8 
\theta_I(\tau,\mu_l) \right) \,.
\label{eq:elliptic_genus_k=1_E-string+defect}
\end{align}
and the details are presented in Appendix \ref{app:elliptic_genus_E-string+defect}.
One may also compute the NS-limit of the normalised 1-string contribution, which becomes
\begin{align}
\begin{aligned}
\lim_{\epsilon_2 \to 0}\widetilde{Z}_{k=1}^{\mathrm{E-str}+\mathrm{def}}  
&\equiv 
\lim_{\epsilon_2 \to 0} \left(
Z_{k=1}^{\mathrm{E-str}+\mathrm{def}}  -   Z_{k=1}^{\mathrm{E-str}}  \right) 
    =
\frac{-1 
}{\eta^6 \theta_1 (\epsilon_1) \theta_1^\prime (0)}
\frac{1}{2}\sum_{I=1}^4 \left[\frac{\theta_I^\prime (x  )}{ \theta_I(x  )} 
\cdot 
\prod_{l=1}^8 
\theta_I(\tau,\mu_l) \right]
\end{aligned}
\label{eq:1-string_normalised}
\end{align}
where $\theta_I^\prime(u,q)$ denotes the first derivative with respect to $u$.
\paragraph{2-string.}
The computation of the 2-string elliptic genus of the theory with defect proceeds analogously to the theory without defect. In contrast to the 1-string case, there are new poles in the JK-residue prescription as there are continuous moduli. Since the resulting expression are rather involved, the details are deferred to Appendix \ref{app:elliptic_genus_E-string+defect}. Finally, one evaluates the NS-limit of the normalised 2-string genus and finds
\begin{align}
\lim_{\epsilon_2\to0} \widetilde{Z}_{2}^{\mathrm{E-str+def}} &\equiv
\lim_{\epsilon_2 \to 0}
\left( 
 Z_{2}^{\mathrm{E-str+def}} -Z_{2}^{\mathrm{E-str}}   
-Z_{1}^{\mathrm{E-str}} \left(Z_{1}^{\mathrm{E-str+def}} 
-Z_{1}^{\mathrm{E-str}} \right) 
\right) \notag
\\
&=\frac{\prod_{l=1}^8 \theta_1(\pm 
x+\mu_l) }{4 \eta^{12} \theta_1(\epsilon_1) 
\theta_1(\pm 2x+\epsilon_1)\theta_1(\pm 2x) } 
 + \frac{1}{8 \eta^{12} (\theta_1^\prime(0))^2 \theta_1(\epsilon_1)^2} 
\sum_{I=1}^4 \frac{(\theta_I^\prime(x))^2 }{ \theta_I(x)^2  }
 \prod_{l=1}^8 \theta_I(\mu_l)^2
 \label{eq:2-string_normalised}\\
&+ \frac{\theta_1^\prime(\epsilon_1)}{4 \eta^{12} 
(\theta_1^\prime(0))^2 \theta_1(\epsilon_1)^3} \sum_{I=1}^4 \frac{ 
\theta_I^\prime(x)  }{  \theta_I(x)  }  \prod_{l=1}^8 \theta_I(\mu_l)^2
\notag \\
&+
\frac{1}{ 4 \eta^{12} (\theta_1^\prime(0))^2 (\theta_1(\epsilon_1))^2}
\frac{\theta_2^\prime (\epsilon_1)}{\theta_2( \epsilon_1)}
\bigg[
  \frac{ \theta_1^\prime (x)   }{ \theta_1(x) }
\prod_{l=1}^8 \theta_1(\mu_l) \theta_2(\mu_l)
+
  \frac{ \theta_2^\prime (x)   }{ \theta_2(x) }
\prod_{l=1}^8 \theta_1(\mu_l) \theta_2(\mu_l) 
\notag \\
&\qquad \qquad \qquad  \qquad \qquad \qquad \qquad +
\frac{ \theta_3^\prime (x)   }{ \theta_3(x) }
\prod_{l=1}^8 \theta_3(\mu_l) \theta_4(\mu_l)
+
\frac{ \theta_4^\prime (x)   }{ \theta_4(x) }
\prod_{l=1}^8 \theta_3(\mu_l)\theta_4(\mu_l) 
\bigg]
 \notag \\
&+
\frac{1}{ 4 \eta^{12} (\theta_1^\prime(0))^2  (\theta_1(\epsilon_1))^2}
\frac{\theta_3^\prime (\epsilon_1) }{\theta_3(\epsilon_1)}
\bigg[
  \frac{  \theta_1^\prime (x)   }{\theta_1(x)  }
  \prod_{l=1}^8 \theta_1(\mu_l)\theta_3(\mu_l) 
  +
  \frac{  \theta_2^\prime (x)   }{\theta_2(x)  }
  \prod_{l=1}^8 \theta_2(\mu_l)\theta_4(\mu_l) 
  \notag \\
&\qquad \qquad \qquad  \qquad \qquad \qquad \qquad +
    \frac{  \theta_3^\prime (x)   }{\theta_3(x)  }
  \prod_{l=1}^8 \theta_1(\mu_l)\theta_3(\mu_l) 
  +
  \frac{  \theta_4^\prime (x)   }{\theta_4(x)  }
  \prod_{l=1}^8 \theta_2(\mu_l)\theta_4(\mu_l) 
  \bigg]
\notag \\ 
&+
 \frac{1}{4 \eta^{12} (\theta_1^\prime(0))^2  (\theta_1(\epsilon_1))^2}
 \frac{ \theta_4^\prime (\epsilon_1)}{ \theta_4(\epsilon_1)  }
 \bigg[
\frac{\theta_1^\prime (x)}{ \theta_1(x)  }
 \prod_{l=1}^8 \theta_1(\mu_l)\theta_4(\mu_l)  
 +
 \frac{\theta_2^\prime (x)}{ \theta_2(x)  }
 \prod_{l=1}^8 \theta_2(\mu_l)\theta_3(\mu_l) 
 \notag \\
&\qquad \qquad \qquad  \qquad \qquad \qquad \qquad +
 \frac{\theta_3^\prime (x)}{ \theta_3(x)  }
 \prod_{l=1}^8 \theta_2(\mu_l)\theta_3(\mu_l)
 +
 \frac{\theta_4^\prime (x)}{ \theta_4(x)  }
 \theta_1(\mu_l)\theta_4(\mu_l)
 \bigg]
\notag \\
  &+ \frac{1}{4 \eta^{12} (\theta_1^\prime(0))^2 \theta_1(\epsilon_1)^2} 
  \bigg[ 
  \frac{ \theta_1^\prime(x) \theta_2^\prime(x) }{ \theta_1(x) \theta_2(x)  }
 \prod_{l=1}^8 \theta_1(\mu_l) \theta_2(\mu_l)
 +
  \frac{ \theta_1^\prime(x) \theta_3^\prime(x) }{ \theta_1(x) \theta_3(x)  }
 \prod_{l=1}^8 \theta_1(\mu_l) \theta_3(\mu_l)
 \notag \\
&\qquad \qquad \qquad  \qquad \qquad  +
  \frac{ \theta_1^\prime(x) \theta_4^\prime(x) }{ \theta_1(x) \theta_4(x)  }
 \prod_{l=1}^8 \theta_1(\mu_l) \theta_4(\mu_l)
 +
  \frac{ \theta_2^\prime(x) \theta_3^\prime(x) }{ \theta_2(x) \theta_3(x)  }
 \prod_{l=1}^8 \theta_2(\mu_l) \theta_3(\mu_l)
 \notag \\
&\qquad \qquad \qquad  \qquad \qquad  +
  \frac{ \theta_2^\prime(x) \theta_4^\prime(x) }{ \theta_2(x) \theta_4(x)  }
 \prod_{l=1}^8 \theta_2(\mu_l) \theta_4(\mu_l)
 +
  \frac{ \theta_3^\prime(x) \theta_4^\prime(x) }{ \theta_3(x) \theta_4(x)  }
 \prod_{l=1}^8 \theta_3(\mu_l) \theta_4(\mu_l)
\bigg]
\notag \\
&+\frac{1}{4 \eta^{12} \theta_1^\prime(0) \theta_1(\pm 
\epsilon_1)\theta_1(2\epsilon_1)} 
\sum_{I=1}^4 
\left[
\frac{ \theta_I^\prime (x+\frac{1}{2} \epsilon_1) }{
\theta_I(x+\frac{1}{2} \epsilon_1) }
+
\frac{ \theta_I^\prime (x-\frac{1}{2} \epsilon_1) }{
\theta_I(x-\frac{1}{2} \epsilon_1) }
\right]
\prod_{l=1}^8 \theta_I(\mu_l\pm \frac{1}{2} \epsilon_1)
 \notag 
\end{align}
where $\theta_I^\prime(u,q)$ and $\theta_I^{\prime\prime}(u,q)$ denote first and second derivatives with respect to $u$.
%
%
\section{Wilson surface/line}
\label{sec:Wilson_surface_line}
In this section, a Wilson surface defect is introduced to the E-string theory either via a co-dimension four D4 brane or via double Higgsing of two co-dimension two defects. The resulting defect partition function is evaluated up to 2-instanton order. 
It is well-known that the E-string theory compactified on a circle is dual to the 5d $\Ncal=1$ $\sprm(1)$ theory with 8 fundamental flavours. The Wilson surface in 6d reduces to the Wilson loop observable of the corresponding 5d gauge theory. However, due to the non-trivial holonomy along the  circle\cite{Kim:2014dza}, the 5d Wilson loop only preserves a $\sorm(16)$ flavour symmetry, which is the maximal subgroup of the $E_8$ symmetry of the E-string. Here, the partition function for the 5d $\sprm(1)$ theory  with a Wilson loop is computed up to 2-instanton order. The Wilson surface/line computation are  important ingredients for quantisation of the E-string curve in next section.
\subsection{Wilson surface in 6d}
Expectation values of Wilson surfaces \cite{Ganor:1996nf,Chen:2007ir} are natural candidates for co-dimension 4 defects. In lack of a field-theoretic formulation, a convenient method to formulate Wilson surface defects is to embed these into string theory. For example, Wilson surface defects on an $\Omega$-deformed $\T^2 \times \R^4$ background have been studied in \cite{Nekrasov:2015wsu,Kim:2016qqs,Bullimore:2014upa,Tong:2014cha,Agarwal:2018tso,Chen:2020jla}.
\paragraph{Wilson surface via D4 brane.}
Analogous to \cite{Tong:2014cha,Agarwal:2018tso,Chen:2020jla}, one can compute the VEV of a Wilson surface introduced by an additional D${4^\prime}$ brane. The Type IIA brane configuration is summarised in Table \ref{Wilson brane setup}, where the D${4^\prime}$ brane, filling $\T^2$ and directions $x^{7, 8, 9}$, is clearly a co-dimension four defect. In presence of the D4' brane on the left of the half-NS5 brane, there is a single D2 brane stretched between the half-NS5 and D4' branes. In the wolrdvolume theory of the NS5 brane, the D2 brane ending on creates a Wilson surface defect in the fundamental representation. To study the 2d field contents respect to above brane set-ups, one can use the Hanany-Witten transition to move the D4' to the right of the NS5 brane. In this frame, the 2d worldsheet theory has new degrees of freedom from open strings stretched between the D4' brane and the stack of $k$ D2 branes describing the $k$-instanton strings. They are (0, 4) twisted hypers $(\phi, \phi')$ in fundamental $\orm(k)\times \urm(1)_v$ , and (0, 4) Fermi $\surm(2)_L$ doublet $(\psi, \psi')$ in bifundamental of $\orm(k)\times U(1)_v$, where $\urm(1)_v$ comes from the symmetry associated to the D4' brane. In more details they, combined with adjoint hypers $(B, B')$ arised from D2-D2 modes, amount to additional cubic $E$-term and $J$-term superpotentials as: $E=B \phi$, $J=\phi' B'$, $E'=-B' \phi$ and $J'=\phi' B$. The $\urm(1)_v$ charges are assigned as $(\psi, \psi')=(-1, -1)$, $(\phi, \phi')=(1, -1)$ and $(B, B')=(0, 0)$. Their contributions to the one-loop determinant are thus given by
\begin{align}
W_D(u, v)\equiv \prod_{\rho\in\mathrm{fund.}}\frac{\theta_1(\pm\epsilon_-+\rho(u)- v)}{\theta_1(\pm\epsilon_++\rho(u)- v)}\,.
\label{Wilson defect}
\end{align}
The corresponding $k$-th instanton string correction to the VEV of Wilson surface is
\begin{align}
W_k(v)=\oint [\diff u ]\, Z_{\mathrm{1-loop}}^{\orm(k)}\cdot W_D(u, v)
=\oint [\diff u ]\,    Z_{\mathrm{1-loop}}^{\orm(k)}\cdot\prod_{\rho\in\mathrm{fund.}}\frac{\theta_1(\pm\epsilon_-+\rho(u)- v)}{\theta_1(\pm\epsilon_++\rho(u)- v)}\,,
\label{Wilson surface pf}
\end{align}
where $v$ is the fugacity of the D$4^\prime$ brane. Therefore, the partition function of the theory with  Wilson surface is  given by
\begin{align}
    \mathcal W^S(v)=\sum_{n=0}^{\infty}q_\phi^n\, W_n(v)
\end{align}

\begin{table}
\centering
\begin{tabular}{c|cc|cccc|c|cccc}
 \toprule
\multirow{2}{*}{IIA} & \multicolumn{2}{c|}{$\T^2$} & 
\multicolumn{4}{c|}{$\R^4_{\epsilon_1,\epsilon_2}$} & & \\
  & 0 & 1 & 2 & 3& 4 & 5 & 6  & 7 & 8 & 9 \\ \midrule
  NS5 & $\bullet$ & $\bullet$  & $\bullet$ & $\bullet$& $\bullet$ & 
$\bullet$&  \\
  D6 & $\bullet$ & $\bullet$ & $\bullet$ & $\bullet$ & $\bullet$ & 
$\bullet$ & $\bullet$
 \\
  D8+O$8^-$& $\bullet$ & $\bullet$ & $\bullet$ & $\bullet$ & $\bullet$ & 
$\bullet$ & & $\bullet$ & $\bullet$ & $\bullet$
 \\
  D2 & $\bullet$ & $\bullet$ & & & & & $\bullet$ & \\
  D$4^\prime$ & $\bullet$ & $\bullet$ & & & & & & 
 $\bullet$ & $\bullet$& $\bullet$
 \\
  \bottomrule 
\end{tabular}
\caption{Wilson surface via a D4 brane.}
\label{Wilson brane setup}
\end{table}
\paragraph{Wilson surface via double Higgings.}
It has been suggested in \cite{Kimura:2017auj} that a co-dimension 4 defect can be constructed from two co-dimension 2 defects. Hence, one may be tempted to introduce the Wilson surface defect via defect Higgings of the 6d $\sprm(2)$ theory with $12$ fundamental flavours; the resulting E-string theory with two co-dimension 2 defects should correspond to the E-string in the presence of a Wilson surface. To make the statement precise, one studies the Higging of the mesonic operators in more detail. For a $\sprm(n)$ gauge theory  with $2n+8$ fundamental hypermultiplets
\begin{align}
H_\alpha^m=(Y^m_\alpha, (\tilde Y^\alpha_m)^\dagger)\,,\ \ \ \ {\rm with}\ \ \  \alpha=1,\dots, n\,\ \ \ \ {\rm and}\ \ \ m=1,\dots, 2n+8\,,
\end{align}
the $\urm(2n+8)$ global symmetry is enhanced to $\sorm(4n+16)$. Since $\urm(2n+8)\subset \sorm(4n+16)$, one can group the meson operators into three types of representations with respect to $\urm(2n+8)$, i.e.\
\begin{align}
M^{mn}\equiv J^{\alpha\beta}Y^m_\alpha Y^n_\beta\,,\ \ \ \widetilde M_{mn}\equiv J_{\alpha\beta}\tilde Y^\alpha_m \tilde Y^\beta_n\,\ \ \ {\rm and}\ \ \ N^m_n\equiv Y^m_\alpha \tilde Y^\alpha_n\,.
\end{align}
The operators $M$, $\widetilde M$ and $N$ correspond to the anti-symmetric, complex conjugate anti-symmetric, and adjoint representations of $\urm(2n+8)$, respectively. All together these three $\urm(2n+8)$ representations form the adjoint representation of $\sorm(4n+16)$. Their fugacities are labelled as
\begin{align}
[M^{mn}]=pq e^{-\mu_m-\mu_n}\,,\ \ \ [\widetilde M_{mn}]=pq e^{\mu_m+\mu_n}\,,\ \ \ [N^m_n]=pq e^{-\mu_m+\mu_n}\,.
\end{align}
One may choose the operator $M$ and $\widetilde M$ to Higgs the $\sprm(2)$ theory. More specifically, for $\sprm(2)$ with 12 hypermultiplets, one assigns VEVs to the following meson operators
\begin{align}
\langle M^{9,\,10}\rangle=\langle \widetilde M_{11,\, 12}\rangle=z\,,
\end{align}  
where $z=x^4+i x^5$ is the complexified coordinate of $x^{4,5}$ direction, which coincides with the positions of the two $D4$ branes that trigger the Higgs mechanism.
The corresponding conditions on the fugacities are
\begin{subequations}
\label{eq:double_Higgs_1}
\begin{alignat}{4}
&{\rm for}\ \ \ M^{9,\,10}\,: & \quad  &q  \cdot pq \ e^{-\mu_{9} - \mu_{10}} = 1 &
\qquad 
&\Rightarrow \qquad &
&\begin{cases}
    \mu_{10} = 2 \epsilon_+ - \mu_{9} + \epsilon_2  \,,\\
    a_{1} = \mu_{9}-\epsilon_+   \,,  
\end{cases} \\
&{\rm for}\ \ \ \widetilde M^{11,\,12}\,: & \quad  &q  \cdot pq \ e^{\mu_{11} + \mu_{12}} = 1 &
\qquad 
&\Rightarrow \qquad & 
&\begin{cases}
    \mu_{12} = -2 \epsilon_+ - \mu_{11} - \epsilon_2 \,, \\
    a_{2} =- \mu_{11}-\epsilon_+     \,.
\end{cases}
\end{alignat}
\end{subequations}
Applying the Higgsing \eqref{eq:double_Higgs_1} to the one-loop determinants of the elliptic genera, one finds
\begin{align}
\prod_{\rho\in\mathrm{fund.}}&\frac{\theta_1(\rho(u)+\mu_9)\theta_1(\rho(u)+\mu_{10})}{\theta_1(\epsilon_++\rho(u)\pm a_1)}\cdot\frac{\theta_1(\rho(u)+\mu_{11})\theta_1(\rho(u)+\mu_{12})}{\theta_1(\epsilon_++\rho(u)\pm a_2)}\notag\\
&=\prod_{\rho\in\mathrm{fund.}}\frac{\theta_1(\rho(u)-\mu_9+2\epsilon_++\epsilon_2)}{\theta_1(\rho(u)-\mu_9+2\epsilon_+)}\cdot\frac{\theta_1(\rho(u)-\mu_{11}-2\epsilon_+-\epsilon_2)}{\theta_1(\rho(u)+\mu_{11}+2\epsilon_+)}\notag\\
&=\prod_{\rho\in\mathrm{fund.}}\frac{\theta_1(\rho(u)-\mu_9+2\epsilon_++\epsilon_2)}{\theta_1(\rho(u)-\mu_9+2\epsilon_+)}\cdot\frac{\theta_1(\rho(u)-\mu_{11}-2\epsilon_+-\epsilon_2)}{\theta_1(\rho(u)-\mu_{11}-2\epsilon_+)}\,,
\end{align}
and the last equality follows from the fact that $\rho(u)$ evaluates to $\{\pm u_i\}$ as $\rho$ takes values in the weights of the fundamental representation of the orthogonal group. Define $\chi_1\equiv -\mu_9+2\epsilon_+$ and $\chi_2\equiv -\mu_{11}-2\epsilon_+$, then the defect contribution from double defect Higgsing reads
\begin{align}
\mathcal V_{\rm D4 {D4}}(\chi_1,\,\chi_2)=\prod_{\rho\in\mathrm{fund.}}\frac{\theta_1(\rho(u)+\chi_1+\epsilon_2)}{\theta_1(\rho(u)+\chi_1)}\cdot\frac{\theta_1(\rho(u)+\chi_2-\epsilon_2)}{\theta_1(\rho(u)+\chi_2)}\,.
\end{align}
If one further assigns 
\begin{align}
\begin{cases}
   \chi_1=-v-\epsilon_+\\
   \chi_2=-v+\epsilon_+\,,
\end{cases}
\end{align}
the defect $\mathcal V_{\rm D4 {D4}}$ from double Higgsing coincides with the Wilson surface defect \eqref{Wilson defect}, i.e.
\begin{align}
V_{\rm D4 {D4}}(u, \chi_1,\,\chi_2)=\prod_{\rho\in\mathrm{fund.}}\frac{\theta_1(\rho(u)-v\pm \epsilon_-)}{\theta_1(\rho(u)-v\pm \epsilon_+)}= W_D(u, v)\,.
\end{align}
\paragraph{1-instanton.}
Now, compute the 1-instanton order of \emph{normalised} Wilson surface defect in the NS-limit. The normalised Wilson surface is defined as follows:
\begin{align}
\mathcal{W}^S_{\hbar=\epsilon_1}=\sum_{n=0}^\infty q^n_\phi\, \widetilde W_n\equiv \lim_{\epsilon_2\rightarrow 0}\frac{\mathcal W^S}{Z^{\mathrm{E-str}}_{\mathrm{str}}}\,.
\end{align}
At one-instanton string order, one finds
\begin{align}
\widetilde W_1=\lim_{\epsilon_2\rightarrow 0}(W_1-Z_1^{\mathrm{E-str}})\,.
\label{1-loop Wilson surface}
\end{align}
At this order for the elliptic genus, the gauge group for the 2d theory is $\orm(1)=\mathbb Z_2$. Therefore, the discrete holonomies of $\orm(1)$ are  $\rho(u)=u_I=\{0,\,\frac{1}{2},\,\frac{\tau}{2},\,\frac{1+\tau}{2}\}$. Further recall that for discrete holonomy, the $\theta_1$ functions from both fermionic and bosonic contributions to elliptic genus should be understood as
\begin{align}
``\theta_1(u_I+z)"\equiv\sqrt{\theta_1(u_I+z)\theta_1(-u_I+z)}\,,
\end{align}
see also Appendix \ref{app:theta_fct}.
Bearing this in mind, one can show that
\begin{align}
W_D(u_I,v)=\frac{\theta_1(\pm \epsilon_-+u_I-v)}{\theta_1(\pm \epsilon_++u_I-v)}=\frac{\theta_I(\pm \epsilon_--v)}{\theta_I(\pm\epsilon_+-v)}\,,
\end{align}
where $I=\{1,\, 2,\, 3,\, 4\}$ corresponding to the choice of $u_I=\{0,\,\frac{1}{2},\,\frac{1+\tau}{2},\,\frac{\tau}{2}\}$. Applying \eqref{eq:ell_genus_E-string_k=1}, \eqref{Wilson surface pf}, and \eqref{1-loop Wilson surface}, one finds
\begin{align} 
\label{eq:E-W1}
\widetilde{W}_1(v)
=-\frac{1}{2}\sum_{I=0}^4\frac{\prod_{l=1}^8\theta_I(\mu_l)}{\eta^6\theta_1(\epsilon_1)\theta_1^\prime(0)}\left(\frac{\theta_I^\prime(v-\frac{\epsilon_1}{2})}{\theta_I(v-\frac{\epsilon_1}{2})}-\frac{\theta_I^\prime(v+\frac{\epsilon_1}{2})}{\theta_I(v+\frac{\epsilon_1}{2})}\right)\,.
\end{align}
In contrast to the $\Ncal=(2, 0)$ $A_1$ SCFT\cite{Chen:2020jla}, the Wilson surface for the E-string theory is $v$-dependent. 
\paragraph{2-instanton.}
Next, the Wilson surface contribution $\widetilde{W}_2$ at 2-instanton string order is evaluated. The computational details are summarised in Appendix \ref{app: Wilson_surface_at_two_instanton}. The elliptic genus $W_2$ for the Wilson surface defect is composed of the sum of continuous and discrete holonomies
\begin{align}
W_{k=2}=
\frac{1}{2}\sum_{i=1}^6 W_{k=2, i}^{\mathrm{cont}}
+\frac{1}{4}\sum_{i=1}^6 W_{k=2, i}^{\mathrm{dis}}\,.
\end{align}
The normalised Wilson surface contribution, $\widetilde{W}_2$, in the NS-limit $\epsilon_2\rightarrow 0$ reads
\begin{align}
\widetilde W_2&=\lim_{\epsilon_2\rightarrow 0}\left(W_{k=2}-Z_{k=2}-Z_{k=1}\left(W_{k=1}-Z_{k=1}\right)\right) \notag\\
&=\frac{\prod_{l=1}^{8}
 \theta_1(v\pm\mu_l+\frac{\epsilon_1}{2})}{2\eta^{12}\theta_1(2v)\theta_1^2(2v+\epsilon_1)\theta_1(2v+2\epsilon_1)}
+\frac{\prod_{l=1}^{8}
 \theta_1(v\pm\mu_l-\frac{\epsilon_1}{2})}{2\eta^{12}\theta_1(2v)\theta_1^2(2v-\epsilon_1)\theta_1(2v-2\epsilon_1)}\notag\\  
&+\sum_{I=1}^4\frac{\prod_{l=1}^8\theta_I(\mu_l\pm\frac{\epsilon_1}{2})}{4\eta^{12}\theta_1^2(\epsilon_1)\theta_1(2\epsilon_1)\theta_1^\prime(0)}\cdot\left(\frac{\theta_I^\prime(v+\epsilon_1)}{\theta_I(v+\epsilon_1)}-\frac{\theta_I^\prime(v-\epsilon_1)}{\theta_I(v-\epsilon_1)}\right)
\notag\\
&+\sum_{I=1}^4\frac{\prod_{l=1}^8\theta_I^2(\mu_l)}{8\eta^{12}\theta^2_1(\epsilon_1)\theta_1^{\prime 2}(0)}\cdot\theta_I^\Delta(v)^2
+\sum_{I=1}^4\frac{\prod_{l=1}^8\theta^2_I(\mu_l)}{4\eta^{12}\theta_1^2(\epsilon_1)\theta_1^{\prime 2}(0)}\cdot\frac{\theta_1^\prime(\epsilon_1)}{\theta_1(\epsilon_1)}\cdot\theta_I^{\Delta}(v)\notag\\
&+\frac{\prod_{l=1}^8\theta_1(\mu_l)\theta_2(\mu_l)}{4\eta^{12}\theta_1^2(\epsilon_1)\theta_1^{\prime 2}(0)}\cdot\frac{\theta_2^\prime(\epsilon_1)}{\theta_2(\epsilon_1)}
\cdot\left(\theta_1^\Delta(v)+\theta_2^\Delta(v)\right)+\frac{\prod_{l=1}^8\theta_3(\mu_l)\theta_4(\mu_l)}{4\eta^{12}\theta_1^2(\epsilon_1)\theta_1^{\prime 2}(0)}\cdot\frac{\theta_2^\prime(\epsilon_1)}{\theta_2(\epsilon_1)}
\cdot\left(\theta_3^\Delta(v)+\theta_4^\Delta(v)\right)\notag\\
&+\frac{\prod_{l=1}^8\theta_1(\mu_l)\theta_3(\mu_l)}{4\eta^{12}\theta_1^2(\epsilon_1)\theta_1^{\prime 2}(0)}\cdot\frac{\theta_3^\prime(\epsilon_1)}{\theta_3(\epsilon_1)}
\cdot\left(\theta_1^\Delta(v)+\theta_3^\Delta(v)\right)
+\frac{\prod_{l=1}^8\theta_2(\mu_l)\theta_4(\mu_l)}{4\eta^{12}\theta_1^2(\epsilon_1)\theta_1^{\prime 2}(0)}\cdot\frac{\theta_3^\prime(\epsilon_1)}{\theta_3(\epsilon_1)}
\cdot\left(\theta_2^\Delta(v)+\theta_4^\Delta(v)\right)
\notag\\
&+\frac{\prod_{l=1}^8\theta_1(\mu_l)\theta_4(\mu_l)}{4\eta^{12}\theta_1^2(\epsilon_1)\theta_1^{\prime 2}(0)}\cdot\frac{\theta_4^\prime(\epsilon_1)}{\theta_4(\epsilon_1)}
\cdot\left(\theta_1^\Delta(v)+\theta_4^\Delta(v)\right)
+\frac{\prod_{l=1}^8\theta_2(\mu_l)\theta_3(\mu_l)}{4\eta^{12}\theta_1^2(\epsilon_1)\theta_1^{\prime 2}(0)}\cdot\frac{\theta_4^\prime(\epsilon_1)}{\theta_4(\epsilon_1)}
\cdot\left(\theta_2^\Delta(v)+\theta_3^\Delta(v)\right)
\notag\\
&+\sum_{I<J}^4\frac{\prod_{l=1}^8\theta_I(\mu_l)\theta_J(\mu_l)}{4\eta^{12}\theta_1^2(\epsilon_1)\theta_1^{\prime 2}(0)}
\theta_I^\Delta(v)\cdot \theta_J^\Delta(v)\,,
\label{eq:normalized_Wilson_surface_at_two_instanton}
\end{align}
where 
\begin{align}
\theta_I^{\Delta}(v)\equiv\frac{\theta_I^\prime(v-\frac{\epsilon_1}{2})}{\theta_I(v-\frac{\epsilon_1}{2})}-\frac{\theta_I^\prime(v+\frac{\epsilon_1}{2})}{\theta_I(v+\frac{\epsilon_1}{2})}\,.
\end{align}
Even though \eqref{eq:normalized_Wilson_surface_at_two_instanton} appears to be $v$-dependent, a careful expansion of $\widetilde W_2$ with respect to torus moduli $Q$ shows that it is in fact independent on brane fugacity $v$ and can be assembled in terms of $E_8$ characters,
\begin{align}
\widetilde W_2\ \ =&\ \ \ \bigg(\chi_{2}^{\mathfrak{su}_2}+(\chi_\mathbf{248}+2)\chi_{1}^{\mathfrak{su}_2}+(\chi_\mathbf{3875}+2\chi_\mathbf{248}+4)\bigg)Q \notag\\
&+\bigg(\chi_{4}^{\mathfrak{su}_2}+(\chi_\mathbf{248}+3)\chi_{3}^{\mathfrak{su}_2}+(\chi_\mathbf{3875}+4\chi_\mathbf{248}+9)\chi_{2}^{\mathfrak{su}_2} +(\chi_\mathbf{30380}+4\chi_\mathbf{3875}+12\chi_\mathbf{248}+20)\chi_{1}^{\mathfrak{su}_2}\notag\\
&\ \ \ \ +(\chi_\mathbf{147250}+3\chi_\mathbf{30380}+8\chi_\mathbf{3875}+17\chi_\mathbf{248}+20) \bigg)Q^2\notag\\
&+\bigg(\chi_{6}^{\mathfrak{su}_2}+(\chi_\mathbf{248}+3)\chi_{5}^{\mathfrak{su}_2}+(\chi_\mathbf{3875}+4\chi_\mathbf{248}+11)\chi_{4}^{\mathfrak{su}_2} +(\chi_\mathbf{30380}+4\chi_\mathbf{3875}+15\chi_\mathbf{248}+28)\chi_{3}^{\mathfrak{su}_2}\notag\\
&\ \ \ \ +(\chi_\mathbf{147250}+4\chi_\mathbf{30380}+\chi_\mathbf{27000}+15\chi_\mathbf{3875}+41\chi_\mathbf{248}+66)\chi_{2}^{\mathfrak{su}_2} \notag\\
&\ \ \ \ +(\chi_\mathbf{779247}+4\chi_\mathbf{147250}+14\chi_\mathbf{30380}+3\chi_\mathbf{27000}+37\chi_\mathbf{3875}+88\chi_\mathbf{248}+111)\chi_{1}^{\mathfrak{su}_2} \notag\\
&\ \ \ \ +(\chi_\mathbf{2450240}+3\chi_\mathbf{779247}+9\chi_\mathbf{147250}+21\chi_\mathbf{30380}+7\chi_\mathbf{27000}+47\chi_\mathbf{3875}+90\chi_\mathbf{248}+96)\bigg)Q^3\cdots\,,
\end{align}
where
\begin{align}
\chi^{\mathfrak{su}_2}_{n}\equiv\sum_{j=-n}^{n} p^j\,,
\label{eq:su2_characters}
\end{align}
is the $\surm(2)$ characters respect to $p=e^{\epsilon_1}$. 
\subsection{Wilson line in 5d}
\label{sec:E-Wloop}
Following \cite{Gaiotto:2015una, Hwang:2014uwa}, the Wilson line in the 5d $\sprm(1)$ gauge theory with 8 fundamental flavours is considered in this section. The 5d Wilson loop observable only preserves an $\sorm(16)$ flavour symmetry, as subgroup of $E_8$. 

The 5d instanton partition function is given by
\begin{align}
    Z_{\mathrm{inst}}^{5d\,\,\sprm(1)}=\sum_{k=0}^{\infty}U^k Z_{k}=\sum_{k=0}^{\infty}U^k\cdot\frac{1}{|W_k|}\int \prod_{i=1}^k\frac{\diff \phi_i}{2\pi i}\,Z_{\mathrm{inst}, k}(a, \phi_i, m_l)\,,
\end{align}
where $Z_{\mathrm{inst}, k}(a, \phi_i, m_l)$ denotes the integrand of the 5d instanton partition function modelled by an 1d quantum mechanics via the ADHM construction to parametrise the moduli space of the 5d $k$ instantons. Furthermore $U$, $a$, and $m_l$ denote the 5d instanton fugacity, $\sprm(1)$ gauge fugacity, and $\sorm(16)$ flavour fugacities, respectively. Moreover, $\phi_i$ are the fugacities of the dual gauge group $\orm(k)$ of the 1d quantum mechanics for $k$ instantons, and $|W_k|$ is the order of Weyl group of $\orm(k)$. The Wilson loop is computed from the 5d instanton partition function with the insertion of the equivariant Chern characters for the fundamental Wilson loop of $\sprm(1)$,
\begin{align}
    \mathcal W^L=\sum_{k=0}^\infty U^k\, W_k(a, m_l)=
\sum_{k=0}^\infty U^k\cdot\frac{1}{|W_k|}\int \left(\prod_{i=1}^k\frac{\diff \phi_i}{2\pi i}\right)\cdot \mathrm{Ch}_k(\alpha, e^{\phi_i})\cdot Z_{\mathrm{inst},\, k}(a, \phi_i, m_l)\,,
\end{align}
where $\mathrm{Ch}_k(\alpha, e^{\phi_i})$ is the equivariant Chern character of $\sprm(1)$ and $\sorm(k)$, with $\alpha=e^a$. 
The normalised Wilson line in the NS-limit is defined by
\begin{align}
    \mathcal W^L_{\hbar=\epsilon_1}=\sum_{k=1}^{\infty}U^k\, \widetilde W_k\equiv\lim_{\epsilon_2\rightarrow 0}\frac{\mathcal W^L}{Z^{5d\,\, \sprm(1)}_{\mathrm{inst}}}\,.
    \label{eq:normalized_Wilson_loop}
\end{align}
By expanding the right-hand side of \eqref{eq:normalized_Wilson_loop} in terms of $U$, one finds
\begin{align}
    \widetilde W_0=W_0=\alpha+\frac{1}{\alpha}\,, \quad \widetilde W_1=\lim_{\epsilon_2\rightarrow 0}\left(W_1-\left(\alpha+\frac{1}{\alpha}\right)Z_1\right)\,.
\end{align}

Next, the computation of the 5d Wilson loop partition function up to two instanton order is briefly discussed. More computational details are presented in Appendix \ref{app:Wilson_loop}.  
\paragraph{1-instanton.}
For 1-instanton order, the 5d $\sprm(1)$ partition function and Wilson line are determined by the discrete holonomies of $\orm(1)_\pm$ of the 1d quantum mechanics in the ADHM construction. One can easily read off their contributions. Further, \eqref{eq:normalized_Wilson_loop} implies that the normalised Wilson loop in the NS-limit, $\epsilon_2\rightarrow 0$ or $q\rightarrow 1$, reads
\begin{align}
\widetilde W_1
&=\lim_{q\rightarrow 1}\left(W_1-\left(\alpha+\frac{1}{\alpha}\right)Z_1\right)\notag\\
&=\chi_\mathbf{128}\cdot\alpha-\chi_\mathbf{128^\prime}\,\chi^{\mathfrak{su}_2}_{1/2}\cdot \alpha^2
+\chi_\mathbf{128}\,\chi^{\mathfrak{su}_2}_{1}\cdot \alpha^3
+\chi_\mathbf{128^\prime}\,\chi^{\mathfrak{su}_2}_{3/2}\cdot \alpha^4+\ldots\,,
\label{eq:5d_Wilson_line_one_instanton}
\end{align}
where $\chi^{\mathfrak{su}_2}_{n}$ is defined in \eqref{eq:su2_characters} and $\chi_\mathbf{128}$, $\chi_\mathbf{128^\prime}$ are the $\sorm(16)$ characters.
\paragraph{2-instanton.} 
Next, consider the 2-instanton contribution of the Wilson loop. The 5d instanton moduli space is parametrised by one complex moduli parameter, that is integrated over in the JK-residue prescription. The computational subtlety here is that due to the large number of flavours, both the integrand of the 5d instanton partition function and Wilson loop suffer from higher order poles at infinity. Therefore, one has to introduce anti-symmetric matter fields in order to control the divergence and then perform the computation. In the end, one has to subtract the extra contributions from these additional matter fields and set their mass to infinity to decouple them from the theory. The reader is referred to \cite{Gaiotto:2015una, Hwang:2014uwa} for a thorough discussion; see also Appendix \ref{app:Wilson_loop} for the specific example of $\sprm(1)$ theory with 8 flavours. Here, only the main results of the normalised Wilson loop in the NS-limit at 2-instanton order are summarised
\begin{align}
\widetilde W_2&=\lim_{q\rightarrow 1}\left(W_2-W_1\cdot Z_1+\left(\alpha+\frac{1}{\alpha}\right)\cdot(Z_1^2-Z_2)\right)\notag\\
&=\bigg(\chi_{2}^{\mathfrak{su}_2}+(\chi_\mathbf{120}+3)\chi_{1}^{\mathfrak{su}_2}+(\chi_\mathbf{1820}+3\chi_\mathbf{120}+6)\bigg)\cdot\alpha \notag\\
&-\bigg(\chi_\mathbf{16}\,\chi_{5/2}^{\mathfrak{su}}+(\chi_\mathbf{560}+3\chi_\mathbf{16})\chi_{3/2}^{\mathfrak{su}_2}+(\chi_\mathbf{4368}+3\chi_\mathbf{560}+7\chi_\mathbf{16})\chi_{1/2}^{\mathfrak{su}_2}\bigg)\cdot\alpha^2 \notag\\
&+\bigg(\chi_{4}^{\mathfrak{su}_2}+(\chi_\mathbf{120}+3)\chi_{3}^{\mathfrak{su}_2}+(\chi_\mathbf{1820}+3\chi_\mathbf{120}+7)\chi_{2}^{\mathfrak{su}_2}\notag\\
&\ \ \ \ +\left(\chi_\mathbf{8008}+3\chi_\mathbf{1820}+7\chi_\mathbf{120}+13\right)\chi_{1}^{\mathfrak{su}_2}+(\chi_\mathbf{1820}+2\chi_\mathbf{120}+4)\bigg)\cdot\alpha^3 \notag\\
&-\bigg(\chi_\mathbf{16}\,\chi_{9/2}^{\mathfrak{su}_2}+(\chi_\mathbf{560}+3\chi_\mathbf{16})\chi_{7/2}^{\mathfrak{su}_2}+(\chi_\mathbf{4368}+3\chi_\mathbf{560}+7\chi_\mathbf{16})\chi_{5/2}^{\mathfrak{su}_2}\notag\\
&\ \ \ \ +\left(\chi_\mathbf{11440}+3\chi_\mathbf{4368}+7\chi_\mathbf{560}+13\chi_\mathbf{16}\right)\chi_{3/2}^{\mathfrak{su}_2}+(\chi_\mathbf{4368}+3\chi_\mathbf{560}+6\chi_\mathbf{16})\chi_{1/2}^{\mathfrak{su}_2}\bigg)\cdot\alpha^4 \notag\\
&+\bigg(\chi_{6}^{\mathfrak{su}_2}+(\chi_\mathbf{120}+3)\chi_{5}^{\mathfrak{su}_2}+(\chi_\mathbf{1820}+3\chi_\mathbf{120}+7)\chi_{4}^{\mathfrak{su}_2}+(\chi_\mathbf{8008}+3\chi_\mathbf{1820}+7\chi_\mathbf{120}+13)\chi_{3}^{\mathfrak{su}_2}\notag\\
&\ \ \ \ +\left(\chi_\mathbf{6435}+\chi_\mathbf{6435^\prime}+3\chi_\mathbf{8008}+7\chi_\mathbf{1820}+13\chi_\mathbf{120}+22\right)\chi_{2}^{\mathfrak{su}_2}+\left(\chi_\mathbf{8008}+3\chi_\mathbf{1820}+7\chi_\mathbf{120}+12\right)\chi_{1}^{\mathfrak{su}_2}\notag\\
&\ \ \ \ +\left(\chi_\mathbf{1820}+2\chi_\mathbf{120}+4\right)\bigg)\cdot\alpha^5 \notag\\
&-\bigg(\chi_\mathbf{16}\,\chi_{13/2}^{\mathfrak{su}_2}+(\chi_\mathbf{560}+3\chi_\mathbf{16})\chi_{11/2}^{\mathfrak{su}_2}+(\chi_\mathbf{4368}+3\chi_\mathbf{560}+7\chi_\mathbf{16})\chi_{9/2}^{\mathfrak{su}_2}\notag\\
&\ \ \ \ +\left(\chi_\mathbf{11440}+3\chi_\mathbf{4368}+7\chi_\mathbf{560}+13\chi_\mathbf{16}\right)\chi_{7/2}^{\mathfrak{su}_2}+\left(4\chi_\mathbf{11440}+7\chi_\mathbf{4368}+13\chi_\mathbf{560}+22\chi_\mathbf{16}\right)\chi_{5/2}^{\mathfrak{su}_2}\notag\\
&\ \ \ \ +\left(\chi_\mathbf{11440}+3\chi_\mathbf{4368}+7\chi_\mathbf{560}+13\chi_\mathbf{16}\right)\chi_{3/2}^{\mathfrak{su}_2}+\left(\chi_\mathbf{4368}+3\chi_\mathbf{560}+6\chi_\mathbf{16}\right)\chi_{1/2}^{\mathfrak{su}_2}\bigg)\cdot\alpha^6 +\cdots\,,
\label{eq:5d_Wilson_line_two_instanton}
\end{align}
where the Wilson loop is expressed in terms of the characters of $\sorm(16)$ and $\surm(2)$ with respect to $\{e^{m_i}\}$ and $p=e^{\epsilon_1}$, respectively.

Overall, the normalised Wilson loop up to two-instanton order is given by
\begin{align}
    \mathcal W^L_{\hbar=\epsilon_1}=\alpha+\frac{1}{\alpha}+\widetilde W_1\,U+\widetilde W_2\,U^2+\ldots\,.
    \label{eq:5d_Wilson_line}
\end{align}
\subsection{From Wilson surface to Wilson line}
As mentioned in the beginning of the section, the circle compactification of the 6d E-string theory gives rise to a 5d $\sprm(1)$ gauge theory with 8 flavours, and correspondingly the Wilson surface gives rise to the Wilson line. Based on the previous computations, the statement can be made more precise.  

The relation between the parameters of the 5d $\sprm(1)$ gauge theory and the 6d E-string theory can be derived by comparing their respective partition functions \cite{Hwang:2014uwa,Kim:2014dza},
\begin{align}
    Q=U^{2}\,, \quad q_\phi=\alpha\, U e^{-m_8}\,, \quad M_8=e^{\mu_8}=e^{m_8}\,U^{-2}\,, \quad \mathrm{and} \quad \mu_l= m_l\,, \quad \mathrm{for} \quad l=1,\ldots,7\,.
    \label{eq:5d_6d_parameters_map}
\end{align}
The non-trivial relation for the eighth mass parameter of the 5d and 6d theories reflects the fact that the $E_8$ flavour symmetry is broken to its maximal subgroup $\sorm(16)$. From \eqref{eq:5d_6d_parameters_map}, one expects to obtain the 5d Wilson line \eqref{eq:5d_Wilson_line} from the 6d Wilson surface \eqref{eq:normalized_Wilson_surface_at_two_instanton}. It is worth noticing that the one-instanton correction to the Wilson surface \eqref{eq:E-W1} depends on brane fugacity $v$. Therefore, one should remove this piece and consider
\begin{align}
    \mathcal W^{\prime\, S}\equiv 1+q_\phi^2\, \widetilde W_2+\mathcal O(q_\phi^3)\,.
\end{align}
Further from \eqref{eq:5d_6d_parameters_map}, one assigns
\begin{align}
    \mu_8=m_8-\tau\,, \quad \mathrm{and} \quad \mu_l=m_l=0\,, \quad \mathrm{for} \quad l=1,\ldots,7\,,
\end{align}
where the remaining seven mass parameters have been turned off for simplicity. With this parametrisation, one finds 
\begin{align}
    \frac{1}{\alpha}\cdot\mathcal W^{\prime\, S}&=\frac{1}{\alpha}+\alpha+\left(\frac{64}{\sqrt{\lambda}}+64\sqrt\lambda\right)\alpha\cdot U\notag\\
    &+\left(\chi_{2}^{\mathfrak{su}_2}+\left(92+\frac{14}{\lambda}+14\lambda\right)+3\right)\chi_{1}^{\mathfrak{su}_2}+\left(\left(1092+\frac{364}{\lambda}+364\lambda\right)+3\left(92+\frac{14}{\lambda}+14\lambda\right)+6\right)\alpha\cdot U^2\notag\\
    &+\mathcal O(\alpha^2,\, U^3)\,,
    \label{eq:Wilson_surface_to_Wilson_loop}
\end{align}
with $\lambda\equiv e^{m_8}$. One realises that the coefficients in \eqref{eq:Wilson_surface_to_Wilson_loop} reproduce the branching $\sorm(16)\rightarrow\sorm(14)\times \urm(1)_\lambda$ and, thus, agree with the 5d Wilson loop results \eqref{eq:5d_Wilson_line_one_instanton}, \eqref{eq:5d_Wilson_line_two_instanton} and \eqref{eq:5d_Wilson_line}. 

It would be interesting to verify if higher order instanton-string corrections to the 6d Wilson surface are still independent of the brane fugacity $v$, and if these descend to the 5d Wilson line given by \eqref{eq:5d_6d_parameters_map}. On the other hand, in next section the Wilson surface/line results reappear via a different approach.
\section{Elliptic quantum curves}
\label{sec:quantum_curves}
To begin with, the classical geometry associated with Seiberg-Witten arising from compactifications of 6d SCFTs on elliptic curves is reviewed in this section. Thereafter, one proceeds to quantise these curves and describe the resulting operators and their relation to defect partition functions. Lastly, the van Diejen operator is proposed as the quantised version of the E-string classical curve. This identification is confirmed by several non-trivial consistency checks. 
\subsection{Classical curves}
It is expected that for compactifications of 6d SCFTs on elliptic curves, the resulting Seiberg-Witten curves can be expressed in terms of polynomials in a $\mathbb{C}^*$-variable $t$ with coefficients being Jacobi-forms depending on an elliptic variable $x$. Such curves were derived for M5 branes probing $\mathbb{C}^2/\mathbb{Z}_k$-singularities as well as for E-strings and $\mathcal{O}(-4)$-strings \cite{Haghighat:2017vch,Haghighat:2018dwe,Chen:2020jla}, and take the general form
\begin{equation}
    H(w,z) = t^N + v_1(x)t^{N-1} + \ldots + v_l(x) t^{N-l} + \ldots + v_N(x) =0, \quad t=e^{2\pi i y},
\end{equation}
where the $v_l$ are (meromorphic) Jacobi-forms. In the case of two M5 branes probing $\mathbb{C}^2/\mathbb{Z}_k$ singularity, the resulting Riemann surface is a two-fold covering of the elliptic curve branched over $2k$ points and as such has, according to the Riemann-Hurwitz formula, a genus $g = k + 1$. This Riemann surface shall be denoted by $\Sigma_{S_k}$. Following the original Seiberg-Witten result \cite{Seiberg:1994rs}, one can define a meromorphic differential on $\Sigma_{S_k}$ 
\begin{equation}
    \lambda_{\textrm{SW}} = y \diff x = \log t~\diff x, 
\end{equation}
which can be successively used to compute IR expectation values of physical observables on the Coulomb branch. In particular, the geometry admits $k$ independent A-cycles\footnote{The $k{+}1$-th A-cycle is fixed by the condition that the sum of all A-cycles gives the corresponding cycle of the elliptic curve.} and the integrals of $\lambda_{\textrm{SW}}$ over these,
\begin{align}
\begin{aligned}
    a_1 &=  \oint_{A_1} \lambda_{\textrm{SW}} \,,  \\
    a_2 &=  \oint_{A_2} \lambda_{\textrm{SW}}\,,   \\
       &\vdots    \\
    a_k &=  \oint_{A_k} \lambda_{\textrm{SW}}\,,  
    \end{aligned}
\end{align}
can be interpreted as expectation values of gauge fugacities $a_i$, $i=1,\ldots,k$ of the corresponding 5d KK affine quiver theory obtained after circle compactification, see also \cite{Braun:2021lzt}. To give a more concrete example, restrict to the $k=1$ case  which corresponds to the M-string and which shares certain features with the E-string whose classical curve is given in Section \ref{sec:van_Diejen}. In the case of the M-string, the 6d theory is the $\Ncal=(2, 0)$ $A_1$ SCFT, and the corresponding 5d theory is the $\Ncal=1^\ast$ $\surm(2)$ SYM with adjoint matter. The classical curve of the theory is, for example, given in \cite{Haghighat:2017vch,Chen:2020jla} as
\begin{eqnarray}
    t^2 - \mathcal{W}^S(x) t + \frac{\theta_1(x-m)\theta_1(x+m)}{\theta_1(x)^2} = 0 \,,
\end{eqnarray}
where $\mathcal{W}^S(x)$ admits the interpretation of a Wilson surface expectation value. By dividing both sides of the equation by the $\mathbb{C}^*$ variable $t$, this curve can be brought to the more suggestive form
\begin{eqnarray}
    t + \frac{\theta_1(x-m)\theta_1(x+m)}{\theta_1(x)^2} t^{-1} = \mathcal{W}^S(x) \,.
\end{eqnarray}
In the following subsection, the above curve is reinterpreted as an operator identity in a process which is known as quantisation of a curve. Before proceeding to the quantisation, the curve is further analysed purely from the classical viewpoint. Note that in the case of the M-string, $\mathcal{W}^S(x)$ is actually independent of $x$ and one can reinterpret it as the complex structure of the algebraic curve and write
\begin{eqnarray} \label{eq:M-curve}
    t + \frac{\theta_1(x-m)\theta_1(x+m)}{\theta_1(x)^2} t^{-1} = \frac{1}{z},
\end{eqnarray}
where $z$ is a free complex parameter. Now, \eqref{eq:M-curve} can be solved in terms of $t$ and one obtains
\begin{eqnarray}
    t = \frac{1 \pm \sqrt{1-4 \omega z^2}}{2\omega z}, \quad \textrm{with} \quad \omega \equiv \frac{\theta_1(x-m)\theta_1(x+m)}{\theta_1(x)^2}\,.
\end{eqnarray}
Taking the series-expansion in $z$, one therefore has the following: 
\begin{align}
\log t=\log z +\omega z^2+\frac{3}{2} \omega^2 z^4+\mathcal O(z^5)\,.
\end{align}
Defining $X \equiv e^x$, the A-period is given by
\begin{align}
a_1(z)= \oint \lambda_{\textrm{SW}} = \frac{1}{2\pi i}\oint \frac{\log t}{X} dX=\log z+\left(1+\frac{(\eta-1)^4}{\eta^2}Q+\mathcal O(Q^2)\right) z^2+\mathcal O (z^3) \,,
\end{align}
from which one can compute the inverse map
\begin{align}
z=\alpha-\left(1+(6-4\eta-4\eta^{-1}+\eta^2+\eta^{-2})Q+\mathcal O(Q^2)\right)\alpha^3+\mathcal O(\alpha^4)\,,
\end{align}
where $\eta \equiv e^m$ and $\alpha\equiv e^{a_1}$, and $a_1$ can be interpreted as the $\surm(2)$ gauge fugacity of the resulting 5d SYM KK theory. To see this more explicitly, note that from the above one obtains the following series expansion of the Wilson line expectation value:
\begin{align}
\begin{aligned}
\mathcal{W}^S
=\frac{1}{z}
=\frac{1}{\alpha }+
\bigg(1+\left(  \eta ^2+\frac{1 }{\eta ^2}-4  \eta
   -\frac{4  }{\eta }+6  \right)Q
   &+ \left(6\eta ^2+\frac{6}{\eta ^2}-24   \eta -\frac{24}{\eta}+36\right)Q^2+\mathcal O(Q^3)\bigg)\alpha \\
   &+\mathcal{O}(\alpha^2)\,,
   \end{aligned}
\end{align}
which for $\eta=1$, where the 5d SYM has maximal $\mathcal{N}=2$ supersymmetry, simplifies to
\begin{align}
\mathcal{W}^S\Big\vert_{\eta=1}=\frac{1}{\alpha}+\alpha
=\chi_{\boldsymbol{1/2}}^{\surm(2)},
\end{align}
which is the Weyl character of $\surm(2)$.

\subsection{Quantisation}
Quantisation of the curve \eqref{eq:M-curve} amounts to promoting the variables $y$ and $x$ to operators satisfying the commutation relation
\begin{eqnarray}
    [\hat y, \hat x] = \hbar\,.
\end{eqnarray}
Defining $Y \equiv e^{-\hat y}$, the exponentiated variables satisfy the following relation
\begin{eqnarray}
    Y \cdot X = p^{-1} X \cdot Y, \quad p = e^{\hbar},
\end{eqnarray}
and in this setup $\hbar$ can be identified with the Omega-deformation parameter $\epsilon_1$. The quantum version of the classical curve \eqref{eq:M-curve} becomes \cite{Chen:2020jla}
\begin{align} \label{eq:M-qcurve}
\left(Y^{-1}+q_\phi\frac{\theta_1(x - m)\theta(- \hbar + x +m)}{\theta_1(x)\theta_1(-\hbar + x)}Y\right)\cdot \widetilde \Psi(x)=q_{\phi}^{1/2} \mathcal{W}^S_{\hbar}(x)\cdot \widetilde\Psi(x)\,,
\end{align}
where one performed the substitutions
\begin{eqnarray}
    t \mapsto Y^{-1}, \quad t^{-1} \mapsto Y,
\end{eqnarray}
and shifted the quantum operator $\hat y$ by the classical value $\phi/2$ as follows:
\begin{eqnarray}
    \hat y \mapsto \hat y - \frac{\phi}{2},
\end{eqnarray}
resulting in factors $q_{\phi}^{1/2}$ (where $q_{\phi} \equiv e^{\phi}$ with $\phi$ being the M-string tension) and $q_{\phi}^{-1/2}$ in front of $Y$ and $Y^{-1}$. Note that \eqref{eq:M-qcurve} is now an operator equation which takes the form of a Hamiltonian operator acting on a wave-function $\widetilde \Psi$ with energy eigenvalue $q_{\phi}^{1/2} \mathcal{W}^S_{\hbar}$. In the limit $\hbar \rightarrow 0$, \eqref{eq:M-qcurve} gives back the classical curve \eqref{eq:M-curve}. The function $\widetilde \Psi$ admits an interpretation in terms of the expectation value of a co-dimension $2$ defect operator \cite{Chen:2020jla} with the perturbative contribution stripped off. Furthermore, note that the 5d gauge fugacity $\alpha$ is related to the 6d tensor VEV as
\begin{eqnarray}
    a_1 = \frac{\phi}{2} \quad \textrm{or} \quad \alpha = q_{\phi}^{1/2},
\end{eqnarray}
which implies that $q_{\phi}^{1/2}\mathcal W$ admits a series expansion in terms of $q_{\phi}$. Next, one can divide by $\widetilde \Psi(x)$ on both sides of \eqref{eq:M-qcurve}, which results in  
\begin{eqnarray} \label{eq:iWeyl}
    \mathcal{Y}(x+\hbar)^{-1} + \mathcal{P}(x) \mathcal{Y}(x) = q_{\phi}^{1/2} \mathcal{W}^S_{\hbar},
\end{eqnarray}
with the definitions
\begin{eqnarray}
    \mathcal{Y}(x) \equiv \frac{\widetilde \Psi(x-\hbar)}{\widetilde \Psi(x)}, \quad \mathcal{P} \equiv q_\phi\frac{\theta_1(x - m)\theta_1(- \hbar + x +m)}{\theta_1(x)\theta_1(-\hbar + x)}.
\end{eqnarray}
Using \eqref{eq:iWeyl}, one can recursively solve for $\mathcal{Y}(x)$ as a series expansion in $q_{\phi}$ order by order. 

Before proceeding, it is worth commenting that the difference equation \eqref{eq:M-qcurve} can be recast as an equation involving the full normalised defect partition function including the perturbative contribution,
\begin{equation}
    \Psi(x) \equiv \widetilde{Z}^{\textrm{def}}_{\textrm{pert}} \cdot \widetilde \Psi(x).
\end{equation}
This can be achieved by letting the Hamiltonian operator acting on $\widetilde \Psi$ undergo the following similarity transformation:
\begin{align}
\begin{aligned}
    &\left(Y^{-1}+q_\phi\frac{\theta_1(x - m)\theta_1(- \hbar + x +m)}{\theta_1(x)\theta_1(-\hbar + x)}Y\right) \\
    &\rightarrow  \widetilde{Z}^{\textrm{def}}_{\textrm{pert}} \left(Y^{-1}+q_\phi\frac{\theta_1(x - m)\theta_1(- \hbar + x +m)}{\theta_1(x)\theta_1(-\hbar + x)}Y\right) \left(\widetilde{Z}^{\textrm{def}}_{\textrm{pert}}\right)^{-1} \\
     &=  \frac{\widetilde{Z}^{\textrm{def}}_{\textrm{pert}}}{Y^{-1} \widetilde{Z}^{\textrm{def}}_{\textrm{pert}}} Y^{-1} + q_\phi\frac{\theta_1(x - m)\theta_1(- \hbar + x +m)}{\theta_1(x)\theta_1(-\hbar + x)} \frac{\widetilde{Z}^{\textrm{def}}_{\textrm{pert}}}{Y \widetilde{Z}^{\textrm{def}}_{\textrm{pert}}} Y.
    \end{aligned}
\end{align}
Using two results of \cite{Chen:2020jla}
\begin{equation}
    \frac{\widetilde{Z}^{\textrm{def}}_{\textrm{pert}}}{Y \widetilde{Z}^{\textrm{def}}_{\textrm{pert}}} = \frac{\theta_1(-\hbar + x)}{\theta_1(-\hbar + x + m)}
    \,, \qquad 
    \frac{\widetilde{Z}^{\textrm{def}}_{\textrm{pert}}}{Y^{-1}\widetilde{Z}^{\textrm{def}}_{\textrm{pert}}} = \frac{\theta_1(x + m + \hbar)}{\theta_1(x + \hbar)},
\end{equation}
one can rewrite \eqref{eq:M-qcurve}, now as an operator acting on the full normalised defect partition function $\Psi(x)$, as follows:
\begin{equation} \label{eq:M-qcurveb}
    \left(\frac{\theta_1(x + m + \hbar)}{\theta_1(x+\hbar)} Y^{-1} + q_{\phi} \frac{\theta_1(x-m)}{\theta_1(x)} Y\right) \Psi(x) = q_{\phi}^{1/2} \mathcal{W}^S_{\hbar}(x) \Psi(x). 
\end{equation}
Both forms, \eqref{eq:M-qcurve} and \eqref{eq:M-qcurveb}, of the difference equation, are encountered when considering the E-string in the next subsection.

\subsection{E-string curve and van Diejen difference operator}
\label{sec:van_Diejen}
Having reviewed the M-string quantum curve, one can now consider the E-string theory, its Seiberg-Witten curve and the relation to the van Diejen model. 
\subsubsection{Classical curve}
The Seiberg-Witten curve for the E-string theory compactified on an elliptic curve was derived for the case where a $\sorm(8)\times \sorm(8)$ subgroup of the flavour symmetry is preserved in \cite{Haghighat:2018dwe}. This corresponds to the case where an M5 brane is probing a $D_4$ singularity, also known as $D_4$ conformal matter, and corresponds to the following restriction of E-string flavour fugacities
\begin{eqnarray} \label{eq:D4-fugacities}
    \mu_l \mapsto - \mu_{l-4}, \quad l = 5,\ldots, 8.
\end{eqnarray}
In \cite[eqs. (4.41) \& (4.47)]{Haghighat:2018dwe}, it has been shown that the curve for this parameter choice takes the following form:
\begin{eqnarray}\label{classicalcurve}
    \frac{\prod_{l=1}^4 \theta_1(x \pm \mu_l)}{\theta_1(2x)^2} t^2 - \left(2 \frac{\prod_{l=1}^4\theta_1(x\pm \mu_l)}{\theta_1(2x)^2} + c \right) t + \frac{\prod_{l}^4 \theta_1(x\pm \mu_l)}{\theta_1(2x)^2} = 0,
\end{eqnarray}
where $c$ is an $x$-independent modular form to be determined. Dividing both sides of the equation by $t$ and bringing the $t$-independent pieces to the right-hand side, gives the following modified form of the curve:
\begin{eqnarray} \label{eq:E-curve}
    \frac{\prod_{l=1}^4 \theta_1(x \pm \mu_l)}{\theta_1(2x)^2} (t + t^{-1}) =  2 \frac{\prod_{l=1}^4\theta_1(x\pm \mu_l)}{\theta_1(2x)^2} + c.
\end{eqnarray}
In analogy to the discussion of M-string above, the aim is to identify the right-hand side with the Wilson surface expectation value 
\begin{eqnarray}
    \mathcal{W}^S(x) = 2 \frac{\prod_{l=1}^4\theta_1(x\pm \mu_l)}{\theta_1(2x)^2} + c.
\end{eqnarray}
As it turns out, the unknown $c$ can be identified with the 5d Wilson loop expectation value
\begin{eqnarray}
    c = \mathcal{W}^{L}.
\end{eqnarray}
This can be shown, similarly to the case of the M-string discussed above, by identifying the constant $c$ with the complex structure $1/z$ of the curve and defining a Seiberg-Witten differential 
\begin{equation}
    \lambda_{\textrm{SW}} \equiv y\, \diff x = \log t \frac{\diff X}{X},
\end{equation}
whose A-period admits a series expansion in terms of $z$. By inverting this series, one can express $c = 1/z$ in terms of a series in the parameter $q_{\phi}$ and compare with the result from the Wilson loop expectation value. This procedure is performed in the next subsection after discussing quantisation and finding a form of the curve where the full $E_8$-symmetry is manifest.

As a comment, \eqref{eq:E-curve} can be recast, by defining the variable
\begin{equation}
    \tilde t \equiv \frac{\prod_{l=1}^4 \theta_1(x \pm \mu_l)}{\theta_1(2x)^2} t,
\end{equation}
to the following form:
\begin{equation} \label{eq:E-curveb}
    \tilde t + \left(\frac{\prod_{l=1}^4 \theta_1(x \pm \mu_l)}{\theta_1(2x)^2}\right)^2 \tilde{t}^{-1} = \mathcal{W}^S.
\end{equation}
Both forms, \eqref{eq:E-curve} and \eqref{eq:E-curveb}, are useful when one proceeds to the quantisation.
\subsubsection{Quantisation}
Now, one proceeds to quantise the curve \eqref{eq:E-curve} by promoting the variables $y$ and $x$ to operators such that $Y = e^{-\hat y}$ acts as a shift operator $Y: x\rightarrow x-\epsilon_1$, or equivalently
\begin{align}
    Y\cdot X=p^{-1}X\cdot Y\,,
\end{align}
holds for the exponentials $Y$ and $X$. The strategy for quantising the classical curve \eqref{eq:E-curve}, following the observations made in the M-string case, is to interpret the factors in front of $t$ and $t^{-1}$ as ratios between the shifted and unshifted perturbative part of the defect expectation value. To this end, it turns out to be useful to define the quantum wave-function as a shifted co-dimension $2$ defect partition function as follows:
\begin{equation}
    \Psi(x) \equiv \widetilde{Z}^{\mathrm{E-str+def}}(x+\epsilon_1/2).
\end{equation}
Similarly, one can define the pure normalised instanton wave-function by
\begin{equation}
    \widetilde \Psi(x) \equiv \widetilde{Z}_{\mathrm{str}}^{\mathrm{E-str+def}}(x+\epsilon_1/2).
\end{equation}
Acting with $Y$ on the perturbative part of the defect partition function \eqref{eq:6d_perturbative_partition_function_with_defect}, and using
\begin{align}
    \Gamma_e(pz;\,p, Q)=\theta(z;\, Q)\,\Gamma_e(z;\, p, Q)\,,
\end{align}
where $\theta(X;\, Q)=\prod_{n=1}^{\infty}(1-X Q^{n-1})(1-X^{-1}Q^{n})$ is the $Q$-theta function, and by further using the identity
\begin{align}
    \theta(X; Q)=-i Q^{-\frac{1}{12}}\eta^{-1}\sqrt{X}\theta_1(x),
\end{align}
one arrives at the following ratio of shifted defect partition functions:
\begin{equation} \label{eq:pertratio}
    \frac{\widetilde{Z}_{\mathrm{pert}}^{\mathrm{E-str+def}}(Xp^{\frac{1}{2}})}{Y \cdot \widetilde{Z}_{\mathrm{pert}}^{\mathrm{E-str+def}}(Xp^{\frac{1}{2}})} = q_{\phi} \mathcal{V}(x) \,.
\end{equation}
For convenience and for later use, the function $\mathcal{V}(x)$ has been introduced, which is defined as 
\begin{equation} 
\label{eq:Vx}
    \mathcal{V}(x) \equiv -p X^{-2}\frac{\prod_{n=1}^8\theta_1(x-\mu_n-\ep_1/2)}{\eta^6\theta_1(2x)\theta_1(2x-\ep_1)} \,.
\end{equation}
In the limit $\epsilon_1 \rightarrow 0$ and for the choice of fugacities given in \eqref{eq:D4-fugacities}, the above takes the form of the factor in front of $(t + t^{-1})$ in \eqref{eq:E-curve}, up to $X^2$-factors which can be absorbed into a redefinition of $t$.
This suggests, in analogy to the M-string case, that the quantum version of the curve \eqref{eq:E-curve}, with all $E_8$-fugacities restored, takes the form
\begin{equation} 
\label{eq:E-qcurve}
    \left(\mathcal{V}(x) Y + \mathcal{V}(-x) Y^{-1}\right) \Psi(x) = \Lambda(x) \Psi(x)\,,
\end{equation}
where the arguments of the function $V(x)$ are chosen such as to respect the symmetry $x \mapsto -x$ as originally observed in \cite{Haghighat:2018dwe}. Although the eigenvalue $\Lambda(x)$ is not yet specified, the left-hand side of the operator equation \eqref{eq:E-qcurve} is known as part of what is called \textit{van Diejen}-operator in the mathematics literature and which has already appeared in the physics literature in connection to the E-string in \cite{Nazzal:2018brc}. The next paragraph serves as an introduction to the van Diejen operator and shows that this operator provides indeed the correct framework to quantise the E-string curve.

\paragraph{Van Diejen operator.}
The van Diejen difference operator was first proposed by van Diejen \cite{MR1275485} as a $BC_N$ generalisation of the $A$-type relativistic Calogero-Moser  model. For rank one case, the $BC_1$ Hamiltonian has two different types, the hyperbolic and the elliptic type, which contain four and eight parameters, respectively. It has been proven in \cite{MR2153340} that these Hamiltonians have $D_4$/$D_8$ invariance, while the spectrum of the elliptic case has $E_8$ Weyl invariance \cite{MR3313680}. In this paper, the focus lies on the elliptic 8-parameter $BC_1$ Hamiltonian, whose connection with the elliptic difference Painlev\'e equation \cite{MR1882403} is discussed in \cite{MR4120359}.

Following the conventions of \cite{MR2153340,MR3313680}, one introduces
\begin{subequations}
\begin{align}
a&= \frac{1}{2}(\ap+\am) \,,
\\
R(r,a;x)&=\prod_{k=1}^{\infty}
\left(1-\exp[2irx-(2k-1)ra]\right)
\left(1-\exp[-2irx-(2k-1)ra] \right)\,, \quad a>0\, ,
\\
R_{\pm}(x)&=R(r,a_{\pm};x) \,.
\end{align}
\end{subequations}
The $BC_1$ analytic difference operator is defined as
\begin{subequations}
\begin{align}
\hat{H}=V(x)\exp(-i \am\px)+V(-x)\exp(i \am\px)+V_b(x),
\end{align}
with 
\begin{align}
V(x)=\frac{\prod_{n=1}^8R_{+}(x-h_n-i\am/2)}{R_{+}(2x+i\ap/2)R_{+}(2x-i\am+i\ap/2)},
\end{align}
\end{subequations}
and it is shown below that $V(x)$ as defined above is equivalent to the form given in \eqref{eq:Vx} after a suitable identification of parameters. To define $V_b(x)$, one needs to introduce a few more definitions. Firstly, one introduces half-periods $\omega_I$, $I=1,\ldots,4$:
\begin{align}
\omega_1=0
\,,\qquad 
\omega_2=\frac{\pi}{2r}
\,,\qquad 
\omega_3=-\frac{ia_+}{2}-\frac{\pi}{2r}
\,,\qquad 
\omega_4=\frac{ia_+}{2}\,.
\end{align}
Secondly, one defines four functions that involve a new parameter $\sigma\in\mathbb{C}$:
\be
\mathcal{E}_I(\sigma;z)\equiv \frac{R_+(z+\sigma-ia-\omega_I)R_+(z-\sigma+ia-\omega_I)}{R_+(z-ia-\omega_I)R_+(z+ia-\omega_I)} \,,\qquad I=1,\ldots,4.
\ee
Thirdly, one defines
\begin{subequations}
\begin{align}
p_1&\equiv\prod_n R_+(h_n)\,,\qquad p_2\equiv\prod_n R_+(h_n-\omega_1)\,,
\\
p_3&\equiv\exp(-2ra_+)\prod_n \exp(i r h_n)R_+(h_n-\omega_3)
\,,
\\
p_4&\equiv\exp(-2ra_+)\prod_n \exp(-i r h_n)R_+(h_n-\omega_2)\,.
\end{align}
\end{subequations}
Such that $V_b(x)$ can be defined as
\be
V_b(z)\equiv \frac{\sum_{I=1}^4p_I[\mathcal{E}_I(\sigma;z)-\mathcal{E}_I(\sigma;z_I)]}{2R_+(\sigma-ia_+/2)R_+(\sigma-ia_--ia_+/2)},
\ee
with\footnote{Note that the definition for $z_I$ in \cite{MR2153340} and \cite{MR3313680} are different. If one follows the notation of \cite{MR3313680}, then $z_I=2\pi i r \omega_I$. These two choices only differ in the constant terms $c$ in \eqref{V0a1}, which are not important in the discussion here.}
\be
z_1=z_4=\omega_2,\quad z_2=z_3=0.
\ee
In order to connect with the case at hand, one defines $h_n\equiv\mu_n/(2 i r )-ia_+/2$, together with the identifications $Q_{}=e^{-2r a_+}$, $\epsilon_1=ia_-$, $x=2 i r z$ and the notation of theta functions, one arrives at
\begin{subequations}
\begin{align}
\label{Vpx}
V(x)&=-Q^{-\frac{1}{2}}p^{\frac{3}{2}}X^{-2}\frac{\prod_{n=1}^8\exp(\frac{1}{2}\mu_n)\theta_1(x-\mu_n-\ep_1/2)}{\eta^6\theta_1(2x)\theta_1(2x-\ep_1)}
\\
\label{Vmx}
V(-x)&=-Q^{-\frac{1}{2}}p^{\frac{3}{2}}X^{2}\frac{\prod_{n=1}^8\exp(\frac{1}{2}\mu_n)\theta_1(x+\mu_n+\ep_1/2)}{\eta^6\theta_1(2x)\theta_1(2x+\ep_1)}
\end{align}
\end{subequations}
and
\begin{subequations}
\be\label{Vb}
V_b(x)= p^{\frac{1}{2}}Q^{-\frac{1}{2}}\frac{\sum_{I=1}^4[\tilde{\mathcal{E}}_I(\sigma;x)-\tilde{\mathcal{E}}_I(\sigma;x_I)]\prod_{n=1}^8\exp(\frac{1}{2}\mu_n){\theta_I(\mu_n)}}{-2\eta^{6}\theta_{1}(\sigma)\theta_1(\sigma-\ep_1)},
\ee
where
\be
\tilde{\mathcal{E}}_{I}( \sigma; X)=\frac{\theta_I(x+\ep_1/2-\sigma)\theta_I(x-\ep_1/2+\sigma)}{\theta_I(x-\ep_1/2)\theta_I(x+\ep_1/2)},\quad I=1,\ldots,4,
\ee
and
\be
x_1=x_4=i\pi\,,\qquad x_2=x_3=0.
\ee
\end{subequations}
With all ingredients at hand, one can define the quantum curve as a time-independent Schr\"odinger equation which, after multiplication by $\exp\left(-\frac{1}{2}\mu_n\right) p^{-\frac{1}{2}} Q^{\frac{1}{2}}$, takes the following form:
\begin{equation} \label{eq:E-diffeq}
\frac{1}{z}\Psi(x)=
\mathcal{V}(x) \Psi(x-\ep_1) + \mathcal{V}(-x) \Psi(x+\ep_1)+V_0(X)\Psi(x),
\end{equation}
where $\mathcal{V}(x)$ was defined in \eqref{eq:Vx} and
\be
V_0(X)=\frac{\sum_{I=1}^4[\tilde{\mathcal{E}}_I(\sigma;x)-\tilde{\mathcal{E}}_I(\sigma;x_I)]\prod_{n=1}^8{\theta_I(\mu_n)}}{-2\eta^{6}\theta_{1}(\sigma)\theta_1(\sigma-\ep_1)}\,.
\ee
Therefore, $\Psi(x)$ is regarded as the eigenfunction and $\frac{1}{z}$ the eigenvalue of the Hamiltonian.

As a remark, note that even though $V_0(X)$ has an extra parameter $\sigma$, it is actually independent of $\sigma$, as proven in \cite{MR3313680}. More precisely, one can set $\sigma$ to any arbitrary value, and the results are the same up to an $x$ independent term. For example, if one sets $\sigma=0$
\begin{subequations}
\be\label{V0a1}
V_0(X)=\frac{1}{2}\sum_{I=1}^4\frac{\prod_{n=1}^8{\theta_I(\mu_n)}}{\eta^6\theta_{1}'(0)\theta_1( \ep_1)}\left(\frac{\theta_I'(x-\ep_1/2)}{\theta_I(x-\ep_1/2)}-\frac{\theta_I'(x+\ep_1/2)}{\theta_I(x+\ep_1/2)}\right)+c,
\ee
where
\be\label{mirrormapa1}
c=\sum_{I=1}^4\frac{\prod_{n=1}^8{\theta_I(\mu_n)}}{\eta^6\theta_{1}'(0)\theta_1( \ep_1)}\frac{\theta_I'(x_I+\ep_1/2)}{\theta_I(x_I+\ep_1/2)} \,.
\ee
\end{subequations}
In the following discussion, the constant term $c$ is absorbed in a redefinition of $\frac{1}{z}$ as it is $X$-independent. A remarkable outcome of rewriting the van Diejen operator in the above form is that the function $V_0(X)$ (with the constant $c$-term now stripped off) is identical to the negative of the Wilson surface expectation value at instanton order one, see \eqref{eq:E-W1}:
\begin{equation}
    V_0(X) = - \widetilde W_1(X).
\end{equation}
As a next step, introduce the function $\mathcal{Y}(x)$ given by the ratio
\be \label{eq:Ydef}
\mathcal{Y}(x)=-q_{\phi}pX^{-2}
\frac{\prod_{n=1}^8\theta_1(x-\mu_n-\ep_1/2)}{\eta^6\theta_1(2x)\theta_1(2x-\ep_1)}\frac{\Psi(x-\ep_1)}{\Psi(x)} = \frac{\widetilde \Psi(x-\epsilon_1)}{\widetilde \Psi(x)},
\ee
where the identity \eqref{eq:pertratio} has been used to show that $\mathcal{Y}(x)$ can be written in terms of the ratio of the pure instanton part of the defect partition function
\begin{equation}
    \widetilde \Psi(x) = \widetilde{Z}_{\mathrm{pert}}^{\mathrm{E-str+def}}(X p^{\frac{1}{2}})^{-1} \Psi(x)\,.
\end{equation}
Then the difference equation \eqref{eq:E-diffeq} can be rewritten in terms of $\mathcal{Y}$ as follows: 
\be
\frac{q_{\phi}}{z}=\mathcal{Y}(x)+\frac{\prod_{n=1}^8\theta_1(x\pm\mu_n+\ep_1/2)}{\eta^{12}\theta_1(2x)\theta_1(2x+\ep_1)^2\theta_1(2x+2\ep_1)}\frac{q_{\phi}^2}{\mathcal{Y}(x+\ep_1)}+q_{\phi}V_0(x),
\label{eq:van_Diejen_equation_version_2}
\ee
where $q_{\phi}$ is defined from
\be\label{Qmap}
\log q_{\phi} = -\mathrm{Res}_{x\rightarrow \infty} {\log(q_{\phi}^{-1} \mathcal{Y}(x)})=- \mathrm{Res}_{X\rightarrow 0} \frac{\log(q_{\phi}^{-1} \mathcal{Y}(x))}{X}=\log z+\mathcal{O}(z).
\ee
As customary, \eqref{Qmap} is referred to as the mirror map, from which one can compute the inverse series. The $q_\phi$ expansion of the inverse series becomes
\be \label{eq:E-Wloop}
\frac{1}{z}=\frac{1}{q_{\phi}}+c+\sum_{i=2}^{\infty} a_i q_{\phi}^{i-1},
\ee
where the expression of $c$ can be found in \eqref{mirrormapa1}, and 
\be\begin{split}
a_2|_{\ep_{1,2}\rightarrow 0}=&-\frac{1}{576 Q\, \eta^{12}}\left(2A_1^2(E_2^2-E_4) +5B_2(E_2E_4-E_6)+3A_2(E_2 E_6-E_4^2)\right),\\
a_3|_{\ep_{1,2}\rightarrow 0}=&-\frac{1}{2^73^5 Q^3\, \eta^{36}}\left(12A_1^3 (E_2^3-3 E_4 E_2+2 E_6)+27 A_1 A_2 (E_6 E_2^2+2 E_4^2 E_2-3 E_4 E_6)\right.\\
&\left.+45A_1B_2 (E_4 E_2^2+2 E_6 E_2-3 E_4^2)+14 A_3E_2 (E_4^3-E_6^2)+40B_3 (E_4^3-E_6^2)\right),\\
&\vdots
\end{split}\ee 
Here $A_n,\, B_n$ are elements in the ring of $E_8$ invariant Jacobi forms proposed in \cite{Eguchi:2002nx}, and proven in \cite{Wang:2018fil}. $E_n$ denotes the Eisenstein series. Note that $c$, the constant term of $\frac{1}{z}$, does not have $E_8$ symmetry. 

In order to compare with the elliptic genera of the co-dimension 2 defect, it is useful to define the $q_{\phi}$ expansion of $\mathcal{Y}(x)$ as $\mathcal{Y}(x)=\sum_{i=0}^{\infty}\mathcal{Y}_i(x)q_{\phi}^i$ such that one can recursively solve for the expansion coefficients. One finds
\begin{subequations}
\begin{align}
\mathcal{Y}_0(x)&=1 \,, \\
\mathcal{Y}_1(x)&=-c-V_0(x) \,, \\
\mathcal{Y}_2(x)&=-a_2-\frac{\prod_{n=1}^8\theta_1(x\pm\mu_n+\ep_1/2)}{\eta^{12}\theta_1(2x)\theta_1(2x+\ep_1)^2\theta_1(2x+2\ep_1)}\,.
\end{align}
\end{subequations}
Thus, it is straightforward to obtain 
\begin{align} 
\label{eq:Y1}
\mathcal{Y}_1(x)&=\widetilde{Z}_{k=1}^{\mathrm{def}}(x-\ep_1/2)-\widetilde{Z}_{k=1}^{\mathrm{def}}(x+\ep_1/2)\,,
\\
\label{eq:Y2}
\mathcal{Y}_2(x)&=\widetilde{Z}_{k=2}^{\mathrm{def}}(x-\ep_1/2)-\widetilde{Z}_{k=2}^{\mathrm{def}}(x+\ep_1/2)-\widetilde{Z}_{k=1}^{\mathrm{def}}(x+\ep_1/2)\left(\widetilde{Z}_{k=1}^{\mathrm{def}}(x-\ep_1/2)-\widetilde{Z}_{k=1}^{\mathrm{def}}(x+\ep_1/2)\right)\,.
\end{align}
The expressions \eqref{eq:Y1} and \eqref{eq:Y2} can be equivalently derived from and are consistent with the definition of $\mathcal{Y}$ as a ratio of defect partition functions given in \eqref{eq:Ydef}, which finally shows that the co-dimension $2$ defect partition function indeed satisfies the difference equation \eqref{eq:E-diffeq}.

Finally, the claim is that $\frac{1}{z}$ can be identified with the 5d supersymmetric Wilson loop $\mathcal{W}^L$ in the fundamental representation, as discussed in Section \ref{sec:E-Wloop}. Here the 5d theory can be $\surm(2)$ theory or $\sprm(1)$ theory with 8 fundamental flavours. The first 5d theory has an enhanced $E_8$ global symmetry, and the latter 5d theory has $\sorm(16)$ global symmetry, i.e.\ the maximal subgroup of $E_8$. The global symmetry is broken due to the effect of circle holonomy \cite{Kim:2014dza}, with the following shift
\be
\mu_8 \rightarrow \mu_8-{\tau},\quad {\phi}\rightarrow {\phi}+\frac{1}{2}\tau-\mu_8,
\ee
In the massless limit, the series expansion for $\frac{1}{z}$ becomes
\begin{align}
\begin{aligned}
\frac{1}{z} &=
q_{\phi}
+\frac{1}{q_\phi}+\text{const.}+\left(128 \QQ-256 \QQ^2+384 \QQ^3-512 \QQ^4+640 \QQ^5-768 \QQ^6+\mathcal{O}(\QQ^7)\right) {Q}^{\frac{1}{2}}\\
   &\quad +\left(2560 \QQ-14848 \QQ^2+56832 \QQ^3-168960 \QQ^4+422400 \QQ^5-929280 \QQ^6+\mathcal{O}(\QQ^7)\right) Q\\
   &\quad +\ldots
\end{aligned}
\label{eq:Wilson_loop_from_van_Diejen}
\end{align}
which agrees with the Wilson loops calculation in Section \ref{sec:E-Wloop}. The agreement at quantum level and with non-vanishing mass parameters has been verified as well.

In summary, the results of this section show that the van Diejen operator can be recast as an eigenvalue equation with eigenvector given by $\Psi(x)$ and eigenvalue the Wilson surface expectation value:
\begin{eqnarray}
    \left(\mathcal{V}(x) Y +\mathcal{V}(-x) Y^{-1}\right) \Psi(x) = \left(-V_0(x) + \mathcal{W}^L_{\hbar}\right) \Psi(x) = q_{\phi}^{-1} \mathcal{W}^S_{\hbar}(x) \Psi(x),
\end{eqnarray}
which is exactly the claim of \eqref{eq:E-qcurve}.

\paragraph{Classical limit.}
Lastly, one is now in the position to check the classical limit and compare with the classical curve \eqref{classicalcurve} derived in \cite{Haghighat:2018dwe} from thermodynamic limit. The curve in \cite{Haghighat:2018dwe} has been derived from the $D_4$ conformal matter theory with manifest $D_4\times D_4$ symmetry, and with the following identifications:
\be\label{D4D4constraint}
\mu_l \rightarrow -\mu_{l-4}, \quad l=5,\ldots,8.
\ee
With the constraints \eqref{D4D4constraint} and $p=1$, define $\Sigma(x)=X^2 V(x)+X^{-2}V(-x)$, such that the following holds:
\be\label{Vbsigma}
V_b(x)=-\Sigma(x)+ X\text{ independent terms}.
\ee
The proof of \eqref{Vbsigma} proceeds by verifying that the residues of $V_b(x)+\Sigma(x)$ at simple poles $x=0$ and at $i\pi$ vanish. 
It is straightforward to verify that in the classical limit $p=1$ together with the constraints \eqref{D4D4constraint} one finds
\be
X^2V(x)=X^{-2}V(x)=-Q^{-\frac{1}{2}}\frac{\prod_{n=1}^4\theta_1(x\pm\mu_n)}{\eta^6\theta_1(2x)^2}
\ee
such that the curve becomes
\be
\frac{Q^{\frac{1}{2}}}{z}=\frac{\prod_{n=1}^4\theta_1(x\pm\mu_n)}{\eta^6\theta_1(2x)^2}(Y+Y^{-1}+2),
\ee
which matches the curve given in \cite{Haghighat:2018dwe} after some normalisation of $z$.

In the end of the section, we make a few remarks on the role of the Wilson surface defect in the quantum curve. First we want to clarify that the co-dimension four surface defect introduced in the beginning of Section \ref{sec:Wilson_surface_line}, is not exactly the same as Wilson surface/loop in 6d/5d. The defect, probed by the heavy string stretched between the NS5 and D4', genuinely depends on the fugacity $v$ labeling the $\urm(1)$ symmetry associated to the D4' brane. In computation of Wilson surface defect in rank $m$ anti-symmetric representations, for example, the Wilson surface defect will in general depend on a collection of fugacities labeling the symmetries of the stack of D4' branes \cite{Agarwal:2018tso}. Therefore one has to extract the Wilson surface expectation value from the Wilson surface defect. Now, for the case of the E-string theory, an observation to not collect the one-instanton contribution of the Wilson surface defect, is that the 6d/5d duality implies the 5d $\surm(2)$ coulomb moduli $\alpha$ is mapped to the 6d tensor moduli $q_\phi$. The 5d Wilson loop starting from the SU(2) character thus implies the 6d Wilson surface should be written as $q_\phi^{-1}(1+q_\phi^2+\dots)$. The prefactor $q_\phi^{-1}$ is due to the perturbative contribution to the Wilson surface, while the $1+q_\phi^2$ in the bracket implies that the one-instanton correction to the Wilson surface defect should be considered as a part of the quantum curve, rather than Wilson surface vev as the eigenvalue of the curve. Furthermore, in this section, we indeed verified that the one-instanton correction is precisely the 4-theta potential term in the van Diejen Hamiltonian.
\section{A Path integral derivation}
\label{sec:path_integral}
In this section, the van Diejen equation for the E-string instanton partition function is derived via the path integral formalism.
\subsection{Path integral representation}
\label{sec:path_integral_representation}
Since the $k$-th instanton strings are modelled by the $\orm(k)$ elliptic genera, they receive various contributions from both continuous and discrete holonomy sectors as discussed in Section \ref{sec:partition_function_elliptic_genus}. In \cite{Haghighat:2018dwe}, it has been shown that the partition function is dominated by the contribution from the continuous holonomy sector in the thermodynamic limit $\epsilon_{1,2}\rightarrow 0$. As shown below, this statement still holds in the NS-limit $\epsilon_2\rightarrow 0$. To begin with, the $2n$ string contribution is evaluated in the continuous holonomy sector and then shown that instanton strings from other sectors give the same contribution as the continuous ones at $\mathcal{O}(\epsilon_2^{-1})$. 
\paragraph{Continuous holonomy sector.}
The $2n$ string contributions in the continuous holonomy sector reads
\begin{align}
Z_{2n,\,\mathrm{cont.}}^{\sorm(2n)}
&=\frac{1}{2^n n!}
 \int \prod_{i=1}^n \diff u_i \left( \frac{ 2\pi \eta^2 \vartheta_1(2 
\epsilon_+)}{\vartheta_1( \epsilon_1) \vartheta_1( \epsilon_2)} \right)^n
\cdot \prod_{i=1}^n \left(\frac{1}{\vartheta_1 (\pm 2u_i+\epsilon_{1,2})}\prod_{l=1}^8\vartheta_1(\pm u_i+\mu_l)\right)\notag\\
&\quad \cdot\prod_{1\leq i< j\leq n}
\frac{\vartheta_1 (\pm u_i\pm u_j)  \vartheta_1 
(\pm u_i\pm u_j+2\epsilon_+)}{\vartheta_1(\pm u_i\pm u_j+\epsilon_1) 
\vartheta_1(\pm u_i\pm u_j+\epsilon_2)}\,,\notag\\
&=\frac{1}{2^n n!}
 \int \left(\prod_{i=1}^n \frac{\diff u_i}{2\pi i}\right) \left( \frac{ 2\pi \eta^2 \vartheta_1(2 
\epsilon_+)}{\vartheta_1( \epsilon_1) \vartheta_1( \epsilon_2)} \right)^n
\cdot \prod_{i=1}^n \left(\frac{1}{\vartheta_1 (\pm 2u_i+\epsilon_{1,2})}\prod_{l=1}^8\vartheta_1(\pm u_i+\mu_l)\right)\notag\\
&\quad \cdot\prod_{i\neq j}^n\left(
\frac{\vartheta_1 (\pm u_i\pm u_j)  \vartheta_1 
(\pm u_i\pm u_j+2\epsilon_+)}{\vartheta_1(\pm u_i\pm u_j+\epsilon_1) 
\vartheta_1(\pm u_i\pm u_j+\epsilon_2)}\right)^{\frac{1}{2}}\notag\\
&\equiv\frac{1}{2^n n!}\int \prod_{i=1}^n \diff u_i \,\Delta^n
\cdot \prod_{i=1}^n Q(u_i)\cdot \prod_{i\neq j}D(u_i, u_j)^{\frac{1}{2}}
\end{align}
where $2^n n!$ is the order of the Weyl group of $\sorm(2n)$ and $\vartheta_1(z)\equiv \frac{\theta_1(z)}{\eta}$, see Appendix \ref{app:theta_fct}. Moreover, the following notation has been introduced: 
\begin{subequations}
\begin{align}
D(u, u^\prime) &=\frac{\vartheta_1 (\pm u\pm u^\prime)  \vartheta_1 
(\pm u\pm u^\prime+2\epsilon_+)}{\vartheta_1(\pm u\pm u^\prime+\epsilon_1) 
\vartheta_1(\pm u\pm u^\prime+\epsilon_2)}\,,\\
Q(u) &=\frac{1}{\vartheta_1 (\pm 2u+\epsilon_{1,2})}\prod_{l=1}^8\vartheta_1(\pm u+\mu_l)\,,\\
\Delta &=\frac{2\pi \eta^2 \vartheta_1(2 
\epsilon_+)}{\vartheta_1( \epsilon_1) \vartheta_1( \epsilon_2)} \,.
\end{align}
\end{subequations}
Next, introduce a density function
\begin{align}
\bar\rho(u)\equiv \frac{1}{\Delta}\sum_{i=1}^n\delta(u-u_i)\,,
\label{eq:density_function}
\end{align}
such that the integrand in the partition function can be recast as
\begin{align}
&\prod_{i=1}^n Q(u_i)\cdot \prod_{i\neq j}D(u_i, u_j)^{\frac{1}{2}}=\exp\left(\sum_{i=1}^n\log Q(u_i)+\frac{1}{2}\sum_{i\neq j}\log D(u_i, u_j)\right)\notag\\
&=\int\!\mathcal D\rho\,\delta\left(\rho(u)-\bar\rho\right)\cdot\exp\left(\frac{\Delta^2}{2}\int\! \diff u \, \diff u^\prime\,\rho(u^\prime)\log D(u, u^{\prime})\rho(u)+\Delta\int\!\diff u\, \rho(u)\log Q(u)\right)\notag\\
&=\int\!\mathcal D\rho\,\mathcal D\lambda\,\exp\bigg(\frac{\Delta^2}{2}\int\! \diff u \, \diff u^\prime\,\rho(u^\prime)\log D(u, u^{\prime})\rho(u)+\Delta\int\!\diff u\,\rho(u)\log Q(u)\notag\\
&\quad +i\int\!\diff u\,\lambda(u)\cdot\left(\rho(u)-\bar\rho\right)\bigg)\,,
\end{align}
where the Fourier transform of the delta distribution
\begin{align}
\delta(\rho(u)-\bar\rho)=\int\!\mathcal D\lambda\,\exp\left(i\int\! \diff u\, \lambda(u)\left(\rho(u)-\bar\rho\right)\right)
\end{align}
has been used. Therefore, for the partition function from the continuous sector, one finds
\begin{align}
Z_{\mathrm{str,\, cont.}}&=\sum_{n=0}^{\infty}q_\phi^{2n}Z_{2n,\,\mathrm{cont.}}^{\sorm(2n)}\notag\\
&=\int\!\mathcal D\rho\,\mathcal D\lambda\,\sum_{n=0}^{\infty}\frac{q_\phi^{2n}}{2^n n!}\int\!\left(\prod_{i=1}^n \frac{\diff u_i}{2\pi i}\right)\cdot\Delta^n e^{-i\int\! \diff u\,\lambda(u)\cdot\bar\rho(u)}\\
&\cdot \exp\left(\frac{\Delta^2}{2}\int\! \diff u \, \diff u^\prime\,\rho(u^\prime)\log D(u, u^{\prime})\rho(u)+\Delta\int\!\diff u\, \rho(u)\log Q(u)+i\!\int\!\diff u\,\lambda(u)\cdot\bar\rho(u)\right)\,.\notag
\end{align}
The sum over $n$ factor can be evaluated
\begin{align}
\begin{aligned}
\sum_{n=0}^{\infty}\frac{q_\phi^{2n}}{2^n n!}\int\!\left(\prod_{i=1}^n \frac{d u_i}{2\pi i}\right)\cdot\Delta^n e^{-i\int\!\diff u\,\lambda(u)\cdot\bar\rho(u)}&=\sum_{n=0}^\infty\frac{1}{n!}\left(\frac{q_\phi^2\Delta}{2}\int\!\frac{\diff u}{2\pi i}e^{-\frac{i}{\Delta}\lambda(u)}\right)^n \\
&=\exp\left(\frac{q_\phi^2\Delta}{2}\int\!\frac{\diff u}{2\pi i}e^{-\frac{i}{\Delta}\lambda(u)}\right)
\end{aligned}
\end{align}
such that the expression becomes
\begin{align}
\begin{aligned}
Z_{\mathrm{str,\, cont.}}=\int\!\mathcal D\rho\,\mathcal D\lambda\,\exp&\bigg(\frac{\Delta^2}{2}\int\! \diff u \, \diff u^\prime\,\rho(u^\prime)\log D(u, u^{\prime})\rho(u) \\
&+\Delta\int\!\diff u\, \rho(u)\log Q(u)+i\!\int\!\diff u\,\lambda(u)\cdot\rho(u)+\frac{q_\phi^2\Delta}{4\pi i}\int\! \diff u\,e^{-\frac{i}{\Delta}\lambda(u)}\bigg)\,.
\end{aligned}
\end{align}
One can further employ a shift in the auxiliary variable $\lambda(u)$ as
\begin{align}
\lambda(u)=\lambda^\prime(u)-i\Delta\log(-q_\phi^2)\,,
\end{align}
to remove the explicit $q_\phi^2$ dependence of the last term. Finally, one ends up with
\begin{align}
\begin{aligned}
Z_{\mathrm{str,\, cont.}}=&\int\!\mathcal D\rho\,\mathcal D\lambda^\prime\,\exp\bigg(\frac{\Delta^2}{2}\int\! \diff u \,\diff u^\prime\,\rho(u^\prime)\log D(u, u^{\prime})\rho(u)\\
&+\Delta\int\!\diff u\, \rho(u)\log\left(-q_\phi^2 Q(u)\right)+i\!\int\!\diff u\,\lambda^\prime(u)\cdot\rho(u)-\frac{\Delta}{4\pi i}\int\! \diff u\,e^{-\frac{i}{\Delta}\lambda^\prime(u)}\bigg)\,.
\end{aligned}
\label{eq:path_integral_reprensentation_of_instanton_partition_function}
\end{align}
Now, consider the leading order approximation of \eqref{eq:path_integral_reprensentation_of_instanton_partition_function} in the NS-limit $\epsilon_2\rightarrow 0$. It is useful to evaluate the small $\epsilon_2$ expansions of the following terms: 
\begin{subequations}
\label{eq:leading_order_approximations}
\begin{align}
\Delta &=\epsilon_2^{-1}+\mathcal O(1)\,,\\
\log D(u, u^\prime) &=\left(K(u, u^\prime;\, \epsilon_1)-K(u, u^\prime;\, 0)\right)\cdot\epsilon_2+\mathcal O(\epsilon_2^2)\,,\\
Q(u) &=Q_0(u)+\mathcal O(\epsilon_2)\,,
\end{align}
\end{subequations}
where the following conventions have been used:
\begin{subequations}
\begin{align}
K(u, u^\prime;\, \beta) &\equiv \frac{\partial}{\partial\beta}\log\vartheta_1(\pm u\pm u^\prime+\beta)\notag\\
&=\frac{\theta_1^{\prime}(u+u^\prime+\beta)}{\theta_1(u+u^\prime+\beta)}-\frac{\theta_1^{\prime}(u+u^\prime-\beta)}{\theta_1(u+u^\prime-\beta)}+\frac{\theta_1^{\prime}(u-u^\prime+\beta)}{\theta_1(u-u^\prime+\beta)}-\frac{\theta_1^{\prime}(u-u^\prime-\beta)}{\theta_1(u-u^\prime-\beta)} \,,\\
Q_0(u)&\equiv Q(u)|_{\epsilon_2=0} =\frac{\prod_{l=1}^8\vartheta_1(\pm u+\mu_l)}{\vartheta_1(\pm 2u)\vartheta_1(\pm 2u+\epsilon_1)}=\frac{\prod_{l=1}^8\theta_1(u\pm\mu_l)}{\eta^{12}\theta_1(2u)^2\theta_1(2u\pm\epsilon_1)}\,.
\end{align}
\end{subequations}
It is worth noticing that $K(u, u^\prime; 0)=0$ by the virtue of the odd and even properties of $\vartheta_1$ and $\vartheta_1^\prime$. Therefore, in leading order $\mathcal O(\epsilon_2^{-1})$, one has
\begin{align}
\begin{aligned}
Z_{\mathrm{str,\, cont.}}=\int\!\mathcal \mathcal D\lambda^\prime\ldots D\rho\,\exp &\bigg(\frac{1}{2\epsilon_2}\int\! \diff u \,, \diff u^\prime\,\rho(u^\prime)K(u,u^\prime;\,\epsilon_1)\rho(u) \\
&+\frac{1}{\epsilon_2}\int\!\diff u\, \rho(u)\log\left(-q_\phi^2 Q_0(u)\right)+\mathcal O(1)\bigg)\,,
\end{aligned}
\label{eq:leading_order_of_instanton_partition_function}
\end{align}
where $\ldots$ denotes terms of $\lambda^\prime(u)$ that are irrelevant to the subsequent saddle point analysis.

\paragraph{Discrete holonomy sectors.} 
Now, string contributions from the discrete holonomy sectors have to be considered. As discussed in Section \ref{sec:partition_function_elliptic_genus}, there are seven discrete sectors for $\sorm(2n)$ and eight for $\sorm(2n+1)$,
\begin{alignat}{4}
\begin{aligned}
&\sorm(2n): & \quad  &\{\pm u_1,\dots, \pm u_{n-1}; u_I, u_J\} &\quad &\mathrm{and}& \quad  &\{\pm u_1,\dots, \pm u_{n-2}; 0, \tfrac{1}{2}, \tfrac{\tau+1}{2}, \tfrac{\tau}{2}\} \\
&\sorm(2n+1): & \quad  &\{\pm u_1,\dots, \pm u_{n}; u_I\} & \quad &\mathrm{and} & \quad  &\{\pm u_1,\dots, \pm u_{n-1}; u_I, u_J, u_K\}\,,
\end{aligned}
\end{alignat}
where $u_i$ denote, as before, the continuous moduli, and $u_I\in \{0, \frac{1}{2}, \frac{\tau+1}{2}, \frac{\tau}{2}\}$ are the discrete holonomies. In the following, the focus is placed on the $\{\pm u_1,\dots, \pm u_{n}; u_I\}$ sector of $\sorm(2n+1)$ in order to demonstrate that its contribution gives the same contribution as $Z_{\mathrm{str,\, cont.}}$ at $\mathcal{O} (\epsilon_2^{-1})$ in the NS-limit. The instanton string contribution in this sector reads
\begin{align}
Z_{2n,\,\{u_I\}}^{SO(2n+1)}&=\frac{1}{2^{n+1} n!}
 \int \prod_{i=1}^n \diff u_i \left( \frac{ 2\pi \eta^2 \vartheta_1(2 
\epsilon_+)}{\vartheta_1( \epsilon_1) \vartheta_1( \epsilon_2)} \right)^n
\cdot \prod_{i=1}^n \left(\frac{1}{\vartheta_1 (\pm 2u_i+\epsilon_{1,2})}\prod_{l=1}^8\vartheta_1(\pm u_i+\mu_l)\right)\notag\\
&\quad \cdot\prod_{1\leq i< j\leq n}
\frac{\vartheta_1 (\pm u_i\pm u_j)  \vartheta_1 
(\pm u_i\pm u_j+2\epsilon_+)}{\vartheta_1(\pm u_i\pm u_j+\epsilon_1) 
\vartheta_1(\pm u_i\pm u_j+\epsilon_2)}\notag\\
&\quad \cdot \left(\prod_{i=1}^{n}\frac{\vartheta_1 (\pm u_i+u_I)  \vartheta_1 
(\pm u_i+u_I+2\epsilon_+)}{\vartheta_1(\pm u_i+u_I+\epsilon_1) 
\vartheta_1(\pm u_i+u_I+\epsilon_2)}\cdot\frac{\prod_{l=1}^8\vartheta_1(u_I+\mu_l)}{\vartheta_1(2u_I+\epsilon_{1,2})}\right) \notag \\
&=\frac{1}{2^{n+1} n!}\int \prod_{i=1}^n \diff u_i \,\Delta^n
\cdot \prod_{i=1}^n Q(u_i)D(u_i, u_I)^{\frac{1}{2}}\cdot \prod_{i\neq j}D(u_i, u_j)^{\frac{1}{2}}\cdot
\frac{\prod_{l=1}^8\vartheta_1(u_I+\mu_l)}{\vartheta_1(2u_I+\epsilon_{1,2})}\,,
\end{align}
where the periodicity of $D(u_i, u_I)$ with respect to $u_i\sim u_i+2u_I$ has been used. Similarly, one can use the density function \eqref{eq:density_function} to rewrite the full order contribution from this discrete sector. After some algebra, one finds
\begin{align}
Z_{\mathrm{str},\, \{u_I\}}&=\sum_{n=1}^{2n+1}q_\phi^{2n+1} Z_{2n,\,\{u_I\}}^{SO(2n+1)}\notag\\
&=\left(\frac{q_\phi}{2}\cdot\frac{\prod_{l=1}^8\vartheta_1(u_I+\mu_l)}{\vartheta_1(2u_I+\epsilon_{1,2})}\right)\int\!\mathcal D\rho\,\mathcal D\lambda^\prime\,\exp\bigg(\frac{\Delta^2}{2}\int\! \diff u\, \diff u^\prime\,\rho(u^\prime)\log D(u, u^{\prime})\rho(u)\\
&\quad +\Delta\int\!\diff u\, \rho(u)\log\left(-q_\phi^2 Q(u)D(u, u_I)^{\frac{1}{2}}\right)+i\!\int\!\diff u\,\lambda^\prime(u)\cdot\rho(u)-\frac{\Delta}{4\pi i}\int\! \diff u\,e^{-\frac{i}{\Delta}\lambda^\prime(u)}\bigg)\,. \notag
\end{align}
Note that \eqref{eq:leading_order_approximations} implies 
\begin{align}
\Delta \cdot\log D(u, u_I)^{\frac{1}{2}}\sim \frac{1}{2}K(u, u_I; \epsilon_1)\cdot\mathcal{O}(1)
\qquad \text{as }\epsilon_2\rightarrow 0
\,.
\end{align}
Compared to \eqref{eq:path_integral_reprensentation_of_instanton_partition_function}, one finds
\begin{align}
\frac{Z_{\mathrm{str},\, \{u_I\}}}{Z_{\mathrm{str,\, cont.}}}\sim\left(\frac{q_\phi}{2}\cdot\frac{\prod_{l=1}^8\vartheta_1(u_I+\mu_l)}{\vartheta_1(2u_I+\epsilon_{1})\vartheta_1(2u_I)}\right)\cdot \mathcal O(1)\,.
\end{align}
Consequently, $Z_{\mathrm{str},\, \{u_I\}}$ has the same contribution as $Z_{\mathrm{str,\, cont.}}$ at order $\mathcal{O}(\epsilon_2^{-1})$.

Similarly, one can show that this argument holds for all other discrete holonomy sectors $Z_{\mathrm{str},\, \{u_I\}}$,  $Z_{\mathrm{str},\, \{u_I, u_J\}}$,  $Z_{\mathrm{str},\, \{u_I, u_J, u_K\}} $, and $Z_{\mathrm{str},\, \{0, \frac{1}{2}, \frac{\tau+1}{2}, \frac{\tau}{2}\}}$. Therefore, the full partition function in the NS-limit behaves as
\begin{align}
Z_{\mathrm{str.}}&=Z_{\mathrm{str,\, cont.}}+\sum_{I}Z_{\mathrm{str},\, \{u_I\}}+\sum_{I, J}Z_{\mathrm{str},\, \{u_I,u_J\}}+\sum_{I,J,K}Z_{\mathrm{str},\, \{u_I,u_J,u_K\}}+Z_{\mathrm{str},\, \{0, \frac{1}{2}, \frac{\tau+1}{2}, \frac{\tau}{2}\}}\notag\\
&=\mathcal{C}\cdot Z_{\mathrm{str,\, cont.}}|_{\epsilon_2^{-1}}+\mathcal O(1)\,,
\end{align}
where $\mathcal C$ is some irrelevant constant. For the purpose of the saddle point analysis, it is therefore sufficient to only focus on the contribution from continuous holonomy sector in \eqref{eq:path_integral_reprensentation_of_instanton_partition_function} and \eqref{eq:leading_order_of_instanton_partition_function}.

\subsection{Saddle point equation as van Diejen equation}
After the preliminary considerations, one can proceed to study the saddle point of  \eqref{eq:leading_order_of_instanton_partition_function}. Recall that
\begin{align}
\begin{aligned}
Z_{\mathrm{str}}=\int\!\mathcal \mathcal D\lambda^\prime\ldots D\rho\,
\exp&\bigg( \frac{1}{2\epsilon_2}\int\! \diff u \, \diff u^\prime \,\rho(u^\prime)K(u,u^\prime;\,\hbar)\rho(u) \\
&\quad +\frac{1}{\epsilon_2}\int\!\diff u\, \rho(u)\log\left(-q_\phi^2 Q_0(u)\right)+\mathcal O(1)\bigg)\,,
\end{aligned}
\end{align}
with the convention $\hbar\equiv \epsilon_1$. One can obtain the saddle point equation by varying $\rho(u)$,
\begin{align}
\int\! \diff u^\prime\, K(u,u^\prime; \hbar)\rho(u^\prime)+\log\left(-q_\phi^2 Q_0(u)\right)=0\,.
\label{eq:saddle_point_equation_raw}
\end{align}
More concretely, recall 
\begin{align}
K(u, u^\prime;\,\hbar)=\frac{\theta_1^{\prime}(u+u^\prime+\hbar)}{\theta_1(u+u^\prime+\hbar)}-\frac{\theta_1^{\prime}(u+u^\prime-\hbar)}{\theta_1(u+u^\prime-\hbar)}+\frac{\theta_1^{\prime}(u-u^\prime+\hbar)}{\theta_1(u-u^\prime+\hbar)}-\frac{\theta_1^{\prime}(u-u^\prime-\hbar)}{\theta_1(u-u^\prime-\hbar)}\,.
\end{align}
and define
\begin{align}
\omega(x)\equiv\exp\left(\int\!\diff u^\prime\,\rho(u^\prime)\frac{\theta_1^\prime(u^\prime+x)}{\theta_1(u^\prime+x)}\right)\,, \quad \mathrm{and} \quad 
\Phi(x)\equiv\frac{\omega(x)}{\omega(-x)}.
\label{eq:defect_partition_function_in_path_integral_formalism}
\end{align}
The saddle point equation can be rewritten as
\begin{align}
\log\left(\frac{\omega(u+\hbar)}{\omega(-u-\hbar)}\cdot\frac{\omega(-u+\hbar)}{\omega(u-\hbar)}\right)+\log\left(-q_\phi^2 Q_0(u)\right)=\log\left(-q_\phi^2 Q_0(u)\cdot\frac{\Phi(u+\hbar)}{\Phi(u-\hbar)}\right)=0\,,
\end{align}
or, using the short-hand notation, this can also be expressed as 
\begin{align}
\Phi(u-\hbar)+q_\phi^2Q_0(u)\cdot\Phi(u+\hbar)=0\,.
\end{align}
One may apply a shift
\begin{align}
u\rightarrow u+\frac{\hbar}{2}\,,
\end{align}
and further define
\begin{align}
\mathcal Y(u)\equiv \frac{\Phi(u-\frac{\hbar}{2})}{\Phi(u+\frac{\hbar}{2})}\,,
\end{align}
such that one finally arrives at
\begin{align}
\mathcal Y(u)+\frac{\prod_{l=1}^8\theta_1(u\pm\mu_l+\hbar/2)}{\eta^{12}\theta_1(2u)\theta_1(2u+\hbar)^2\theta_1(2u+2\hbar)}\frac{q_\phi^2}{\mathcal Y(u+\hbar)}=0\,.
\label{eq:saddle_point_equation_of_Estring}
\end{align}
One would immediately recognise \eqref{eq:saddle_point_equation_of_Estring} as the van Diejen equation in \eqref{eq:van_Diejen_equation_version_2} without the external potential term $V_0(x)$. The absence of $V_0(x)$ is of course not a problem, because the saddle point equation is \emph{only} satisfied for certain fixed saddle point(s) ``$u$". For generic defect parameter $x$, one has to equate the LHS of saddle point equation \eqref{eq:saddle_point_equation_of_Estring} to a Wilson surface expectation value, i.e.
\begin{align}
\mathcal Y(x)+\frac{\prod_{l=1}^8\theta_1(x\pm\mu_l+\hbar/2)}{\eta^{12}\theta_1(2x)\theta_1(2x+\hbar)^2\theta_1(2x+2\hbar)}\frac{q_\phi^2}{\mathcal Y(x+\hbar)}=\mathcal W_{\hbar}^{S}(x)\,.
\end{align}
On the other hand, recall that $\mathcal W_{\hbar}^S(x)$ only depends on the defect parameter $x$ at one-instanton order, which is precisely the external potential $V_0(x)$ with a minus sign, while the remaining terms in $\mathcal W_{\hbar}^S(x)$ are $x$-independ, and form the Wilson loop $\mathcal W_{\hbar}^L$ from $5d$ perspective,
\begin{align}
\mathcal W_{\hbar}^S(x)=1+q_\phi W_1(x)+q_\phi^2 W_2+\cdots=-q_\phi V_0(x)+q_\phi\mathcal W_{\hbar}^L\,.
\end{align}
Therefore, the van Diejen difference equation is established 
\begin{align}
\mathcal Y(x)+\frac{\prod_{l=1}^8\theta_1(x\pm\mu_l+\hbar/2)}{\eta^{12}\theta_1(2x)\theta_1(2x+\hbar)^2\theta_1(2x+2\hbar)}\frac{q_\phi^2}{\mathcal Y(x+\hbar)}+q_\phi V_0(x)=q_\phi\mathcal W_{\hbar}^{L}\,,
\end{align}
from the saddle point approach of the instanton partition function of the E-string.
\subsection{Defect partition function in path integral formalism}
In this subsection, the aim is to show that $\Phi(x)$, as defined in \eqref{eq:defect_partition_function_in_path_integral_formalism}, coincides with the partition function of the theory in the presence of a co-dimension two defect. Recall from \eqref{eq:1-loop_det_E-string+defect} that the defect induces extra contributions to the one-loop determinant, i.e.
\begin{align}
Z_D(u, x)=\prod_{\rho\in\mathrm{fund.}}\frac{\theta_1(\rho(u)+x+\epsilon_2)}{\theta_1(\rho(u)+x)}=\prod_{i=1}^n\frac{\theta_1(\pm u_i+x+\epsilon_2)}{\theta_1(\pm u_i+x)}\,,
\end{align}
where, due to similar arguments as in subsection \ref{sec:path_integral_representation}, it is sufficient to focus on the contribution from the continuous holonomy sector of $\sorm(2n)$.
The above saddle point analysis can equally well be applied to the defect partition function. In this case, one starts with
\begin{align}
\begin{aligned}
Z_{\mathrm{str}}^{\mathrm{def}}&=\int\!\mathcal D\rho\,\mathcal D\lambda^\prime\,\exp\bigg(\frac{\Delta^2}{2}\int\! \diff u \, \diff u^\prime\,\rho(u^\prime)\log D(u, u^{\prime})\rho(u)\\
&\quad+\Delta\int\!\diff u\, \rho(u)\log\left(-q_\phi^2 Q(u)\cdot V_D(u, x)\right)+i\!\int\!\diff u\,\lambda^\prime(u)\cdot\rho(u)-\frac{\Delta}{4\pi i}\int\! \diff u\,e^{-\frac{i}{\Delta}\lambda^\prime(u)}\bigg)\,,
\end{aligned}
\end{align}
with  
\begin{align}
    V_D(u, x)\equiv\frac{\theta_1(\pm u+x+\epsilon_2)}{\theta_1(\pm u+x)} \,.
\end{align}
One notes in particular that
\begin{align}
\Delta\cdot\log V_D(u, x)=\left(\frac{\theta_1^\prime(u+x)}{\theta_1(u+x)}-\frac{\theta_1^\prime(u-x)}{\theta_1(u-x)}\right)+\mathcal O(\epsilon_2)\,.
\end{align}
Therefore, the defect does not contribute to the saddle point equation \eqref{eq:saddle_point_equation_raw}, but rather gives corrections at order $\mathcal{O}(1)$. Consequently, the normalised defect partition function is given by\footnote{One notes that $\Phi(x)$ has an additional symmetry: $x\rightarrow -x$, $\Phi(x)\rightarrow \Phi(x)^{-1}$. It has been verified explicitly that the defect partition function up to two instanton order obeys this symmetry except for the contributions from additional poles of the defect. The reason is likely that the saddle point equation is insensitive to the presence of the defect.}
\begin{align}
\widetilde{Z}_{\mathrm{str}}^{\mathrm{E-str+def}}(x)=\lim_{\epsilon_2\rightarrow 0}\frac{Z_{\mathrm{str}}^{\mathrm{def}}}{Z_{\mathrm{str}}^{\mathrm{E-str}}}=\exp\left(\int\! \diff u\, \rho(u)\left(\frac{\theta_1^\prime(u+x)}{\theta_1(u+x)}-\frac{\theta_1^\prime(u-x)}{\theta_1(u-x)}\right)\right)=\frac{\omega(x)}{\omega(-x)}=\Phi(x)
\end{align}
due to the  saddle point equation. By expanding $\mathcal Y(x)=\frac{\Phi(x-\hbar/2)}{\Phi(x+\hbar/2)}$ on both sides with respect to $q_\phi$, one obtains
\begin{align}
    \mathcal{Y}_1(x)
    &=\widetilde{Z}_{k=1}^{\mathrm{E-str}+\mathrm{def}} (x-\hbar/2)
    -\widetilde{Z}_{k=1}^{\mathrm{E-str}+\mathrm{def}}(x+\hbar/2) \,,  \\
    \mathcal{Y}_2(x)&=\widetilde{Z}_{k=2}^{\mathrm{def}}(x-\hbar/2)-\widetilde{Z}_{k=2}^{\mathrm{def}}(x+\hbar/2)-\widetilde{Z}_{k=1}^{\mathrm{def}}(x+\hbar/2)\left(\widetilde{Z}_{k=1}^{\mathrm{def}}(x-\hbar/2)-\widetilde{Z}_{k=1}^{\mathrm{def}}(x+\hbar/2)\right)\,,
\end{align}
which matches \eqref{eq:Y1} and \eqref{eq:Y2}.
\paragraph{Ground state degeneracy.}
The Hamiltonian
\begin{equation} \label{eq:Hamiltonian}
    \widehat{H} \equiv \mathcal{V}(x) Y + \mathcal{V}(-x) Y^{-1} - q_{\phi}^{-1}\mathcal{W}^S_{\hbar}(x),
\end{equation}
commutes with the parity operator $\widehat{P} : x \mapsto -x$, i.e.\
\begin{equation}
    [\widehat{H}, \widehat{P}] = 0,
\end{equation}
which can be seen from the fact that the $x$-dependent part of $\mathcal{W}^S_{\hbar}$ is given by $V_0(x)$, see \eqref{V0a1}, which is an even function in $x$. As a consequence  
\begin{equation}
    \widehat{P}^2 = \mathbbm{1},
\end{equation}
and the ground state of the Hamiltonian \eqref{eq:Hamiltonian} is twice degenerate, namely
\begin{equation}
    \widehat{H} ~\psi_1 = 0, \quad \widehat{H} ~\psi_2 = 0. 
\end{equation}
Here $\psi_1$ and $\psi_2$ are two linearly independent wave-functions, which are permuted by the parity operator $\widehat{P}$. In fact, one can immediately write down these two wave-functions. Using the identity \eqref{eq:defect_partition_function_in_path_integral_formalism}, one sees that 
\begin{eqnarray}
    \widehat{P} ~\Phi(x) = \Phi(x)^{-1},
\end{eqnarray}
and, hence, one identifies $\psi_1(x) \equiv \Phi(x)$ and $\psi_2(x) \equiv \Phi(x)^{-1}$. Moreover, one can define simultaneous $\widehat{H}$ and $\widehat{P}$ eigenfunctions as follows:
\begin{eqnarray}
    \Phi_-(x) \equiv \Phi(x) - \Phi(x)^{-1}, \quad \Phi_+(x) \equiv \Phi(x) + \Phi(x)^{-1},
\end{eqnarray}
which are odd and even parity eigenstates, respectively.

\section{Conclusions}
\label{sec:conclusion}
In this work, we study the E-string theory in presence of the co-dimension two defect and co-dimension four defect, the Wilson surface. It turns out that these provide important ingredients for the quantisation of the E-string curve. More specifically, it has been shown that the co-dimension two defect partition function of the E-string theory can be understood as eigenfunction of the quantised E-string curve, which is equivalent to the van Diejen difference operator, with eigenvalue given by a Wilson surface expectation value.  
On the one hand, this result is a further step in the systematic analysis of quantisations of 6d Seiberg-Witten curves, as initiated in \cite{Chen:2020jla}. A crucial ingredient is the E-string curve of \cite{Haghighat:2018dwe}.
On the other hand, the outcome shows how the quantum curve is non-trivially connected to the van Diejen model, as already observed in the 4d context in \cite{Nazzal:2018brc}.
The transition between both viewpoints sheds light on a remarkable relationship between the quantum Seiberg-Witten curves associated to 6d SCFTs and elliptic integrable systems. Besides, we also establish the relation between the 6d Wilson surface expectation value and the 5d Wilson loop observable as a byproduct.

It is intriguing to study quantum curves of other 6d SCFTs defined on $-n$ curves. A straightforward generalisation is to extend our method to the D-type minimal conformal matter theories, where the E-string theory can be viewed as the simplest one along this chain. Recently in the work of \cite{Nazzal:2021tiu}, it has been shown that various generalizations of the van Diejen operators can be obtained via the 4d/2d coupled theories resulted from the 6d D-type minimal conformal matter theories compactified onto Riemann surfaces with punctures. It would be curious to see if our approaches to the quantization of the 6d Seiberg-Witten curves in the corresponding conformal matters will also give rise to the difference operators they studied. On the other hand, higher rank E-strings would be also interesting to study from the 6d SCFTs perspective, whose 4d $\mathcal N=1$ compactifcations onto Riemann surfaces with fluxes \cite{Pasquetti:2019hxf} are expected to give a $BC_N$ type generalization of the van Diejen operator. In fact, It has been studied that, certain types of $\mathcal N=2$ 4d theories from torus compactification of the 6d SCFTs have their coulomb branches identified to the Inozemtsev integrable system \cite{Argyres:2021iws}, which can be understood as the non-relativistic limit of the $BC_N$ type van Diejen. It would be interesting to put these clues together and make the whole picture complete. Moreover, the quantum curves of a series of minimal non-Higgsable SCFTs are surely interesting to investigate. Among them, the 6d Seiberg-Witten curve of the minimal $\sorm(8)$ SCFT on $-4$ curve has been derived in \cite{Haghighat:2018dwe}, which is fibre-base dual to the E-string theory before taking the SCFT limit of the corresponding little string theory. Since it also admits a brane construction, we expect that our approach would be directly applicable to this case. In contrast, the minimal $\surm(3)$ SCFT on a $-3$ curve is certainly of interest, but it does not admit a brane construction, nor is the 6d Seiberg-Witten curve known.

Another important and interesting direction for future research is the realisation of 4d $\mathcal N=1$ SCFTs as surface defects inside the E-string theory. Indeed, the co-dimension 2 defect can be regarded as such a 4d theory extended over $\T^2\times_{\epsilon_1}\mathbb R^2$, meanwhile the Wilson surface wrapping over $\T^2$ may be treated as a further co-dimension 2 object of the E-string defect itself. In addition, a wide class of 4d $\mathcal N=1$ SCFTs have been constructed via compactifications of the E-string onto Riemann surfaces with punctures \cite{Kim:2017toz, Kim:2018bpg}. In \cite{Nazzal:2018brc}, it is further shown that the superconformal index of these 4d theories solves the same van Diejen equation and generates their co-dimension 2 defect in the 4d context. It would be interesting to establish the precise dictionary between our 6d co-dimension two/four defects and the 4d/2d coupled system from the E-string compactifications. Also, it would be worthwhile to study the 6d/4d/2d coupled systems from general 6d SCFTs and the relations to various elliptic integrable systems.

\paragraph{Acknowledgements.}
We would like to thank Min-xin Huang, Joonho Kim, Kimyeong Lee, Yongchao L\"u and Yuji Sugimoto for valuable discussions. The work of J.C., B.H., and M.S. is supported by the National Thousand-Young-Talents Program of China.
M.S. is further supported by the National Natural Science Foundation of China (grant no.\ 11950410497), and the China Postdoctoral Science Foundation (grant no.\ 2019M650616). The research of H.K. is supported by the POSCO Science Fellowship of POSCO TJ Park Foundation and the National Research Foundation of Korea (NRF) Grant 2018R1D1A1B07042934. X.W. is supported by KIAS Individual Grant QP079201. 
M.S. thanks Fudan University, Department of Physics for hospitality during the intermediate stage of this work. 
%
\appendix
\section{Background material}
\label{app:background}
\subsection{Theta functions}
\label{app:theta_fct}
The non-perturbative contributions to the partition function of the 6d theory placed on the $\Omega$-deformed $\T^2 \times \R^4$ background are an infinite sum of 2d elliptic genera. As these genera are most naturally written in terms of elliptic modular forms, the relevant definitions are summarised here.
\paragraph{Basic definitions.}
Various resources employ various different definitions of the theta functions. For this work, one may choose to follow \cite{Kim:2014dza}. Using the conventions $Q=e^{2\pi i \tau}$, $x=e^{2\pi i z}$, the Dedekind eta function and the elliptic theta functions are defined as follows:
\begin{subequations}
\begin{align}
\eta(\tau) &= Q^{\frac{1}{24}} \prod_{k=1}^{\infty} (1-Q^k) \,,\\
    \theta_1(\tau|z) &= - i Q^{\frac{1}{8}} x^{\frac{1}{2}} \prod_{k=1}^\infty (1-Q^k) (1-xQ^k)(1-x^{-1}Q^{k-1}) \,, \\
    \theta_2(\tau|z) &=  Q^{\frac{1}{8}} x^{\frac{1}{2}} \prod_{k=1}^\infty (1-Q^k) (1+xQ^k)(1+x^{-1}Q^{k-1}) \,, \\
    \theta_3(\tau|z) &= \prod_{k=1}^\infty (1-Q^k) (1+xQ^{k-\frac{1}{2}})(1+x^{-1}Q^{k-\frac{1}{2}}) \,,\\
    \theta_4(\tau|z) &= \prod_{k=1}^\infty (1-Q^k) 
    ( 1-xQ^{k-\frac{1}{2}} )( 1-x^{-1}Q^{k-\frac{1}{2}} ) \,. 
\end{align}
\end{subequations}
Sometimes, it is more convenient to work with the following function
\begin{align}
    \vartheta_1 (\tau|z) = \frac{\theta_1(\tau|z)}{Q^{\frac{1}{12}} \eta(\tau)} ,
\end{align}
as already used in \cite{Chen:2020jla}.

\paragraph{Square root contribution.}
It has been argued in \cite{Kim:2014dza} that potential prefactors from half-period shifts drop out of the elliptic genus calculation. However, this is only true once the one-loop determinant contributions are understood appropriately:
\emph{Each real scalar or each real fermion contributes a "square root" of $\theta_1$ factor.} 
To exemplify the implications, consider \eqref{eq:elliptic_genus_1-str} in more detail:
\begin{align}
    Z_{k=1,i}^{\mathrm{E-str}}&=  
\frac{\eta^2}{\thone(\epsilon_{1,2}+2u_i )} 
\cdot
\prod_{l=1}^{8} \frac{\thone(\mu_l+u_i) }{\eta} 
\;, \qquad 
u_i \in 
\left\{ 
0,\frac{1}{2}, \frac{\tau}{2}, \frac{1+\tau}{2}
\right\} 
\end{align}
\begin{compactitem}
\item For $u=0$, there is nothing to show.
\item For $u=\tfrac{1}{2}$, one has
\begin{align}
    \thone(\mu_l+\tfrac{1}{2}) = \theta_2(\mu_l) \qquad \text{and} \qquad \thone(\epsilon_{1,2}+2u_i ) = \thone(\epsilon_{1,2}+1) = - \thone(\epsilon_{1,2})
\end{align}
and hence 
\begin{align}
Z_{k=1,2}^{\mathrm{E-str}}=  
\frac{\eta^2}{\thone(\epsilon_{1,2} )} 
\cdot
\prod_{l=1}^{8} \frac{\theta_2(\mu_l) }{\eta}
\end{align}
without any potential extra factors to care about.
\item For $u=\frac{\tau}{2}$, one has
\begin{align}
    \thone(\mu_l+\tfrac{\tau}{2}) &= \sqrt{ \thone(\mu_l+\tfrac{\tau}{2}) \thone(\mu_l-\tfrac{\tau}{2}) } \notag \\
    &= \sqrt{ \im Q^{-\frac{1}{8}} e^{-\im \pi \mu_l} \theta_4(\mu_l) \thone(\mu_l-\tfrac{\tau}{2}) } \notag \\
    &= \sqrt{ \im Q^{-\frac{1}{8}} e^{-\im \pi \mu_l} \theta_4(\mu_l) \thone(\mu_l+\tfrac{\tau}{2} -\tau) } \notag \\
    &= \sqrt{ \im Q^{-\frac{1}{8}} e^{-\im \pi \mu_l} \theta_4(\mu_l)  (-1) e^{2\pi \im \mu_l} \thone(\mu_l+\tfrac{\tau}{2}) }  \notag \\
    &= \sqrt{ \im Q^{-\frac{1}{8}} e^{-\im \pi \mu_l} \theta_4(\mu_l)  (-1) e^{2\pi \im \mu_l} \im Q^{-\frac{1}{8}} e^{-\im \pi \mu_l} \theta_4(\mu_l) } \notag \\
    &= \sqrt{ -\im^2 Q^{-\frac{2}{8}}  \theta_4(\mu_l) \theta_4(\mu_l) } \notag \\
    &\equiv Q^{-\frac{1}{8}}  \theta_4(\mu_l),
\end{align}
where the shift property in $\tau$ has been used. Note that the extra $\mu_l$ dependent factors only drop out once the square root understanding is used. Nevertheless, there is still an extra factor which only cancels once the $\thone(\epsilon_{1,2} )$ terms are taken into account. For this,
\begin{align}
    \thone(\epsilon_{j} + \tau)&= \sqrt{\thone(\epsilon_{j} + \tau) \thone(\epsilon_{j} - \tau)} \notag \\
    &= \sqrt{ (-1) Q^{-\frac{1}{2}} e^{-2\pi \im \epsilon_j} \thone(\epsilon_{j} ) \cdot
    (-1) Q^{-\frac{1}{2}} e^{2\pi \im \epsilon_j} \thone(\epsilon_{j} )} \notag \\
    &=  Q^{-\frac{1}{2}}  \thone(\epsilon_{j} ),  
\end{align}
where the potential $p, q$ factors have cancelled due to the square root interpretation. In total, one finds
\begin{align}
Z_{k=1,3}^{\mathrm{E-str}}
=
\frac{\eta^2}{ Q^{-1} \thone(\epsilon_{1,2} )} 
\cdot
\prod_{l=1}^{8} Q^{-\frac{1}{8}} \frac{\theta_4(\mu_l) }{\eta}
= \frac{\eta^2}{ \thone(\epsilon_{1,2} )} 
\cdot
\prod_{l=1}^{8} \frac{\theta_4(\mu_l) }{\eta}.
\end{align}
and indeed, all extra factors have cancelled.
\item For $u=\frac{\tau+1}{2}$, one proceeds analogously.
\end{compactitem}

\subsection{Normalised partition function}
\label{app:normalised_partition_fct}
For the 6d theory with and without defect, the partition functions are 
\begin{subequations}
\begin{align}
 Z^{6d} &= Z_{\mathrm{pert}}^{6d} \left( 1 + \sum_{k=1}^\infty q_\phi^k\, Z_{k}^{6d} 
\right), \\
Z^{6d+\mathrm{def}} &= Z_{\mathrm{pert}}^{6d+\mathrm{def}} \left( 1 + 
\sum_{k=1}^\infty q_\phi^k\, Z_{k}^{6d+\mathrm{def}} 
\right),
\end{align}
such that the normalised defect partition function is defined as
\begin{align}
 \widetilde{Z}^{6d+\mathrm{def}} = \frac{Z^{6d+\mathrm{def}}}{ Z^{6d} } \,.
\end{align}
A perturbative $q_\phi$ expansion yields
\begin{align}
 \frac{1}{ Z^{6d} }
 =
 \frac{1}{Z_{\mathrm{pert}}^{6d}} \left[ 
 1- Z_{1}^{6d} q_\phi  
 - \left(Z_{2}^{6d}- \left(Z_{1}^{6d}\right)^2 \right) q_\phi^2 
 -\left(Z_{3}^{6d} - 2 Z_{1}^{6d} Z_{2}^{6d} +\left(Z_{1}^{6d} \right)^3 
\right) q_\phi^3
 +\mathcal{O}(q_\phi^4)
 \right].
\end{align}
such that the series expansion of the normalised defect partition function becomes
\begin{align}
 \widetilde{Z}^{6d+\mathrm{def}}=
 \frac{Z_{\mathrm{pert}}^{6d+\mathrm{def}}}{ Z_{\mathrm{pert}}^{6d} }
 \left[ 1 
 + \left(Z_{1}^{6d+\mathrm{def}} {-}Z_{1}^{6d} \right) q_\phi
 + \left(Z_{2}^{6d+\mathrm{def}} {-}Z_{2}^{6d}   
{-}Z_{1}^{6d} \left(Z_{1}^{6d+\mathrm{def}} {-}Z_{1}^{6d} \right)  
 \right) q_\phi^2
 +\mathcal{O}(q_\phi^3)
 \right] \,.
\label{a1e}
\end{align}
\end{subequations}
\subsection{Elliptic genus of E-string theory}
\label{app:elliptic_genus_E-string}
In this appendix, the elliptic genus computation for $k=1,2$ is reviewed for the benefit of the defect partition function computation.
\paragraph{1-String.}
Following \cite{Kim:2014dza}, for a $\orm(1)$ gauge group there are four Wilson lines, labelled by $(1,1)$, $(1,-1)$, $(-1,1)$, $(-1,-1)$. For these discrete numbers, the exponents of the Wilson lines have been assigned to be 
\begin{align}
 (1,1): \; u_i =0
 \;, \qquad 
 (-1,1): \; u_i =\frac{1}{2}
 \;, \qquad 
 (1,-1): \; u_i =\frac{\tau}{2}
 \;, \qquad 
 (-1,-1): \; u_i =\frac{1+\tau}{2}
 \;.
\end{align}
As reviewed in Appendix \ref{app:theta_fct}, the 
$\theta_1$ functions can be replaced by one of the other Jacobi Theta functions under the half-period shifts
\begin{align}
 \theta_1 (z +\tfrac{1}{2}) = \theta_2(z)
 \;,\qquad 
 \theta_1 (z +\tfrac{\tau}{2}) = \theta_4(z)
 \;,\qquad
 \theta_1 (z +\tfrac{1+\tau}{2}) = \theta_3(z)
 \;.
\end{align}
with all potential extra factors dropping out in the elliptic genus. 
The one-loop determinant \eqref{eq:E-string_1-loop_det} for $k=1$ reduces to
\begin{align}
\label{eq:elliptic_genus_1-str}
 Z_{k=1,i}^{\mathrm{E-str}}&=  
\frac{\eta^2}{\thone(\epsilon_{1,2}+2u_i )} 
\cdot
\prod_{l=1}^{8} \frac{\thone(\mu_l+u_i) }{\eta} 
\;, \qquad 
u_i \in 
\left\{ 
0,\frac{1}{2}, \frac{\tau}{2}, \frac{1+\tau}{2}
\right\} \,.
\end{align}
The contribution from the symmetric representation is invariant under 
the half-period shifts, because it has an addition factor of $2$ that renders 
them into full-period shifts. Then, one obtains
\begin{align}
\begin{aligned}
  Z_{k=1,i=1}^{\mathrm{E-str}}&=
\frac{\prod_{l=1}^{8} \thone(\mu_l)}{\thone(\epsilon_{1}) 
\thone(\epsilon_{2}) \eta^6}
\; , \qquad 
Z_{k=1,i=2}^{\mathrm{E-str}} =  
\frac{\prod_{l=1}^{8} \theta_2(\mu_l)}{\thone(\epsilon_{1}) 
\thone(\epsilon_{2}) \eta^6}
\; , \\
Z_{k=1,i=3}^{\mathrm{E-str}}&=  
\frac{\prod_{l=1}^{8} \theta_4(\mu_l)}{\thone(\epsilon_{1}) 
\thone(\epsilon_{2}) \eta^6}
\; , \qquad
Z_{k=1,i=4}^{\mathrm{E-str}} =  
\frac{\prod_{l=1}^{8} \theta_3(\mu_l)}{\thone(\epsilon_{1}) 
\thone(\epsilon_{2}) \eta^6} \,.
\end{aligned}
\end{align}
Lastly, summing up all contributions and dividing by the order of the Weyl 
group in each case yields the desired result \eqref{eq:ell_genus_E-string_k=1}.
\paragraph{2-String.}
Next, review the 2-string case, for which the one-loop determinant \eqref{eq:E-string_1-loop_det} reduces to
\begin{align}
 Z_{\mathrm{1-loop}}^{\mathrm{E-str}}\big|_{k=2}
 &=\frac{1}{2\pi}  2\pi \diff u \,
\thone(2\epsilon_+) 
 \prod_{e\in \mathrm{root}} \frac{\thone(e(u)) \thone(2\epsilon_+ +e(u))  
}{\eta^2}
 \cdot \prod_{\rho\in \mathrm{sym}} \frac{\eta^2}{\thone(\epsilon_{1,2}+\rho(u) 
)} \notag \\
 &\qquad \cdot
 \prod_{\rho\in \mathrm{fund}}  \prod_{l=1}^{8} \frac{\thone(\mu_l 
+\rho(u))}{\eta}. 
\end{align}
As detailed in \cite{Kim:2014dza} there are seven sectors of $\orm(2)$ Wilson lines: one sector with one complex modulus, and six discrete sectors. In 
the discrete sector, there are two eigenvalues $(u_+,u_-)$ of $u$ given by
\begin{alignat}{3}
\begin{aligned}
(1):\; (u_+,u_-)&= (0,\tfrac{1}{2}) & \qquad
(2):\; (u_+,u_-)&= (\tfrac{\tau}{2},\tfrac{1+\tau}{2}) & \qquad 
(3):\; (u_+,u_-)&= (0,\tfrac{\tau}{2}) \\
(4):\; (u_+,u_-)&= (\tfrac{1}{2},\tfrac{1+\tau}{2}) & \qquad 
(5):\; (u_+,u_-)&= (0,\tfrac{1+\tau}{2}) & \qquad 
(6):\; (u_+,u_-)&= (\tfrac{1}{2},\tfrac{\tau}{2})
\end{aligned}
\label{eq:O2_discrete _Wilson}
\end{alignat}
and 
\begin{align}
 e\in \mathrm{root} : \;
 e(u) \in \left\{ u_+ +u_- \right\} 
 \;,\quad 
\rho\in \mathrm{sym} : \;
\rho(u) \in \left\{ 2 u_+, 2u_-,u_++u_- \right\} 
\;, \quad 
\rho\in \mathrm{fund} : \;
\rho(u) \in \left\{ u_+, u_- \right\} \,.
\notag
\end{align}
Then, consider the following function
\begin{align}
 Z_{k=2}^{\mathrm{E-str}}&= 
 \frac{\thone(u_+ + u_-) \thone(2\epsilon_+ +u_+ + u_-)}{\eta^{12}}
 \cdot  \frac{  \prod_{l=1}^{8}  \thone(\mu_l +u_+) \thone(\mu_l +u_-)  }{
 \thone(\epsilon_{1,2}+2u_+)
 \thone(\epsilon_{1,2}+2u_-)
 \thone(\epsilon_{1,2}+u_++u_-)
 },
\end{align}
such that the contributions for the discrete sectors are given by
\begin{subequations}
\begin{align}
 Z_{k=2,{i=1}}^{\mathrm{E-str}}&= 
 \frac{\theta_2(0) \theta_2(2\epsilon_+)}{\eta^{12}}
 \cdot  \frac{  \prod_{l=1}^{8}  \thone(\mu_l ) \theta_2(\mu_l)  }{
 \thone(\epsilon_{1})^2
 \thone(\epsilon_{2})^2
 \theta_2(\epsilon_{1})
  \theta_2(\epsilon_{2})
 } \;,\\
Z_{k=2,{i=2}}^{\mathrm{E-str}}&=
 \frac{\theta_2(0) \theta_2(2\epsilon_+ )}{\eta^{12}}
 \cdot  \frac{  \prod_{l=1}^{8}  \theta_4(\mu_l ) \theta_3(\mu_l )  }{
 \thone(\epsilon_{1})^2
 \thone(\epsilon_{2})^2
 \theta_2(\epsilon_{1})
  \theta_2(\epsilon_{2})
 }
\;, \\
 Z_{k=2,{i=3}}^{\mathrm{E-str}}&= 
 \frac{\theta_4(0) \theta_4(2\epsilon_+ )}{\eta^{12}}
 \cdot  \frac{  \prod_{l=1}^{8}  \thone(\mu_l ) \theta_4(\mu_l )  }{
 \thone(\epsilon_{1})^2
 \thone(\epsilon_{2})^2
 \theta_4(\epsilon_{1})
 \theta_4(\epsilon_{2})
 }
\;, \\
Z_{k=2,{i=4}}^{\mathrm{E-str}}&= 
 \frac{\theta_4(0) \theta_4(2\epsilon_+ )}{\eta^{12}}
 \cdot  \frac{  \prod_{l=1}^{8}  \theta_2(\mu_l ) \theta_3(\mu_l )  }{
 \thone(\epsilon_{1})^2
 \thone(\epsilon_{2})^2
 \theta_4(\epsilon_{1})
 \theta_4(\epsilon_{2})
 }
\;, \\
Z_{k=2,{i=5}}^{\mathrm{E-str}}&= 
 \frac{\theta_3(0) \theta_3(2\epsilon_+ )}{\eta^{12}}
 \cdot  \frac{  \prod_{l=1}^{8}  \thone(\mu_l ) \theta_3(\mu_l )  }{
 \thone(\epsilon_{1})^2
 \thone(\epsilon_{2})^2
 \theta_3(\epsilon_{1})
 \theta_3(\epsilon_{2})
 }
\;, \\
Z_{k=2,{i=6}}^{\mathrm{E-str}}&= 
 \frac{\theta_3(0) \theta_3(2\epsilon_+ )}{\eta^{12}}
 \cdot  \frac{  \prod_{l=1}^{8}  \theta_2(\mu_l ) \theta_4(\mu_l )  }{
 \thone(\epsilon_{1})^2
 \thone(\epsilon_{2})^2
 \theta_3(\epsilon_{1})
 \theta_3(\epsilon_{2})
 }\;.
\end{align}
\label{eq:O2_discrete_Wilson}
\end{subequations}
In the continuous sector, the contour integral becomes
\begin{align}
 Z_{k=2,i=0}^{\mathrm{E-str}}
&= \oint  \diff u \frac{\thone(2\epsilon_+) 
}{i \eta^9}
 \cdot  \frac{\prod_{l=1}^{8}
 \thone(\mu_l \pm u)
 }{
 \thone(\epsilon_{1} )
 \thone(\epsilon_{2} )
 \thone(\epsilon_{1}\pm 2u )
 \thone(\epsilon_{2}\pm 2u )
 } \,.
 \end{align}
 To evaluate the integral, one needs to consider the poles at 
$\thone(\epsilon_{1,2}+2 u )=0$. Hence, one computes the residues at
\begin{align}
 u\in \left\{  
-\tfrac{\epsilon_{1,2}}{2},-\tfrac{\epsilon_{1,2}}{2}+\tfrac{1}{2}, 
-\tfrac{\epsilon_{1,2}}{2} + \tfrac{1+\tau}{2} , -\tfrac{\epsilon_{1,2}}{2} + 
\tfrac{\tau}{2} \right\}
\label{eq:O2_poles_no_defect}
\end{align}
which are given by
\begin{subequations}
\begin{compactitem}
\item for $u=-\tfrac{\epsilon_{1,2}}{2}$
\begin{align}
 &=\frac{1 
}{ \eta^{12}\, 
\thone(\epsilon_{1} )
 \thone(\epsilon_{2} )}
  \cdot \left(  \frac{\prod_{l=1}^{8}
 \thone(\mu_l \pm \tfrac{\epsilon_{1}}{2})
 }{
  \thone(2\epsilon_{1} )
 \thone(\epsilon_{2}- \epsilon_{1} )
 } 
 +  \frac{\prod_{l=1}^{8}
 \thone(\mu_l \pm \tfrac{\epsilon_{2}}{2})
 }{
  \thone(2\epsilon_{2} )
 \thone(\epsilon_{1}  - \epsilon_{2})
 }
 \right)
\end{align}
\item for $u = -\tfrac{\epsilon_{1,2}}{2}+\tfrac{1}{2}$
\begin{align}
&=
\frac{1}{ \eta^{12} \,  
\thone(\epsilon_{1} )
 \thone(\epsilon_{2} )}
 \cdot  
 \left(
 \frac{\prod_{l=1}^{8}
 \theta_2(\mu_l \pm \tfrac{\epsilon_{1}}{2} )
 }{
 \thone(2\epsilon_{1} )
 \thone(\epsilon_{2}- \epsilon_1 )
 }
 + \frac{\prod_{l=1}^{8}
 \theta_2(\mu_l  \pm \tfrac{\epsilon_{2}}{2})
 }{
 \thone(2\epsilon_{2}  )
 \thone(\epsilon_{1}- \epsilon_2 )
 }
 \right)
\end{align}
\item for $u=-\tfrac{\epsilon_{1,2}}{2} + \tfrac{1+\tau}{2}$
\begin{align}
 &=
 \frac{1}{ \eta^{12}\,  
 \thone(\epsilon_{1} )
 \thone(\epsilon_{2} )}
 \cdot 
 \left( \frac{\prod_{l=1}^{8}
 \theta_3(\mu_l \pm \tfrac{\epsilon_{1}}{2})
 }{
  \thone(2\epsilon_{1} )
 \thone(\epsilon_{2}- \epsilon_1 )
 }
 +
 \frac{\prod_{l=1}^{8}
 \theta_3(\mu_l \pm \tfrac{\epsilon_{2}}{2})
 }{
 \thone(2\epsilon_{2})
 \thone(\epsilon_{1}- \epsilon_2)
 }
 \right)
\end{align}
\item for $u = -\tfrac{\epsilon_{1,2}}{2}+\tfrac{\tau}{2}$
\begin{align}
&=
\frac{1}{ \eta^{12} \,  
\thone(\epsilon_{1} )
 \thone(\epsilon_{2} )}
 \cdot  
 \left(
 \frac{\prod_{l=1}^{8}
 \theta_4(\mu_l \pm \tfrac{\epsilon_{1}}{2} )
 }{
 \thone(2\epsilon_{1} )
 \thone(\epsilon_{2}- \epsilon_1 )
 }
 + \frac{\prod_{l=1}^{8}
 \theta_4(\mu_l  \pm \tfrac{\epsilon_{2}}{2})
 }{
 \thone(2\epsilon_{2}  )
 \thone(\epsilon_{1}- \epsilon_2 )
 }
 \right)
\end{align}
\end{compactitem}
\label{eq:O2_continous_Wilson_parts}
\end{subequations}
Therefore, the total contribution of the continuous Wilson line yields
\begin{align}
  Z_{k=2,i=0}^{\mathrm{E-str}}&=
  \frac{1}{{2}
  \eta^{12} \,  
\thone(\epsilon_{1} )
 \thone(\epsilon_{2} )}
 \cdot  \sum_{I=1}^4
 \left(
 \frac{\prod_{l=1}^{8}
 \theta_I(\mu_l \pm \tfrac{\epsilon_{1}}{2} )
 }{
 \thone(2\epsilon_{1} )
 \thone(\epsilon_{2}- \epsilon_1 )
 }
 + \frac{\prod_{l=1}^{8}
 \theta_I(\mu_l  \pm \tfrac{\epsilon_{2}}{2})
 }{
 \thone(2\epsilon_{2}  )
 \thone(\epsilon_{1}- \epsilon_2 )
 }
 \right)  
 \label{eq:k=2_conti_Wilson} \,.
\end{align}
In total, dividing by the order of the Weyl group in each sector leads to
\begin{align}
 Z_{k=2}^{\mathrm{E-str}}&= 
  \frac{1}{2} Z_{k=2,i=0}^{\mathrm{E-str}} + 
\frac{1}{4}\sum_{i=1}^6 Z_{k=2,i}^{\mathrm{E-str}}  \,.
\end{align}
\subsection{Elliptic genus of E-string theory with co-dimension 2 defect}
\label{app:elliptic_genus_E-string+defect}
Having reviewed the E-string elliptic genera in Appendix \ref{app:elliptic_genus_E-string}, this appendix provides the details for the elliptic genera of the E-string theory with a co-dimension 2 defect.
\paragraph{1-String.}
The one-loop determinant \eqref{eq:1-loop_det_E-string+defect} for $k=1$ becomes 
\begin{align}
 Z_{k=1,i}^{\mathrm{E-str}+\mathrm{def}}&=  
\frac{\prod_{l=1}^{8} \thone(\mu_l+u_i)}{\eta^6 \cdot 
\thone(\epsilon_{1,2}+2u_i )} 
\cdot
\frac{  \thone(x +u_i + \epsilon_2) }{ \thone(x +u_i ) }
\;, \qquad 
u_i \in 
\left\{ 
0,\frac{1}{2}, \frac{\tau}{2}, \frac{1+\tau}{2}
\right\} \,.
\end{align}
The individual contributions are given by
\begin{align}
\begin{aligned}
 Z_{k=1,i=1}^{\mathrm{E-str}+\mathrm{def}}&=  
\frac{\prod_{l=1}^{8} \thone(\mu_l)}{\eta^6 \cdot 
\thone(\epsilon_{1})\thone(\epsilon_{2})} 
\cdot
\frac{  \thone(x + \epsilon_2) }{ \thone(x ) }
\;, \qquad 
 Z_{k=1,i=2}^{\mathrm{E-str}+\mathrm{def}} =  
\frac{\prod_{l=1}^{8} \theta_2(\mu_l)}{\eta^6 \cdot 
\thone(\epsilon_{1})\thone(\epsilon_{2})} 
\cdot
\frac{  \theta_2(x + \epsilon_2) }{ \theta_2(x  ) }
\;, \\
Z_{k=1,i=3}^{\mathrm{E-str}+\mathrm{def}}&=  
\frac{\prod_{l=1}^{8} \theta_4(\mu_l)}{\eta^6 \cdot 
\thone(\epsilon_{1})\thone(\epsilon_{2})} 
\cdot
\frac{  \theta_4(x + \epsilon_2) }{ \theta_4(x  ) }
\;, \qquad 
Z_{k=1,i=4}^{\mathrm{E-str}+\mathrm{def}}=  
\frac{\prod_{l=1}^{8} \theta_3(\mu_l)}{\eta^6 \cdot 
\thone(\epsilon_{1})\thone(\epsilon_{2})} 
\cdot
\frac{  \theta_3(x + \epsilon_2) }{ \theta_3(x  ) }
\;.
\end{aligned}
\end{align}
Therefore, summing up the pieces and dividing by the Weyl group yields \eqref{eq:elliptic_genus_k=1_E-string+defect}.
\paragraph{Normalised partition function in NS limit.}
Following Appendix \ref{app:normalised_partition_fct}, consider the normalised 
1-string contribution in the NS-limit
\begin{align}
\widetilde{Z}_{k=1}^{\mathrm{E-str}+\mathrm{def}}   
&\equiv 
Z_{k=1}^{\mathrm{E-str}+\mathrm{def}}  -   Z_{k=1}^{\mathrm{E-str}} 
=-\frac{\widetilde{\Theta}(\tau,\mu_l,x) -\Theta(\tau,\mu_l) 
}{\eta^6 \theta_1 (\epsilon_1) \theta_1 (\epsilon_2)} 
\notag \\
&= 
 -\frac{1 
}{\eta^6 \theta_1 (\epsilon_1) \theta_1 (\epsilon_2)}
\frac{1}{2}\sum_{I=1}^4 \left[ \left( \frac{  
\theta_I(x + \epsilon_2) }{ \theta_I(x  ) } -1\right) \cdot \prod_{l=1}^8 
\theta_I(\tau,\mu_l) \right]
\end{align}
and taking the $\epsilon_2\to 0$ limit yields the claim \eqref{eq:1-string_normalised}.
%
%
\paragraph{2-String.}
Having reviewed the $k=2$ case without defect, one can approach the 
elliptic genus with defect. For the 6 discrete sector \eqref{eq:O2_discrete 
_Wilson} of $\orm(2)$ Wilson lines, one needs to consider the following one-loop determinant
\begin{align}
 Z_{\mathrm{1-loop}}^{\mathrm{E-str+defect}}\big|_{k=2}&= 
 \frac{\thone(u_+ + u_-) \thone(2\epsilon_+ +u_+ + u_-)}{\eta^{12}}
 \cdot  \frac{  \prod_{l=1}^{8}  \thone(\mu_l +u_+) \thone(\mu_l +u_-)  }{
 \thone(\epsilon_{1,2}+2u_1)
 \thone(\epsilon_{1,2}+2u_2)
 \thone(\epsilon_{1,2}+u_1+u_2)
 }
 \notag \\
 &\qquad \cdot
 \frac{  \thone(x +u_+ + \epsilon_2) \thone(x +u_- + \epsilon_2)}{ \thone(x 
+u_+ ) \thone(x +u_- )}
\end{align}
and one finds for the discrete sectors
\begin{subequations}
\begin{align}
 Z_{k=2,{i=1}}^{\mathrm{E-str+def}}&=
 \frac{\theta_2(0) \theta_2(2\epsilon_+)}{\eta^{12}}
 \cdot  \frac{  \prod_{l=1}^{8}  \thone(\mu_l ) \theta_2(\mu_l)  }{
 \thone(\epsilon_{1})^2
 \thone(\epsilon_{2})^2
 \theta_2(\epsilon_{1})
  \theta_2(\epsilon_{2})
 } 
 \cdot
 \frac{  \thone(x + \epsilon_2) \theta_2(x  + \epsilon_2)}{ \thone(x ) 
\theta_2(x  )}
\;, \\
Z_{k=2,{i=2}}^{\mathrm{E-str+def}}&=
 \frac{\theta_2(0) \theta_2(2\epsilon_+ )}{\eta^{12}}
 \cdot  \frac{  \prod_{l=1}^{8}  \theta_4(\mu_l ) \theta_3(\mu_l )  }{
 \thone(\epsilon_{1})^2
 \thone(\epsilon_{2})^2
 \theta_2(\epsilon_{1})
  \theta_2(\epsilon_{2})
 }
 \cdot
 \frac{  \theta_4(x  + \epsilon_2) \theta_3(x  + \epsilon_2)}{ \theta_4(x ) 
\theta_3(x  )}
\;, \\
 Z_{k=2,{i=3}}^{\mathrm{E-str+def}}&= 
 \frac{\theta_4(0) \theta_4(2\epsilon_+ )}{\eta^{12}}
 \cdot  \frac{  \prod_{l=1}^{8}  \thone(\mu_l ) \theta_4(\mu_l )  }{
 \thone(\epsilon_{1})^2
 \thone(\epsilon_{2})^2
 \theta_4(\epsilon_{1})
 \theta_4(\epsilon_{2})
 }
 \cdot
 \frac{  \thone(x + \epsilon_2) \theta_4(x  + \epsilon_2)}{ \thone(x ) 
\theta_4(x  )}
\;, \\
Z_{k=2,{i=4}}^{\mathrm{E-str+def}}&= 
 \frac{\theta_4(0) \theta_4(2\epsilon_+ )}{\eta^{12}}
 \cdot  \frac{  \prod_{l=1}^{8}  \theta_2(\mu_l ) \theta_3(\mu_l )  }{
 \thone(\epsilon_{1})^2
 \thone(\epsilon_{2})^2
 \theta_4(\epsilon_{1})
 \theta_4(\epsilon_{2})
 }
 \cdot
 \frac{  \theta_2(x  + \epsilon_2) \theta_3(x  + \epsilon_2)}{ \theta_2(x ) 
\theta_3(x  )}
\;, \\
Z_{k=2,{i=5}}^{\mathrm{E-str+def}}&= 
 \frac{\theta_3(0) \theta_3(2\epsilon_+ )}{\eta^{12}}
 \cdot  \frac{  \prod_{l=1}^{8}  \thone(\mu_l ) \theta_3(\mu_l )  }{
 \thone(\epsilon_{1})^2
 \thone(\epsilon_{2})^2
 \theta_3(\epsilon_{1})
 \theta_3(\epsilon_{2})
 }
 \cdot
 \frac{  \thone(x + \epsilon_2) \theta_3(x  + \epsilon_2)}{ \thone(x ) 
\theta_3(x  )}
\;, \\
Z_{k=2,{i=6}}^{\mathrm{E-str+def}}&= 
 \frac{\theta_3(0) \theta_3(2\epsilon_+ )}{\eta^{12}}
 \cdot  \frac{  \prod_{l=1}^{8}  \theta_2(\mu_l ) \theta_4(\mu_l )  }{
 \thone(\epsilon_{1})^2
 \thone(\epsilon_{2})^2
 \theta_3(\epsilon_{1})
 \theta_3(\epsilon_{2})
 }
 \cdot
 \frac{  \theta_2(x  + \epsilon_2) \theta_4(x  + \epsilon_2)}{ \theta_2(x ) 
\theta_4(x  )}
\;.
\end{align}
\label{eq:O2_discrete_Wilson_w_defect}
\end{subequations}
For the continuous sector with defect, the integrand becomes
\begin{align}
 Z_{k=2,i=0}^{\mathrm{E-str+def}}&= 
 \oint  \diff u \frac{\thone(2\epsilon_+) 
}{i \eta^9}
 \cdot  \frac{\prod_{l=1}^{8}
 \thone(\mu_l \pm u)
 }{
 \thone(\epsilon_{1} )
 \thone(\epsilon_{2} )
 \thone(\epsilon_{1}\pm 2u )
 \thone(\epsilon_{2}\pm 2u )
 }
 \cdot
 \frac{  \thone(x \pm u + \epsilon_2) }{ \thone(x \pm u ) } 
 \,.
 \end{align}
 Besides the poles \eqref{eq:O2_poles_no_defect} from the part without defect 
there is a single new pole from $\thone(x+u)$ at
\begin{align}
 u=-x \,.
\end{align}
Evaluating each pole, one finds
\begin{subequations}
\begin{compactitem}
\item for $u=-\tfrac{\epsilon_{1,2}}{2}$
\begin{align}
 \frac{1 
}{ \eta^{12}\, 
\thone(\epsilon_{1} )
 \thone(\epsilon_{2} )}
   \left(  \frac{\prod_{l=1}^{8}
 \thone(\mu_l \pm \tfrac{\epsilon_{1}}{2})
 }{
  \thone(2\epsilon_{1} )
 \thone(\epsilon_{2}- \epsilon_{1} )
 } 
 \frac{  \thone(x \pm \tfrac{\epsilon_{1}}{2} + \epsilon_2) }{ \thone(x 
\pm \tfrac{\epsilon_{1}}{2} ) }
 +  \frac{\prod_{l=1}^{8}
 \thone(\mu_l \pm \tfrac{\epsilon_{2}}{2})
 }{
  \thone(2\epsilon_{2} )
 \thone(\epsilon_{1}  - \epsilon_{2})
 }
 \frac{  \thone(x \pm \tfrac{\epsilon_{2}}{2} + \epsilon_2) }{ \thone(x 
\pm \tfrac{\epsilon_{2}}{2} ) }
 \right)
\end{align}
\item for $u = -\tfrac{\epsilon_{1,2}}{2}+\tfrac{1}{2}$
\begin{align}
\frac{1}{ \eta^{12} \,  
\thone(\epsilon_{1} )
 \thone(\epsilon_{2} )}
 \left(
 \frac{\prod_{l=1}^{8}
 \theta_2(\mu_l \pm \tfrac{\epsilon_{1}}{2} )
 }{
 \thone(2\epsilon_{1} )
 \thone(\epsilon_{2}- \epsilon_1 )
 }
 \frac{  \theta_2(x \pm \tfrac{\epsilon_{1}}{2} + \epsilon_2) }{ \theta_2(x \pm 
\tfrac{\epsilon_{1}}{2} ) }
 + \frac{\prod_{l=1}^{8}
 \theta_2(\mu_l  \pm \tfrac{\epsilon_{2}}{2})
 }{
 \thone(2\epsilon_{2}  )
 \thone(\epsilon_{1}- \epsilon_2 )
 }
 \frac{  \theta_2(x \pm \tfrac{\epsilon_{2}}{2} + \epsilon_2) }{ \theta_2(x \pm 
\tfrac{\epsilon_{2}}{2} ) }
 \right)
\end{align}
\item for $u=-\tfrac{\epsilon_{1,2}}{2} + \tfrac{1+\tau}{2}$
\begin{align}
 \frac{1}{ \eta^{12}\,  
 \thone(\epsilon_{1} )
 \thone(\epsilon_{2} )}
 \left( \frac{\prod_{l=1}^{8}
 \theta_3(\mu_l \pm \tfrac{\epsilon_{1}}{2})
 }{
  \thone(2\epsilon_{1} )
 \thone(\epsilon_{2}- \epsilon_1 )
 }
 \frac{  \theta_3(x \pm \tfrac{\epsilon_{1}}{2} + \epsilon_2) }{ \theta_3(x \pm 
\tfrac{\epsilon_{1}}{2} ) }
 +
 \frac{\prod_{l=1}^{8}
 \theta_3(\mu_l \pm \tfrac{\epsilon_{2}}{2})
 }{
 \thone(2\epsilon_{2})
 \thone(\epsilon_{1}- \epsilon_2)
}
\frac{  \theta_3(x \pm \tfrac{\epsilon_{2}}{2} + \epsilon_2) }{ \theta_3(x \pm 
\tfrac{\epsilon_{2}}{2} ) }
 \right)
\end{align}
\item for $u = -\tfrac{\epsilon_{1,2}}{2}+\tfrac{\tau}{2}$
\begin{align}
\frac{1}{ \eta^{12} \,  
\thone(\epsilon_{1} )
 \thone(\epsilon_{2} )}
 \left(
 \frac{\prod_{l=1}^{8}
 \theta_4(\mu_l \pm \tfrac{\epsilon_{1}}{2} )
 }{
 \thone(2\epsilon_{1} )
 \thone(\epsilon_{2}- \epsilon_1 )
 }
 \frac{  \theta_4(x \pm \tfrac{\epsilon_{1}}{2}  + \epsilon_2) }{ \theta_4(x 
\pm \tfrac{\epsilon_{1}}{2}  ) }
 + \frac{\prod_{l=1}^{8}
 \theta_4(\mu_l  \pm \tfrac{\epsilon_{2}}{2})
 }{
 \thone(2\epsilon_{2}  )
 \thone(\epsilon_{1}- \epsilon_2 )
 }
 \frac{  \theta_4(x \pm \tfrac{\epsilon_{2}}{2} + \epsilon_2) }{ \theta_4(x \pm 
\tfrac{\epsilon_{2}}{2} ) }
 \right)
\end{align}
\item for $u=-x$
\begin{align}
 &\frac{\thone(2\epsilon_+) 
}{ \eta^{12}}
 \cdot  \frac{\prod_{l=1}^{8}
 \thone(\mu_l \pm x)
 }{
 \thone(\epsilon_{1} )
 \thone(\epsilon_{2} )
 \thone(\epsilon_{1}\pm 2x )
 \thone(\epsilon_{2}\pm 2x )
 }
 \frac{ \thone( \epsilon_2)   \thone(2x + \epsilon_2)}{ \thone(2x ) }
 \notag \\
&=
  \frac{ 1
}{ \eta^{12}}
 \cdot  \frac{ \thone(2\epsilon_+) \; \prod_{l=1}^{8}
 \thone(\mu_l \pm x)
 }{
 \thone(\epsilon_{1} )
 \thone(2x )
 \thone(\epsilon_{1}\pm 2x )
 \thone(\epsilon_{2}- 2x )
 }
\end{align}
\end{compactitem}
\label{eq:O2_continous_Wilson_parts_w_defect}
\end{subequations}
such that 
\begin{align}
  Z_{k=2,i=0}^{\mathrm{E-str+defect}} =
  \frac{1}{ {{2}}
  \eta^{12} \,  
\thone(\epsilon_{1} )
 \thone(\epsilon_{2} )}
 &\cdot  \sum_{I=1}^4
 \bigg(
 \frac{\prod_{l=1}^{8}
 \theta_I(\mu_l \pm \tfrac{\epsilon_{1}}{2} )
 }{
 \thone(2\epsilon_{1} )
 \thone(\epsilon_{2}- \epsilon_1 )
 }
 \frac{  \theta_I(x \pm \tfrac{\epsilon_{1}}{2}  + \epsilon_2) }{ \theta_I(x 
\pm \tfrac{\epsilon_{1}}{2}  ) }
\notag \\
&\qquad \qquad  + 
\frac{\prod_{l=1}^{8}
 \theta_I(\mu_l  \pm \tfrac{\epsilon_{2}}{2})
 }{
 \thone(2\epsilon_{2}  )
 \thone(\epsilon_{1}- \epsilon_2 )
 }
 \frac{  \theta_I(x \pm \tfrac{\epsilon_{2}}{2} + \epsilon_2) }{ \theta_I(x \pm 
\tfrac{\epsilon_{2}}{2} ) }
 \bigg) \notag \\
&\qquad \qquad  +
\frac{ 1
}{ \eta^{12}}
 \cdot  \frac{ \thone(2\epsilon_+) \; \prod_{l=1}^{8}
 \thone(\mu_l \pm x)
 }{
 \thone(\epsilon_{1} )
 \thone(2x )
 \thone(\epsilon_{1}\pm 2x )
 \thone(\epsilon_{2}- 2x )
 }
 \label{eq:O2_continous_Wilson_w_defect}
\end{align}
and the full $k=2$ elliptic genus is obtained by dividing the continuous 
contribution \eqref{eq:O2_continous_Wilson_w_defect} as well as discrete 
contributions \eqref{eq:O2_discrete_Wilson_w_defect} by the order of the 
respective Weyl group
\begin{align}
  Z_{k=2}^{\mathrm{E-str+def}}
  = \frac{1}{2} Z_{k=2,i=0}^{\mathrm{E-str+def}}
  + \frac{1}{4 }\sum_{i=1}^6 Z_{k=2,i}^{\mathrm{E-str+def}}
  \,.
\end{align} 
\paragraph{Normalised partition function in NS limit.}
Following appendix \ref{app:normalised_partition_fct}, the normalised 2-string contribution is computed as follows:
\begin{align}
  \widetilde{Z}_{2}^{\mathrm{E-str+def}}\equiv
 Z_{2}^{\mathrm{E-str+def}} -Z_{2}^{\mathrm{E-str}}   
-Z_{1}^{\mathrm{E-str}} \left(Z_{1}^{\mathrm{E-str+def}} 
-Z_{1}^{\mathrm{E-str}} \right) \,.
\end{align}
One can verify via direct computation, for instance using \texttt{Mathematica}, 
that the normalised 2-instanton contribution in the NS-limit is given by \eqref{eq:2-string_normalised}.

\subsection{Elliptic genera of Wilson surface}
\label{app: Wilson_surface_at_two_instanton}
In this appendix, the details of the Wilson surface computation at 2-instanton order are provided. Similar to the computation in Section \ref{sec:DH_elliptic_genus_Sp1}, the contributions of the defect partition function can be divided into the discrete and continuous holonomies, respectively.
\paragraph{Discrete sector.}
There are six discrete sectors, where two eigenvalues $(u_+,u_-)$ of $u$ are given by
\begin{alignat}{3}
\begin{aligned}
(1):\; (u_+,u_-)&= (0,\tfrac{1}{2}) & \qquad
(2):\; (u_+,u_-)&= (\tfrac{\tau}{2},\tfrac{1+\tau}{2}) & \qquad 
(3):\; (u_+,u_-)&= (0,\tfrac{\tau}{2}) \\
(4):\; (u_+,u_-)&= (\tfrac{1}{2},\tfrac{1+\tau}{2}) & \qquad 
(5):\; (u_+,u_-)&= (0,\tfrac{1+\tau}{2}) & \qquad 
(6):\; (u_+,u_-)&= (\tfrac{1}{2},\tfrac{\tau}{2})
\end{aligned}
\end{alignat}
and 
\begin{align}
 e\in \mathrm{root} : \;
 e(u) \in \left\{ u_+ +u_- \right\} 
 \;,\quad 
\rho\in \mathrm{sym} : \;
\rho(u) \in \left\{ 2 u_+, 2u_-,u_++u_- \right\} 
\;, \quad 
\rho\in \mathrm{fund} : \;
\rho(u) \in \left\{ u_+, u_- \right\} \,.
\notag
\end{align}
and
\begin{align}
 W_{k=2}^{\mathrm{dis}}&\sim 
 \frac{\thone(u_+ + u_-) \thone(2\epsilon_+ +u_+ + u_-)}{\eta^{12}}
 \cdot  \frac{  \prod_{l=1}^{8}  \thone(\mu_l +u_+) \thone(\mu_l +u_-)  }{
 \thone(\epsilon_{1,2}+2u_+)
 \thone(\epsilon_{1,2}+2u_-)
 \thone(\epsilon_{1,2}+u_++u_-)
 }\notag\\
 &\cdot\frac{\thone(u_+-v\pm\epsilon_-)\thone(u_--v\pm\epsilon_-)}{\thone(u_+-v\pm\epsilon_+)\thone(u_--v\pm\epsilon_+)}\,,
\end{align}
where the term in second line is from the contribution of Wilson surface defect \eqref{Wilson defect}. Therefore, the contributions for the discrete sectors are given by
\begin{subequations}
\begin{align}
 W_{k=2,{i=1}}^{\mathrm{dis}}&= 
 \frac{\theta_2(0) \theta_2(2\epsilon_+)}{\eta^{12}}
 \cdot  \frac{  \prod_{l=1}^{8}  \thone(\mu_l ) \theta_2(\mu_l)  }{
 \thone(\epsilon_{1})^2
 \thone(\epsilon_{2})^2
 \theta_2(\epsilon_{1})
  \theta_2(\epsilon_{2})
 }\cdot \frac{\thone(v\pm\epsilon_-)\theta_2(v\pm\epsilon_-)}{\thone(v\pm\epsilon_+)\theta_2(v\pm\epsilon_+)} \;,\\
W_{k=2,{i=2}}^{\mathrm{dis}}&= 
 \frac{\theta_2(0) \theta_2(2\epsilon_+ )}{\eta^{12}}
 \cdot  \frac{  \prod_{l=1}^{8}  \theta_4(\mu_l ) \theta_3(\mu_l )  }{
 \thone(\epsilon_{1})^2
 \thone(\epsilon_{2})^2
 \theta_2(\epsilon_{1})
  \theta_2(\epsilon_{2})
 }\cdot \frac{\theta_3(v\pm\epsilon_-)\theta_4(v\pm\epsilon_-)}{\theta_3(v\pm\epsilon_+)\theta_4(v\pm\epsilon_+)}
\;, \\
 W_{k=2,{i=3}}^{\mathrm{dis}}&= 
 \frac{\theta_4(0) \theta_4(2\epsilon_+ )}{\eta^{12}}
 \cdot  \frac{  \prod_{l=1}^{8}  \thone(\mu_l ) \theta_4(\mu_l )  }{
 \thone(\epsilon_{1})^2
 \thone(\epsilon_{2})^2
 \theta_4(\epsilon_{1})
 \theta_4(\epsilon_{2})
 }\cdot \frac{\theta_1(v\pm\epsilon_-)\theta_4(v\pm\epsilon_-)}{\theta_1(v\pm\epsilon_+)\theta_4(v\pm\epsilon_+)}
\;, \\
W_{k=2,{i=4}}^{\mathrm{dis}}&= 
 \frac{\theta_4(0) \theta_4(2\epsilon_+ )}{\eta^{12}}
 \cdot  \frac{  \prod_{l=1}^{8}  \theta_2(\mu_l ) \theta_3(\mu_l )  }{
 \thone(\epsilon_{1})^2
 \thone(\epsilon_{2})^2
 \theta_4(\epsilon_{1})
 \theta_4(\epsilon_{2})
 }\cdot \frac{\theta_2(v\pm\epsilon_-)\theta_3(v\pm\epsilon_-)}{\theta_2(v\pm\epsilon_+)\theta_3(v\pm\epsilon_+)}
\;, \\
W_{k=2,{i=5}}^{\mathrm{dis}}&= 
 \frac{\theta_3(0) \theta_3(2\epsilon_+ )}{\eta^{12}}
 \cdot  \frac{  \prod_{l=1}^{8}  \thone(\mu_l ) \theta_3(\mu_l )  }{
 \thone(\epsilon_{1})^2
 \thone(\epsilon_{2})^2
 \theta_3(\epsilon_{1})
 \theta_3(\epsilon_{2})
 }\cdot \frac{\theta_1(v\pm\epsilon_-)\theta_3(v\pm\epsilon_-)}{\theta_1(v\pm\epsilon_+)\theta_3(v\pm\epsilon_+)}
\;, \\
W_{k=2,{i=6}}^{\mathrm{dis}}&= 
 \frac{\theta_3(0) \theta_3(2\epsilon_+ )}{\eta^{12}}
 \cdot  \frac{  \prod_{l=1}^{8}  \theta_2(\mu_l ) \theta_4(\mu_l )  }{
 \thone(\epsilon_{1})^2
 \thone(\epsilon_{2})^2
 \theta_3(\epsilon_{1})
 \theta_3(\epsilon_{2})
 }\cdot \frac{\theta_2(v\pm\epsilon_-)\theta_4(v\pm\epsilon_-)}{\theta_2(v\pm\epsilon_+)\theta_4(v\pm\epsilon_+)}\;.
\end{align}
\end{subequations}
\paragraph{Continuous sector.} In the continuous sector, the contour integral is given by
\begin{align}
 W_{k=2}^{\mathrm{cont}}&= \oint  \diff u \frac{\thone(2\epsilon_+) 
}{i \eta^9}
 \cdot  \frac{\prod_{l=1}^{8}
 \thone(\mu_l \pm u)
 }{
 \thone(\epsilon_{1} )
 \thone(\epsilon_{2} )
 \thone(\epsilon_{1}\pm 2u )
 \thone(\epsilon_{2}\pm 2u )
 }\cdot\frac{\thone(\pm u-v\pm\epsilon_-)}{\thone(\pm u-v\pm\epsilon_+)} \,.
 \end{align}
 The relevant poles for the JK-residue prescription are
 \begin{align}
 u\in \left\{  
-\tfrac{\epsilon_{1,2}}{2},-\tfrac{\epsilon_{1,2}}{2}+\tfrac{1}{2}, 
-\tfrac{\epsilon_{1,2}}{2} + \tfrac{1+\tau}{2} , -\tfrac{\epsilon_{1,2}}{2} + 
\tfrac{\tau}{2},\, v\pm\epsilon_+ \right\}
\end{align}
and the contributions of the individual poles are as follows:
\begin{subequations}
\begin{compactitem}
\item for $u=-\tfrac{\epsilon_{1,2}}{2}$
\begin{align}
 W_{k=2, i=1}^{\mathrm{cont}}&=\frac{1 
}{ 2\eta^{12}\, 
\thone(\epsilon_{1} )
 \thone(\epsilon_{2} )}
  \cdot \left(  \frac{\prod_{l=1}^{8}
 \thone(\mu_l \pm \tfrac{\epsilon_{1}}{2})
 }{
  \thone(2\epsilon_{1} )
 \thone(\epsilon_{2}- \epsilon_{1} )
 }\cdot\frac{\theta_1(v+\epsilon_1-\frac{\epsilon_2}{2}) \theta_1(v-\epsilon_1+\frac{\epsilon_2}{2})}{\theta_1(v+\epsilon_1+\frac{\epsilon_2}{2}) \theta_1(v-\epsilon_1-\frac{\epsilon_2}{2})} 
 +  \left(\epsilon_1\leftrightarrow \epsilon_2\right) \right)
\end{align}
\item for $u = -\tfrac{\epsilon_{1,2}}{2}+\tfrac{1}{2}$
\begin{align}
W_{k=2, i=2}^{\mathrm{cont}}&=
\frac{1}{ 2\eta^{12} \,  
\thone(\epsilon_{1} )
 \thone(\epsilon_{2} )}
 \cdot  
 \left(
 \frac{\prod_{l=1}^{8}
 \theta_2(\mu_l \pm \tfrac{\epsilon_{1}}{2} )
 }{
 \thone(2\epsilon_{1} )
 \thone(\epsilon_{2}- \epsilon_1 )
 }\cdot\frac{\theta_2(v+\epsilon_1-\frac{\epsilon_2}{2}) \theta_2(v-\epsilon_1+\frac{\epsilon_2}{2})}{\theta_2(v+\epsilon_1+\frac{\epsilon_2}{2}) \theta_2(v-\epsilon_1-\frac{\epsilon_2}{2})} 
 +  \left(\epsilon_1\leftrightarrow \epsilon_2\right) \right)
\end{align}
\item for $u=-\tfrac{\epsilon_{1,2}}{2} + \tfrac{1+\tau}{2}$
\begin{align}
 W_{k=2, i=3}^{\mathrm{cont}}&=
 \frac{1}{ 2\eta^{12}\,  
 \thone(\epsilon_{1} )
 \thone(\epsilon_{2} )}
 \cdot 
 \left( \frac{\prod_{l=1}^{8}
 \theta_3(\mu_l \pm \tfrac{\epsilon_{1}}{2})
 }{
  \thone(2\epsilon_{1} )
 \thone(\epsilon_{2}- \epsilon_1 )
 }\cdot\frac{\theta_3(v+\epsilon_1-\frac{\epsilon_2}{2}) \theta_3(v-\epsilon_1+\frac{\epsilon_2}{2})}{\theta_3(v+\epsilon_1+\frac{\epsilon_2}{2}) \theta_3(v-\epsilon_1-\frac{\epsilon_2}{2})}
 +  \left(\epsilon_1\leftrightarrow \epsilon_2\right)
 \right)
\end{align}
\item for $u = -\tfrac{\epsilon_{1,2}}{2}+\tfrac{\tau}{2}$
\begin{align}
W_{k=2, i=4}^{\mathrm{cont}}&=
\frac{1}{ 2\eta^{12} \,  
\thone(\epsilon_{1} )
 \thone(\epsilon_{2} )}
 \cdot  
 \left(
 \frac{\prod_{l=1}^{8}
 \theta_4(\mu_l \pm \tfrac{\epsilon_{1}}{2} )
 }{
 \thone(2\epsilon_{1} )
 \thone(\epsilon_{2}- \epsilon_1 )
 }\cdot\frac{\theta_4(v+\epsilon_1-\frac{\epsilon_2}{2}) \theta_4(v-\epsilon_1+\frac{\epsilon_2}{2})}{\theta_4(v+\epsilon_1+\frac{\epsilon_2}{2}) \theta_4(v-\epsilon_1-\frac{\epsilon_2}{2})}
 +  \left(\epsilon_1\leftrightarrow \epsilon_2\right)
 \right) 
\end{align}
\item for $u = v+\epsilon_+$
\begin{align}
W_{k=2, i=5}^{\mathrm{cont}}&=
\frac{\prod_{l=1}^{8}
 \theta_1(v\pm\mu_l+\epsilon_+)}{\eta^{12}\theta_1(2v+2\epsilon_{1,2})\theta_1(2v+\epsilon_1+\epsilon_2)\theta_1(2v+\epsilon_1+2\epsilon_2) \,}
\cdot\frac{\theta_1(2v+\epsilon_{1,2})}{\theta_1(2v)\theta_1(2v+2\epsilon_+)}\,.
\end{align}
\item for $u = v-\epsilon_+$
\begin{align}
W_{k=2, i=6}^{\mathrm{cont}}&=
\frac{\prod_{l=1}^{8}
 \theta_1(v\pm\mu_l-\epsilon_+)}{\eta^{12}\theta_1(2v-\epsilon_{1,2})\theta_1(2v-2\epsilon_1-\epsilon_2)\theta_1(2v-\epsilon_1-2\epsilon_2) \,}
\cdot\frac{\theta_1(2v-\epsilon_{1,2})}{\theta_1(2v)\theta_1(2v-2\epsilon_+)}
\,.
\end{align}
\end{compactitem}
\end{subequations}
As a result, the 2-string elliptic genus for the Wilson surface is computed by suitably summing up the individual parts as 
\begin{align}
W_{k=2}=\frac{1}{2}\sum_{i=1}^6 W_{k=2, i}^{\mathrm{cont}}+\frac{1}{4}\sum_{i=1}^6 W_{k=2, i}^{\mathrm{dis}}
\end{align}
Lastly, one computes the NS-limit, $\epsilon_2\rightarrow 0$, of the normalised Wilson surface partition function, $\widetilde W_2$, at 2-instanton order
\begin{align}
\widetilde W_2&=\lim_{\epsilon_2\rightarrow 0}\left(W_{k=2}-Z_{k=2}-Z_{k=1}\left(W_{k=1}-Z_{k=1}\right)\right) \notag\\
&=\frac{\prod_{l=1}^{8}
 \theta_1(v\pm\mu_l+\frac{\epsilon_1}{2})}{2\eta^{12}\theta_1(2v)\theta_1^2(2v+\epsilon_1)\theta_1(2v+2\epsilon_1)}
+\frac{\prod_{l=1}^{8}
 \theta_1(v\pm\mu_l-\frac{\epsilon_1}{2})}{2\eta^{12}\theta_1(2v)\theta_1^2(2v-\epsilon_1)\theta_1(2v-2\epsilon_1)}\notag\\  
&+\sum_{I=1}^4\frac{\prod_{l=1}^8\theta_I(\mu_l\pm\frac{\epsilon_1}{2})}{4\eta^{12}\theta_1^2(\epsilon_1)\theta_1(2\epsilon_1)\theta_1^\prime(0)}\cdot\left(\frac{\theta_I^\prime(v+\epsilon_1)}{\theta_I(v+\epsilon_1)}-\frac{\theta_I^\prime(v-\epsilon_1)}{\theta_I(v-\epsilon_1)}\right)
\notag\\
&+\sum_{I=1}^4\frac{\prod_{l=1}^8\theta_I^2(\mu_l)}{8\eta^{12}\theta^2_1(\epsilon_1)\theta_1^{\prime 2}(0)}\cdot\theta_I^\Delta(v)^2
+\sum_{I=1}^4\frac{\prod_{l=1}^8\theta^2_I(\mu_l)}{4\eta^{12}\theta_1^2(\epsilon_1)\theta_1^{\prime 2}(0)}\cdot\frac{\theta_1^\prime(\epsilon_1)}{\theta_1(\epsilon_1)}\cdot\theta_I^{\Delta}(v)\notag\\
&+\frac{\prod_{l=1}^8\theta_1(\mu_l)\theta_2(\mu_l)}{4\eta^{12}\theta_1^2(\epsilon_1)\theta_1^{\prime 2}(0)}\cdot\frac{\theta_2^\prime(\epsilon_1)}{\theta_2(\epsilon_1)}
\cdot\left(\theta_1^\Delta(v)+\theta_2^\Delta(v)\right)+\frac{\prod_{l=1}^8\theta_3(\mu_l)\theta_4(\mu_l)}{4\eta^{12}\theta_1^2(\epsilon_1)\theta_1^{\prime 2}(0)}\cdot\frac{\theta_2^\prime(\epsilon_1)}{\theta_2(\epsilon_1)}
\cdot\left(\theta_3^\Delta(v)+\theta_4^\Delta(v)\right)\notag\\
&+\frac{\prod_{l=1}^8\theta_1(\mu_l)\theta_3(\mu_l)}{4\eta^{12}\theta_1^2(\epsilon_1)\theta_1^{\prime 2}(0)}\cdot\frac{\theta_3^\prime(\epsilon_1)}{\theta_3(\epsilon_1)}
\cdot\left(\theta_1^\Delta(v)+\theta_3^\Delta(v)\right)
+\frac{\prod_{l=1}^8\theta_2(\mu_l)\theta_4(\mu_l)}{4\eta^{12}\theta_1^2(\epsilon_1)\theta_1^{\prime 2}(0)}\cdot\frac{\theta_3^\prime(\epsilon_1)}{\theta_3(\epsilon_1)}
\cdot\left(\theta_2^\Delta(v)+\theta_4^\Delta(v)\right)
\notag\\
&+\frac{\prod_{l=1}^8\theta_1(\mu_l)\theta_4(\mu_l)}{4\eta^{12}\theta_1^2(\epsilon_1)\theta_1^{\prime 2}(0)}\cdot\frac{\theta_4^\prime(\epsilon_1)}{\theta_4(\epsilon_1)}
\cdot\left(\theta_1^\Delta(v)+\theta_4^\Delta(v)\right)
+\frac{\prod_{l=1}^8\theta_2(\mu_l)\theta_3(\mu_l)}{4\eta^{12}\theta_1^2(\epsilon_1)\theta_1^{\prime 2}(0)}\cdot\frac{\theta_4^\prime(\epsilon_1)}{\theta_4(\epsilon_1)}
\cdot\left(\theta_2^\Delta(v)+\theta_3^\Delta(v)\right)
\notag\\
&+\sum_{I<J}^4\frac{\prod_{l=1}^8\theta_I(\mu_l)\theta_J(\mu_l)}{4\eta^{12}\theta_1^2(\epsilon_1)\theta_1^{\prime 2}(0)}
\theta_I^\Delta(v)\cdot \theta_J^\Delta(v)\,,
\end{align}
where the following convention has been used
\begin{align}
\theta_I^{\Delta}(v)\equiv\frac{\theta_I^\prime(v-\frac{\epsilon_1}{2})}{\theta_I(v-\frac{\epsilon_1}{2})}-\frac{\theta_I^\prime(v+\frac{\epsilon_1}{2})}{\theta_I(v+\frac{\epsilon_1}{2})}\,.
\end{align}

\subsection{\texorpdfstring{Wilson loop of 5d $\sprm(1)$ theory with 8 flavours}{Wilson loop of 5d Sp(1) theory with 8 flavours}}
\label{app:Wilson_loop}
\paragraph{1-instanton.}
For 1-instanton order, the $5$d $\sprm(1)$ partition function and Wilson line are determined by the discrete holonomies of $\orm(1)_\pm$ of the $1$d quantum mechanics in the ADHM construction. Therefore, one has
\begin{align}
\begin{aligned}
Z_1 &=\frac{1}{2}(Z_1^++Z_1^-),\\
W_1 &=\frac{1}{2}(W_1^++W_1^-),
\end{aligned}
\end{align}
where
\begin{align}
\begin{aligned}
Z_1^\pm &=Z_{1,\,{\rm vec}}^\pm\cdot Z_{1,\,{\rm fund}}^\pm, \\
W_1^\pm &=\mathrm{Ch}_1^\pm\cdot Z_{1,\,{\rm vec}}^\pm\cdot Z_{1,\,{\rm fund}}^\pm,
\end{aligned}
\end{align}
with
\begin{align}
&Z_{1,\,{\rm vec}}^+\cdot Z_{1,\,{\rm fund}}^+=\frac{\prod_{i=1}^8\sh(m_i)}{\sh(\epsilon_{1,2})\sh(\pm a+\epsilon_+)}\,,\ \ \  Z_{1,\,{\rm vec}}^-=\frac{\prod_{i=1}^8\ch(m_i)}{\sh(\epsilon_{1,2})\ch(\pm a+\epsilon_+)},
\end{align} 
and
\begin{align}
\mathrm{Ch}_1^\pm=\alpha+\frac{1}{\alpha}\mp\frac{(1-p)(1-q)}{\sqrt{pq}}\,,
\end{align}
as the equivariant Chern character of $\sprm(1)$, with $p=e^{\epsilon_1}$ and $q=e^{\epsilon_2}$. Also, it is convenient to define
\begin{align}
\sh(x)\equiv 2\sinh(x/2)=e^{x/2}-e^{-x/2}\,,\ \ \ \ch(x)\equiv 2\cosh(x/2)=e^{x/2}+e^{-x/2}\,.
\end{align}
From \eqref{eq:normalized_Wilson_loop}, one has
\begin{align}
\widetilde W_1
&=\lim_{q\rightarrow 1}\left(W_1-\left(\alpha+\frac{1}{\alpha}\right)Z_1\right)\notag\\
&=\chi_\mathbf{128}\cdot\alpha-\chi_\mathbf{128^\prime}\,\chi^{\mathfrak{su}_2}_{1/2}\cdot \alpha^2
+\chi_\mathbf{128}\,\chi^{\mathfrak{su}_2}_{1}\cdot \alpha^3
+\chi_\mathbf{128^\prime}\,\chi^{\mathfrak{su}_2}_{3/2}\cdot \alpha^4+\ldots\,.
\end{align}
\paragraph{2-instanton.} 
Next, consider the 2-instanton correction to Wilson loop. Note that there are no continuous holonomy contributions to $Z_2^-$. Therefore, one only has
\begin{align}
Z_2^+ &=\oint \! \diff \phi\, \frac{\sh(2\epsilon_+)}{\sh(\epsilon_{1,2})}\frac{\prod_{l=1}^8\sh(\pm \phi+m_l)}{\sh(\epsilon_+\pm \phi\pm \alpha)\sh(\pm 2\phi+\epsilon_{1,2})},\notag\\
Z_2 &=\frac{1}{2}Z_2^+\,,
\end{align}
and the integration is performed via the JK-residue prescription. However, there are higher order poles at infinities, which render the whole computation invalid. Therefore, following \cite{Gaiotto:2015una, Hwang:2014uwa}, one has to introduce additional anti-symmetric matter fields to resolve this singularity. For one anti-symmetric matter field, the extra contribution is
\begin{align}
Z_{\mathrm{asym}}=\frac{\sh(-\epsilon_-\pm \tau)\sh(\pm \phi\pm \alpha-\tau)}{\sh(-\epsilon_+\pm \tau)\sh(-\epsilon_+\pm 2\phi\pm \tau)}\,,
\end{align}
where $\mu=e^\tau$ is the mass of the anti-symmetric matter field. In the large $\phi$ limit, one finds
\begin{align}
Z_{\mathrm{asym}}\rightarrow \frac{e^{2\phi}}{e^{4\phi}}=e^{-2\phi}\,.
\end{align}
Therefore, the integrand with one anti-symmetric matter field is of order $\mathcal{O}(1)$ at infinity, which can be dealt with. In this situation, one can compute the partition function $Z_{\mathrm{QM}}(\mu)$. To extract the genuine 5d partition function, one needs to further strip off an extra factor, $Z_{\mathrm{extra}}$, and take the limit $\mu\rightarrow \infty$ to decouple the anti-symmetric matter field, i.e.
\begin{align}
Z_{\mathrm{5d\,QFT}}=\lim_{\mu\rightarrow \infty}\frac{Z_{\mathrm{QM}}(\mu)}{Z_{\mathrm{extra}}}.
\end{align}
To find the full order of $Z_{\mathrm{extra}}$, the reader is referred to \cite{Hwang:2014uwa}. For the purposes here, one only needs $Z_{\mathrm{extra}}$ up to $\mathcal O(U^2)$. For adding a single anti-symmetric matter field, one finds that
\begin{align}
Z_{\mathrm{extra}}=1+\left(\frac{\sqrt{p q}\mu}{(1-p)(1-q)}-\frac{1+p q}{2(1-p)(1-q)}\right)U^2+\mathcal O(U^2)\,,
\end{align}
where at $\mathcal O(U^2)$ the first term corresponds to the extra contribution from the anti-symmetric matter, and the second term originates from the continuum spectra at infinity.

One can follow a similar procedure to compute the Wilson loop. However, the subtlety is that, for the Wilson loop at 2-instanton order, one has to insert the equivariant Chern character,
\begin{align}
\mathrm{Ch}_2=\alpha+\frac{1}{\alpha}-\frac{(1-p)(1-q)}{\sqrt{p q}}(e^\phi+e^{-\phi})\,, 
\end{align}
into the numerator of the integrand. This modification further \emph{increases} the order of poles at infinity. Therefore, at 2-instanton order, one has to add \emph{two} anti-symmetric matter fields, with masses $\mu_{1, 2}=e^{\tau_{1,2}}$ respectively. With these two additional anti-symmetric matter fields, one can follow the JK-residue prescription to compute the partition function of the 1d quantum mechanic model, say $W_{QM}(\mu_1, \mu_2)$. Similarly to above, it is not the genuine 5d partition function yet. By studying the divergence of $W_{\mathrm{QM}}(\mu_1, \mu_2)$ for $\mu_1, \mu_2$ tending to infinity, one finds that
\begin{align}
W_{\mathrm{extra}}=1+\left(\left(\alpha+\frac{1}{\alpha}\right)\frac{(p q)^{3/2}}{(1-p)(1-q)}-\chi_\mathbf{16}\cdot pq\right)\frac{\mu_1+\mu_2}{(1-pq \mu_1/\mu_2)(1-pq \mu_2/\mu_1)}U^2+\mathcal O(U^2)\,.
\end{align}
Therefore, one obtains
\begin{align}
W_2=\lim_{\mu_1,\mu_2\rightarrow\infty}\frac{W_{\mathrm{QM}}(\mu_1,\mu_2)}{W_{\mathrm{extra}}}\,.
\end{align}
Lastly, computing the normalised Wilson loop and taking the NS-limit, one finds
\begin{align}
\widetilde W_2&=\lim_{q\rightarrow 1}\left(W_2-W_1\cdot Z_1+\left(\alpha+\frac{1}{\alpha}\right)\cdot(Z_1^2-Z_2)\right)\notag\\
&=\bigg(\chi_{4}^{\mathfrak{su}_2}+(\chi_\mathbf{120}+3)\chi_{2}^{\mathfrak{su}_2}+(\chi_\mathbf{1820}+3\chi_\mathbf{120}+6)\bigg)\cdot\alpha \notag\\
&-\bigg(\chi_\mathbf{16}\,\chi_{5}^{\mathfrak{su}}+(\chi_\mathbf{560}+3\chi_\mathbf{16})\chi_{3}^{\mathfrak{su}_2}+(\chi_\mathbf{4368}+3\chi_\mathbf{560}+7\chi_\mathbf{16})\chi_{1}^{\mathfrak{su}_2}\bigg)\cdot\alpha^2 \notag\\
&+\bigg(\chi_{8}^{\mathfrak{su}_2}+(\chi_\mathbf{120}+3)\chi_{6}^{\mathfrak{su}_2}+(\chi_\mathbf{1820}+3\chi_\mathbf{120}+7)\chi_4^{\mathfrak{su}_2}\notag\\
&\ \ \ \ +\left(\chi_\mathbf{8008}+3\chi_\mathbf{1820}+7\chi_\mathbf{120}+13\right)\chi_{2}^{\mathfrak{su}_2}+(\chi_\mathbf{1820}+2\chi_\mathbf{120}+4)\bigg)\cdot\alpha^3 \notag\\
&-\bigg(\chi_\mathbf{16}\,\chi_{9}^{\mathfrak{su}_2}+(\chi_\mathbf{560}+3\chi_\mathbf{16})\chi_{7}^{\mathfrak{su}_2}+(\chi_\mathbf{4368}+3\chi_\mathbf{560}+7\chi_\mathbf{16})\chi_{5}^{\mathfrak{su}_2}\notag\\
&\ \ \ \ +\left(\chi_\mathbf{11440}+3\chi_\mathbf{4368}+7\chi_\mathbf{560}+13\chi_\mathbf{16}\right)\chi_{3}^{\mathfrak{su}_2}+(\chi_\mathbf{4368}+3\chi_\mathbf{560}+6\chi_\mathbf{16})\chi_{1}^{\mathfrak{su}_2}\bigg)\cdot\alpha^4 \notag\\
&+\bigg(\chi_{12}^{\mathfrak{su}_2}+(\chi_\mathbf{120}+3)\chi_{10}^{\mathfrak{su}_2}+(\chi_\mathbf{1820}+3\chi_\mathbf{120}+7)\chi_{8}^{\mathfrak{su}_2}+(\chi_\mathbf{8008}+3\chi_\mathbf{1820}+7\chi_\mathbf{120}+13)\chi_{6}^{\mathfrak{su}_2}\notag\\
&\ \ \ \ +\left(\chi_\mathbf{6435}+\chi_\mathbf{6435^\prime}+3\chi_\mathbf{8008}+7\chi_\mathbf{1820}+13\chi_\mathbf{120}+22\right)\chi_{4}^{\mathfrak{su}_2}+\left(\chi_\mathbf{8008}+3\chi_\mathbf{1820}+7\chi_\mathbf{120}+12\right)\chi_{2}^{\mathfrak{su}_2}\notag\\
&\ \ \ \ +\left(\chi_\mathbf{1820}+2\chi_\mathbf{120}+4\right)\bigg)\cdot\alpha^5 \notag\\
&-\bigg(\chi_\mathbf{16}\,\chi_{13}^{\mathfrak{su}_2}+(\chi_\mathbf{560}+3\chi_\mathbf{16})\chi_{11}^{\mathfrak{su}_2}+(\chi_\mathbf{4368}+3\chi_\mathbf{560}+7\chi_\mathbf{16})\chi_{9}^{\mathfrak{su}_2}\notag\\
&\ \ \ \ +\left(\chi_\mathbf{11440}+3\chi_\mathbf{4368}+7\chi_\mathbf{560}+13\chi_\mathbf{16}\right)\chi_{7}^{\mathfrak{su}_2}+\left(4\chi_\mathbf{11440}+7\chi_\mathbf{4368}+13\chi_\mathbf{560}+22\chi_\mathbf{16}\right)\chi_{5}^{\mathfrak{su}_2}\notag\\
&\ \ \ \ +\left(\chi_\mathbf{11440}+3\chi_\mathbf{4368}+7\chi_\mathbf{560}+13\chi_\mathbf{16}\right)\chi_{3}^{\mathfrak{su}_2}+\left(\chi_\mathbf{4368}+3\chi_\mathbf{560}+6\chi_\mathbf{16}\right)\chi_{1}^{\mathfrak{su}_2}\bigg)\cdot\alpha^6 +\cdots\,.
\end{align}


\section{Open topological strings and refined Ooguri-Vafa invariants}
\label{app:OV_invariant}
In this appendix, we understand the co-dimension 2 defect partition function as an open refined topological string partition function
\be
Z^{\text{top}}_{\text{open}}=Z^{\text{pert}}\left(1+\sum_{k=1}^{\infty}q_{\phi}^k Z_k^{\text{def}}\right),
\ee
and extract the refined Ooguri-Vafa (OV) invariants, as a generalisation of the disk invariants \cite{Aganagic:2000gs,Aganagic:2001nx}, from the free energy $F^{\text{top}}_{\text{open}}({t},x)=\log Z^{\text{top}}_{\text{open}}({t},x)$.   
The all genus refined open string free energy has a close form expression from refined Chern-Simons theory \cite{Ooguri:1999bv}\cite{Aganagic:2012hs} with the insertion of an $\epsilon_2$-brane
\be
F^{\text{top}}_{\text{open}}({t},\ep_{1,2},x)=-\sum_{n=1}^{\infty}\sum_{s_1,s_2,{d}}\sum_{m\neq 0} D^{s_1,s_2}_{m,{d}}\frac{p^{n s_1} q^{ns_2}}{n(p^{-\frac{n}{2}}-p^{\frac{n}{2}})} e^{n{d}\cdot{t}}e^{mn x}.
\ee
In our case, the defect partition function has $\sorm(16)$ symmetry, we can use $\sorm(16)$ characters $\chi$ to expand the free energy
\begin{align}
F^{\text{top}}_{\text{open}}({t},\ep_{1,2},x)&=-\sum_{n=1}^{\infty}\sum_{{d_{b},d_f,\chi}}\sum_{s_1,s_2}\sum_{m\neq 0} {D}^{s_1,s_2}_{m,{{d_{b},d_f,\chi}}}\frac{p^{ns_1}q^{ns_2} }{n(p^{-\frac{n}{2}}-p^{\frac{n}{2}})} e^{n({d_b}{t_b}+d_f\tau)}e^{mn x}\sum_{\vec{\omega}\in \chi}\exp(n\vec{\omega}\cdot\vec{\mu}).
\end{align}
where we use the notation $t=\{\phi,\tau,\vec{\mu}\}$ and $t_b=\phi-\frac{\tau}{2}$. In practice, we observe that there exist an new variable $\widehat{D}^{s_1,s_2}_{m,{{d_{b},d_f,\chi}}}=\mathrm{sgn}(m)(-1)^{2s_1+2s_2+d_b+2(\vec{\lambda}_{\mathbf{128}},\vec{\omega})}{D}^{s_1,s_2}_{m,{{d_{b},d_f,\chi}}}$ always has positive value. Here $\vec{\lambda}_{\mathbf{128}}$ is the highest weight of representation $\chi_{{\mathbf{128}}}$, and $(\vec{\lambda}_{\mathbf{128}},\vec{\omega})$ is the scalar product of two weights of $\sorm(16)$. We list the refined OV invariants in Table \ref{BPStable1}, \ref{BPStable2}, \ref{BPStable3}, \ref{BPStable4}.

\begin{table}[ht]
\begin{center}
\vspace{5mm}
\begin{minipage}{0.26\linewidth}
{\centering\footnotesize
\begin{tabular}{|c|c|} \hline 
$2s_2 \backslash 2s_1$&0\\ \hline -1&1\\ \hline
\multicolumn{2}{c}{$\beta= (t_b+\tau-2x,\chi_{\mathbf{1}})$,} \\  
\multicolumn{2}{c}{$(t_b+\tau-x,\chi_{\mathbf{16}})$,} \\  
\multicolumn{2}{c}{$(t_b+2\tau-4x,\chi_{\mathbf{1}})$,} \\  
\multicolumn{2}{c}{$(t_b+2\tau-3x,\chi_{\mathbf{16}})$,} \\  
\multicolumn{2}{c}{$(t_b+2\tau-2x,\chi_{\mathbf{120}})$,} \\  
\multicolumn{2}{c}{$(t_b+2\tau-x,\chi_{\mathbf{560}})$
}\end{tabular}}
\end{minipage}\begin{minipage}{0.26\linewidth}
{\centering\footnotesize
\begin{tabular}{|c|c|} \hline 
$2s_2 \backslash 2s_1$&0\\ \hline 0&1\\ \hline
\multicolumn{2}{c}{$\beta= (t_b+\tau+x,\chi_{\mathbf{128^\prime}})$,} \\  
\multicolumn{2}{c}{$(t_b+\tau+2x,\chi_{\mathbf{128}})$,} \\  
\multicolumn{2}{c}{$(t_b+\tau+3x,\chi_{\mathbf{128^\prime}})$,} \\  
\multicolumn{2}{c}{$(t_b+2\tau+x,\chi_{\mathbf{1920^\prime}})$,} \\  
\multicolumn{2}{c}{$(t_b+2\tau+2x,\chi_{\mathbf{1920}})$,} \\  
\multicolumn{2}{c}{$(t_b+2\tau+3x,\chi_{\mathbf{1920^\prime}})$
}\end{tabular}}
\end{minipage}\begin{minipage}{0.23\linewidth}
{\centering\footnotesize
\begin{tabular}{|c|c|} \hline 
$2s_2 \backslash 2s_1$&0\\ \hline 1&1\\ \hline
\multicolumn{2}{c}{$\beta= (t_b+\tau+x,\chi_{\mathbf{16}})$,} \\  
\multicolumn{2}{c}{$(t_b+\tau+2x,\chi_{\mathbf{1}})$,} \\  
\multicolumn{2}{c}{$(t_b+2\tau+x,\chi_{\mathbf{560}})$,} \\  
\multicolumn{2}{c}{$(t_b+2\tau+2x,\chi_{\mathbf{120}})$,} \\  
\multicolumn{2}{c}{$(t_b+2\tau+3x,\chi_{\mathbf{16}})$
}\end{tabular}}
\end{minipage}\vspace{5mm}\begin{minipage}{0.26\linewidth}
{\centering\footnotesize
\begin{tabular}{|c|c|} \hline 
$2s_2 \backslash 2s_1$&0\\ \hline -2&1\\ \hline
\multicolumn{2}{c}{$\beta= (t_b+2\tau-x,\chi_{\mathbf{128^\prime}})$}
\end{tabular}}
\end{minipage}
\vspace{5mm}
\begin{minipage}{0.34\linewidth}
{\centering\footnotesize
\begin{tabular}{|c|ccccc|} \hline 
$2s_2 \backslash 2s_1$&-2&-1&0&1&2\\ \hline -1&1& &2& &1\\ \hline
\multicolumn{6}{c}{$\beta= (t_b+2\tau-2x,\chi_{\mathbf{1}})$
}\end{tabular}}
\end{minipage}\begin{minipage}{0.34\linewidth}
{\centering\footnotesize
\begin{tabular}{|c|ccccc|} \hline 
$2s_2 \backslash 2s_1$&-2&-1&0&1&2\\ \hline -3& & &1& & \\ -2& & & & & \\ -1&1& &3& &1\\ \hline
\multicolumn{6}{c}{$\beta= (t_b+2\tau-x,\chi_{\mathbf{16}})$
}\end{tabular}}
\end{minipage}\begin{minipage}{0.34\linewidth}
{\centering\footnotesize
\begin{tabular}{|c|ccccc|} \hline 
$2s_2 \backslash 2s_1$&-2&-1&0&1&2\\ \hline 1&1& &2& &1\\ \hline
\multicolumn{6}{c}{$\beta= (t_b+2\tau+2x,\chi_{\mathbf{1}})$
}\end{tabular}}
\end{minipage}
\vspace{5mm}
\begin{minipage}{0.34\linewidth}
{\centering\footnotesize
\begin{tabular}{|c|ccccc|} \hline 
$2s_2 \backslash 2s_1$&-2&-1&0&1&2\\ \hline 0&1& &3& &1\\ 1& & & & & \\ 2& & &1& & \\ \hline
\multicolumn{6}{c}{$\beta= (t_b+2\tau+x,\chi_{\mathbf{128^\prime}})$
}\end{tabular}}
\end{minipage}\begin{minipage}{0.34\linewidth}
{\centering\footnotesize
\begin{tabular}{|c|ccccc|} \hline 
$2s_2 \backslash 2s_1$&-2&-1&0&1&2\\ \hline 1&1& &3& &1\\ 2& & & & & \\ 3& & &1& & \\ \hline
\multicolumn{6}{c}{$\beta= (t_b+2\tau+x,\chi_{\mathbf{16}})$
}\end{tabular}}
\end{minipage}\begin{minipage}{0.34\linewidth}
{\centering\footnotesize
\begin{tabular}{|c|ccccc|} \hline 
$2s_2 \backslash 2s_1$&-2&-1&0&1&2\\ \hline 0&1& &3& &1\\ \hline
\multicolumn{6}{c}{$\beta= (t_b+2\tau+2x,\chi_{\mathbf{128}})$,} \\  
\multicolumn{6}{c}{$(t_b+2\tau+3x,\chi_{\mathbf{128^\prime}})$
}\end{tabular}}
\end{minipage}
\caption{Refined OV invariants $\widehat{D}^{s_1,s_2}_{m,{{d_{b},d_f,\chi}}}$ for the class $\beta=(d_bt_b+d_f\tau+mx,\chi)$ with $d_b=1,\, d_f\leq 2 $ and $m \leq 3$.}
\label{BPStable1}
\end{center}
\end{table}

\begin{table}[p]
\begin{center}
\begin{minipage}{0.33\linewidth}
{\centering\footnotesize
\begin{tabular}{|c|c|} \hline 
$2s_2 \backslash 2s_1$&-1\\ \hline -2&1\\ \hline
\multicolumn{2}{c}{$\beta= (2t_b+2\tau-4x,\chi_{\mathbf{1}})$,} \\  
\multicolumn{2}{c}{$(2t_b+2\tau-3x,\chi_{\mathbf{16}})$,} \\  
\multicolumn{2}{c}{$(2t_b+2\tau-2x,\chi_{\mathbf{120}})$,} \\  
\multicolumn{2}{c}{$(2t_b+2\tau-x,\chi_{\mathbf{560}})$,} \\  
\multicolumn{2}{c}{$(2t_b+3\tau-4x,\chi_{\mathbf{135}})$,} \\  
\multicolumn{2}{c}{$(2t_b+3\tau-3x,\chi_{\mathbf{1344}})$,} \\  
\multicolumn{2}{c}{$(2t_b+3\tau-2x,\chi_{\mathbf{7020}})$,} \\  
\multicolumn{2}{c}{$(2t_b+3\tau-x,\chi_{\mathbf{24192}})$
}\end{tabular}}\vspace{5mm}
\end{minipage}\begin{minipage}{0.33\linewidth}
{\centering\footnotesize
\begin{tabular}{|c|ccc|} \hline 
$2s_2 \backslash 2s_1$&-3&-2&-1\\ \hline -2&1& &1\\ \hline
\multicolumn{4}{c}{$\beta= (2t_b+2\tau-2x,\chi_{\mathbf{1}})$,} \\  
\multicolumn{4}{c}{$(2t_b+3\tau-6x,\chi_{\mathbf{1}})$
}\end{tabular}}
\end{minipage}\begin{minipage}{0.33\linewidth}
{\centering\footnotesize
\begin{tabular}{|c|ccc|} \hline 
$2s_2 \backslash 2s_1$&-1&0&1\\ \hline -3&1& & \\ -2& & & \\ -1& & &1\\ \hline
\multicolumn{4}{c}{$\beta= (2t_b+2\tau-x,\chi_{\mathbf{128^\prime}})$,} \\  
\multicolumn{4}{c}{$(2t_b+3\tau-3x,\chi_{\mathbf{128^\prime}})$,} \\  
\multicolumn{4}{c}{$(2t_b+3\tau-2x,\chi_{\mathbf{1920}})$,} \\  
\multicolumn{4}{c}{$(2t_b+3\tau-x,\chi_{\mathbf{13312^\prime}})$
}\end{tabular}}
\end{minipage}
\vspace{5mm}
\begin{minipage}{0.33\linewidth}
{\centering\footnotesize
\begin{tabular}{|c|ccccc|} \hline 
$2s_2 \backslash 2s_1$&-3&-2&-1&0&1\\ \hline -4& & &1& & \\ -3& & & & & \\ -2&1& &2& &1\\ -1& & & & & \\ 0& & & & &1\\ \hline
\multicolumn{6}{c}{$\beta= (2t_b+2\tau-x,\chi_{\mathbf{16}})$
}\end{tabular}}
\end{minipage}\begin{minipage}{0.33\linewidth}
{\centering\footnotesize
\begin{tabular}{|c|c|} \hline 
$2s_2 \backslash 2s_1$&1\\ \hline 0&1\\ \hline
\multicolumn{2}{c}{$\beta= (2t_b+2\tau+x,\chi_{\mathbf{4368}})$,} \\  
\multicolumn{2}{c}{$(2t_b+3\tau+x,\chi_{\mathbf{112320}})$
}\end{tabular}}
\end{minipage}\begin{minipage}{0.33\linewidth}
{\centering\footnotesize
\begin{tabular}{|c|ccc|} \hline 
$2s_2 \backslash 2s_1$&-1&0&1\\ \hline -1&1& & \\ 0& & & \\ 1& & &1\\ \hline
\multicolumn{4}{c}{$\beta= (2t_b+2\tau+x,\chi_{\mathbf{1920^\prime}})$,} \\  
\multicolumn{4}{c}{$(2t_b+3\tau+x,\chi_{\mathbf{56320^\prime}})$,} \\  
\multicolumn{4}{c}{$(2t_b+3\tau+x,\chi_{\mathbf{15360^\prime}})$
}\end{tabular}}
\end{minipage}
\vspace{5mm}
\begin{minipage}{0.3\linewidth}
{\centering\footnotesize
\begin{tabular}{|c|ccccc|} \hline 
$2s_2 \backslash 2s_1$&-1&0&1&2&3\\ \hline 0&1& &1& &1\\ 1& & & & & \\ 2& & &1& & \\ \hline
\multicolumn{6}{c}{$\beta= (2t_b+2\tau+x,\chi_{\mathbf{560}})$
}\end{tabular}}
\end{minipage}\begin{minipage}{0.4\linewidth}
{\centering\footnotesize
\begin{tabular}{|c|ccccccc|} \hline 
$2s_2 \backslash 2s_1$&-3&-2&-1&0&1&2&3\\ \hline -1&1& &2& & & & \\ 0& & & & & & & \\ 1& & &1& &2& &1\\ 2& & & & & & & \\ 3& & & & &1& & \\ \hline
\multicolumn{8}{c}{$\beta= (2t_b+2\tau+x,\chi_{\mathbf{128^\prime}})$
}\end{tabular}}
\end{minipage}\begin{minipage}{0.3\linewidth}
{\centering\footnotesize
\begin{tabular}{|c|ccc|} \hline 
$2s_2 \backslash 2s_1$&-3&-2&-1\\ \hline -2&1& &2\\ \hline
\multicolumn{4}{c}{$\beta= (2t_b+3\tau-5x,\chi_{\mathbf{16}})$,} \\  
\multicolumn{4}{c}{$(2t_b+3\tau-4x,\chi_{\mathbf{120}})$,} \\  
\multicolumn{4}{c}{$(2t_b+3\tau-3x,\chi_{\mathbf{560}})$,} \\  
\multicolumn{4}{c}{$(2t_b+3\tau-2x,\chi_{\mathbf{1820}})$
}\end{tabular}}
\end{minipage}
\vspace{5mm}
\begin{minipage}{0.5\linewidth}
{\centering\footnotesize
\begin{tabular}{|c|ccccccccc|} \hline 
$2s_2 \backslash 2s_1$&-3&-2&-1&0&1&2&3&4&5\\ \hline 0&1& &2& &2& &1& &1\\ 1& & & & & & & & & \\ 2& & &1& &2& &1& & \\ 3& & & & & & & & & \\ 4& & & & &1& & & & \\ \hline
\multicolumn{10}{c}{$\beta= (2t_b+2\tau+x,\chi_{\mathbf{16}})$
}\end{tabular}}
\end{minipage}\begin{minipage}{0.5\linewidth}
{\centering\footnotesize
\begin{tabular}{|c|ccccccc|} \hline 
$2s_2 \backslash 2s_1$&-5&-4&-3&-2&-1&0&1\\ \hline -2&1& &3& &4& &1\\ \hline
\multicolumn{8}{c}{$\beta= (2t_b+3\tau-4x,\chi_{\mathbf{1}})$
}\end{tabular}}
\end{minipage}
\vspace{5mm}
\begin{minipage}{0.5\linewidth}
{\centering\footnotesize
\begin{tabular}{|c|ccccccc|} \hline 
$2s_2 \backslash 2s_1$&-5&-4&-3&-2&-1&0&1\\ \hline -4& & & & &1& & \\ -3& & & & & & & \\ -2&1& &4& &7& &2\\ -1& & & & & & & \\ 0& & & & & & &1\\ \hline
\multicolumn{8}{c}{$\beta= (2t_b+3\tau-3x,\chi_{\mathbf{16}})$,} \\  
\multicolumn{8}{c}{$(2t_b+3\tau-2x,\chi_{\mathbf{120}})$
}\end{tabular}}
\end{minipage}\begin{minipage}{0.5\linewidth}
{\centering\footnotesize
\begin{tabular}{|c|ccccc|} \hline 
$2s_2 \backslash 2s_1$&-3&-2&-1&0&1\\ \hline -4& & &1& & \\ -3& & & & & \\ -2&1& &3& &1\\ -1& & & & & \\ 0& & & & &1\\ \hline
\multicolumn{6}{c}{$\beta= (2t_b+3\tau-2x,\chi_{\mathbf{135}})$,} \\  
\multicolumn{6}{c}{$(2t_b+3\tau-x,\chi_{\mathbf{1344}})$
}\end{tabular}}
\end{minipage}
\vspace{5mm}
\begin{minipage}{0.5\linewidth}
{\centering\footnotesize
\begin{tabular}{|c|ccccccc|} \hline 
$2s_2 \backslash 2s_1$&-3&-2&-1&0&1&2&3\\ \hline -3&1& &2& & & & \\ -2& & & & & & & \\ -1& & & & &2& &1\\ \hline
\multicolumn{8}{c}{$\beta= (2t_b+3\tau-2x,\chi_{\mathbf{128}})$
}\end{tabular}}
\end{minipage}\begin{minipage}{0.5\linewidth}
{\centering\footnotesize
\begin{tabular}{|c|ccccccccccc|} \hline 
$2s_2 \backslash 2s_1$&-7&-6&-5&-4&-3&-2&-1&0&1&2&3\\ \hline -4& & & & &1& &2& & & & \\ -3& & & & & & & & & & & \\ -2&1& &3& &8& &10& &4& &1\\ -1& & & & & & & & & & & \\ 0& & & & & & & & &2& &1\\ \hline
\multicolumn{12}{c}{$\beta= (2t_b+3\tau-2x,\chi_{\mathbf{1}})$
}\end{tabular}}
\end{minipage}
\caption{Refined OV invariants $\widehat{D}^{s_1,s_2}_{m,{{d_{b},d_f,\chi}}}$ for the class $\beta=(d_bt_b+d_f\tau+mx,\chi)$ with $d_b=2,\, d_f\leq 3 $ and $m \leq 1$.}
\label{BPStable2}
\end{center}
\end{table}

\begin{table}[p]
\begin{center}
\begin{minipage}{0.39\linewidth}
{\centering\footnotesize
\begin{tabular}{|c|ccccc|} \hline 
$2s_2 \backslash 2s_1$&-3&-2&-1&0&1\\ \hline -4& & &1& & \\ -3& & & & & \\ -2&1& &2& &1\\ \hline
\multicolumn{6}{c}{$\beta= (2t_b+3\tau-x,\chi_{\mathbf{4368}})$
}\end{tabular}}
\end{minipage}\begin{minipage}{0.59\linewidth}
{\centering\footnotesize
\begin{tabular}{|c|ccccccccccc|} \hline 
$2s_2 \backslash 2s_1$&-5&-4&-3&-2&-1&0&1&2&3&4&5\\ \hline -7& & & & &1& & & & & & \\ -6& & & & & & & & & & & \\ -5& & &1& &3& &1& & & & \\ -4& & & & & & & & & & & \\ -3&1& &4& &8& &4& &1& & \\ -2& & & & & & & & & & & \\ -1& & & & &2& &8& &4& &1\\ 0& & & & & & & & & & & \\ 1& & & & & & &1& & & & \\ \hline
\multicolumn{12}{c}{$\beta= (2t_b+3\tau-x,\chi_{\mathbf{128^\prime}})$
}\vspace{5mm}\end{tabular}}
\end{minipage}
\vspace{5mm}
\begin{minipage}{0.44\linewidth}
{\centering\footnotesize
\begin{tabular}{|c|ccccccc|} \hline 
$2s_2 \backslash 2s_1$&-3&-2&-1&0&1&2&3\\ \hline -5& & &1& & & & \\ -4& & & & & & & \\ -3&1& &3& &1& & \\ -2& & & & & & & \\ -1& & & & &3& &1\\ \hline
\multicolumn{8}{c}{$\beta= (2t_b+3\tau-x,\chi_{\mathbf{1920^\prime}})$
}\end{tabular}}
\end{minipage}\begin{minipage}{0.56\linewidth}
{\centering\footnotesize
\begin{tabular}{|c|ccccccccc|} \hline 
$2s_2 \backslash 2s_1$&-5&-4&-3&-2&-1&0&1&2&3\\ \hline -6& & & & &1& & & & \\ -5& & & & & & & & & \\ -4& & &1& &3& &1& & \\ -3& & & & & & & & & \\ -2&1& &4& &8& &4& &1\\ -1& & & & & & & & & \\ 0& & & & & & &2& & \\ \hline
\multicolumn{10}{c}{$\beta= (2t_b+3\tau-x,\chi_{\mathbf{560}})$
}\end{tabular}}
\end{minipage}
\vspace{5mm}
\begin{minipage}{0.65\linewidth}
{\centering\footnotesize
\begin{tabular}{|c|ccccccccccccc|} \hline 
$2s_2 \backslash 2s_1$&-7&-6&-5&-4&-3&-2&-1&0&1&2&3&4&5\\ \hline -8& & & & & & &1& & & & & & \\ -7& & & & & & & & & & & & & \\ -6& & & & &1& &3& &1& & & & \\ -5& & & & & & & & & & & & & \\ -4& & &1& &4& &9& &4& &1& & \\ -3& & & & & & & & & & & & & \\ -2&1& &4& &12& &19& &12& &4& &1\\ -1& & & & & & & & & & & & & \\ 0& & & & & & &2& &7& &3& & \\ 1& & & & & & & & & & & & & \\ 2& & & & & & & & &1& & & & \\ \hline
\multicolumn{14}{c}{$\beta= (2t_b+3\tau-x,\chi_{\mathbf{16}})$
}\end{tabular}}
\end{minipage}\begin{minipage}{0.34\linewidth}
{\centering\footnotesize
\begin{tabular}{|c|ccccc|} \hline 
$2s_2 \backslash 2s_1$&-1&0&1&2&3\\ \hline 0&1& &2& &1\\ 1& & & & & \\ 2& & &1& & \\ \hline
\multicolumn{6}{c}{$\beta= (2t_b+3\tau+x,\chi_{\mathbf{24192}})$
}\end{tabular}}
\end{minipage}
\vspace{5mm}
\begin{minipage}{0.48\linewidth}
{\centering\footnotesize
\begin{tabular}{|c|ccccccc|} \hline 
$2s_2 \backslash 2s_1$&-3&-2&-1&0&1&2&3\\ \hline -1&1& &3& & & & \\ 0& & & & & & & \\ 1& & &1& &3& &1\\ 2& & & & & & & \\ 3& & & & &1& & \\ \hline
\multicolumn{8}{c}{$\beta= (2t_b+3\tau+x,\chi_{\mathbf{13312^\prime}})$
}\end{tabular}}
\end{minipage}\begin{minipage}{0.48\linewidth}
{\centering\footnotesize
\begin{tabular}{|c|ccccc|} \hline 
$2s_2 \backslash 2s_1$&-1&0&1&2&3\\ \hline -2&1& & & & \\ -1& & & & & \\ 0&1& &3& &1\\ 1& & & & & \\ 2& & &1& & \\ \hline
\multicolumn{6}{c}{$\beta= (2t_b+3\tau+x,\chi_{\mathbf{11440}})$
}\end{tabular}}
\end{minipage}
\vspace{5mm}
\begin{minipage}{0.45\linewidth}
{\centering\footnotesize
\begin{tabular}{|c|ccccccccc|} \hline 
$2s_2 \backslash 2s_1$&-3&-2&-1&0&1&2&3&4&5\\ \hline -2& & &1& & & & & & \\ -1& & & & & & & & & \\ 0&1& &4& &7& &4& &1\\ 1& & & & & & & & & \\ 2& & &1& &3& &1& & \\ 3& & & & & & & & & \\ 4& & & & &1& & & & \\ \hline
\multicolumn{10}{c}{$\beta= (2t_b+3\tau+x,\chi_{\mathbf{4368}})$
}\end{tabular}}
\end{minipage}\begin{minipage}{0.52\linewidth}
{\centering\footnotesize
\begin{tabular}{|c|ccccccccccc|} \hline 
$2s_2 \backslash 2s_1$&-5&-4&-3&-2&-1&0&1&2&3&4&5\\ \hline -3& & & & &1& & & & & & \\ -2& & & & & & & & & & & \\ -1&1& &5& &11& &2& & & & \\ 0& & & & & & & & & & & \\ 1& & &1& &5& &11& &5& &1\\ 2& & & & & & & & & & & \\ 3& & & & &1& &4& &1& & \\ 4& & & & & & & & & & & \\ 5& & & & & & &1& & & & \\ \hline
\multicolumn{12}{c}{$\beta= (2t_b+3\tau+x,\chi_{\mathbf{1920^\prime}})$
}\end{tabular}}
\end{minipage}
\caption{Refined OV invariants $\widehat{D}^{s_1,s_2}_{m,{{d_{b},d_f,\chi}}}$ for the class $\beta=(d_bt_b+d_f\tau+mx,\chi)$ with $d_b=2,\, d_f\leq 3 $ and $m \leq 1$, continued from Table \ref{BPStable2}.}
\label{BPStable3}
\end{center}
\end{table}

\begin{table}[p]
\begin{center}
\begin{minipage}{0.48\linewidth}
{\centering\footnotesize
\begin{tabular}{|c|ccccccccc|} \hline 
$2s_2 \backslash 2s_1$&-3&-2&-1&0&1&2&3&4&5\\ \hline 0&1& &3& &3& &2& &1\\ 1& & & & & & & & & \\ 2& & &1& &3& &1& & \\ 3& & & & & & & & & \\ 4& & & & &1& & & & \\ \hline
\multicolumn{10}{c}{$\beta= (2t_b+3\tau+x,\chi_{\mathbf{1344}})$
}\end{tabular}}
\end{minipage}

\vspace{5mm}
\begin{minipage}{0.61\linewidth}
{\centering\footnotesize
\begin{tabular}{|c|ccccccccccccc|} \hline 
$2s_2 \backslash 2s_1$&-5&-4&-3&-2&-1&0&1&2&3&4&5&6&7\\ \hline -2& & & & &1& & & & & & & & \\ -1& & & & & & & & & & & & & \\ 0&1& &4& &11& &13& &9& &4& &1\\ 1& & & & & & & & & & & & & \\ 2& & &1& &4& &9& &4& &1& & \\ 3& & & & & & & & & & & & & \\ 4& & & & &1& &3& &1& & & & \\ 5& & & & & & & & & & & & & \\ 6& & & & & & &1& & & & & & \\ \hline
\multicolumn{14}{c}{$\beta= (2t_b+3\tau+x,\chi_{\mathbf{560}})$
}\vspace{5mm}\end{tabular}}
\end{minipage}
\vspace{5mm}
\begin{minipage}{0.7\linewidth}
{\centering\footnotesize
\begin{tabular}{|c|ccccccccccccccc|} \hline 
$2s_2 \backslash 2s_1$&-7&-6&-5&-4&-3&-2&-1&0&1&2&3&4&5&6&7\\ \hline -3& & & & & & &1& & & & & & & & \\ -2& & & & & & & & & & & & & & & \\ -1&1& &4& &12& &19& &4& & & & & & \\ 0& & & & & & & & & & & & & & & \\ 1& & &1& &4& &12& &20& &12& &4& &1\\ 2& & & & & & & & & & & & & & & \\ 3& & & & &1& &4& &9& &4& &1& & \\ 4& & & & & & & & & & & & & & & \\ 5& & & & & & &1& &3& &1& & & & \\ 6& & & & & & & & & & & & & & & \\ 7& & & & & & & & &1& & & & & & \\ \hline
\multicolumn{16}{c}{$\beta= (2t_b+3\tau+x,\chi_{\mathbf{128^\prime}})$
}\end{tabular}}
\end{minipage}
\vspace{5mm}
\begin{minipage}{0.8\linewidth}
{\centering\footnotesize
\begin{tabular}{|c|ccccccccccccccccc|} \hline 
$2s_2 \backslash 2s_1$&-7&-6&-5&-4&-3&-2&-1&0&1&2&3&4&5&6&7&8&9\\ \hline -2& & & & & & &2& & & & & & & & & & \\ -1& & & & & & & & & & & & & & & & & \\ 0&1& &4& &12& &22& &21& &15& &9& &4& &1\\ 1& & & & & & & & & & & & & & & & & \\ 2& & &1& &4& &12& &20& &12& &4& &1& & \\ 3& & & & & & & & & & & & & & & & & \\ 4& & & & &1& &4& &9& &4& &1& & & & \\ 5& & & & & & & & & & & & & & & & & \\ 6& & & & & & &1& &3& &1& & & & & & \\ 7& & & & & & & & & & & & & & & & & \\ 8& & & & & & & & &1& & & & & & & & \\ \hline
\multicolumn{18}{c}{$\beta= (2t_b+3\tau+x,\chi_{\mathbf{16}})$
}\end{tabular}}
\end{minipage}
\caption{Refined OV invariants $\widehat{D}^{s_1,s_2}_{m,{{d_{b},d_f,\chi}}}$ for the class $\beta=(d_bt_b+d_f\tau+mx,\chi)$ with $d_b=2,\, d_f\leq 3 $ and $m \leq 1$, continued from Table \ref{BPStable3}.}
\label{BPStable4}
\end{center}
\end{table}

\clearpage
%
%
\newpage
 \bibliographystyle{JHEP}     
 {\footnotesize{\bibliography{references}}}
\end{document}